\documentclass[aps,prd,twocolumn,superscriptaddress,showpacs,preprintnumbers,nofootinbib,notitlepage,floatfix]{revtex4-2}
\usepackage{graphicx}
\usepackage{bm}
\usepackage[utf8]{inputenc}
\usepackage{multirow}
\usepackage{amsmath}
\usepackage{amsfonts}
\usepackage{amssymb}
\usepackage{color}
\usepackage{xcolor}
\usepackage{xspace}
\usepackage{braket}
\usepackage{units}
\usepackage{slashed}
\usepackage{multirow}
\usepackage[normalem]{ulem}
\usepackage{hyperref}
\usepackage{dcolumn}
\usepackage{float}

\graphicspath{ {plots/} }

\newcommand{\be}{\begin{equation}}
\newcommand{\ee}{\end{equation}}

\newcommand{\im}{\mathrm{Im}\,}
\newcommand{\re}{\mathrm{Re}\,}

\usepackage{bm} 

\renewcommand*{\d}[1][]{\mathop{\mathrm{d}^{#1}}\mkern-3mu}

\renewcommand{\Re}{\ensuremath{\textrm{Re}}}
\renewcommand{\Im}{\ensuremath{\textrm{Im}}}

\newcommand{\mevnospace}{\ensuremath{{\mathrm{\,Me\kern -0.1em V}}}}
\newcommand{\gevnospace}{\ensuremath{{\mathrm{\,Ge\kern -0.1em V}}}}
\newcommand{\tevnospace}{\ensuremath{{\mathrm{\,Te\kern -0.1em V}}}}
\newcommand{\mev}{\mevnospace\xspace}
\newcommand{\gev}{\gevnospace\xspace}

\newcommand{\sig}
{\ensuremath{\sigma/f_0(500)}\xspace}

\newcommand{\pipi}
{\ensuremath{\pi \pi \to \pi \pi}\xspace}

\usepackage{mfirstuc} 
\newcommand{\addReviewer}[2]{
  \expandafter\newcommand\csname #1\endcsname[1]{{\bf \color{#2} \capitalisewords{#1}:\,##1}}
  \expandafter\newcommand\csname #1cor\endcsname[2]{{\color{#2} \capitalisewords{#1}:\,\st{##1}{\,\bf ##2}}}
  \expandafter\newcommand\csname #1color\endcsname{\,#2}
}

\usepackage{tikz, marginnote} 
\newcommand{\checkedby}[1]{
\ifdefined\CROSSCHECKS
  \marginnote{
    \begin{tikzpicture}
      \foreach \x [count=\xi] in {#1} {
         \node[shape=circle,inner sep=0mm,
         minimum size=2mm,
         fill=\csname \x color\endcsname] at (\xi*3mm,0) {};
       }
    \end{tikzpicture}
  }
\else
\fi
}

\usepackage{soul,color}
\definecolor{chromeyellow}{rgb}{1.0, 0.65, 0.0}
\definecolor{DodgeBlue}{rgb}{0.118, 0.565,1.000}
\definecolor{asparagus}{rgb}{0.53, 0.66, 0.42}
\definecolor{cadmiumgreen}{rgb}{0.0, 0.42, 0.24}

\addReviewer{AR}{asparagus}
\addReviewer{JR}{red}
\addReviewer{JRE}{blue}

\newcommand{\ucm}{Departamento de F\'isica Te\'orica and IPARCOS, 
Universidad Complutense de Madrid, 
E-28040 Madrid, Spain.}

\begin{document}
\title{Global parametrizations of $\pi \pi$ scattering with dispersive constraints: \\
Beyond the S0 wave}
\preprint{IPARCOS-UCM-24-062}
\author{J.~R.~Pel\'aez}
\email{jrpelaez@ucm.es}\affiliation{\ucm}
\affiliation{Helmholtz-Institut für Strahlen- und Kernphysik,
Universität Bonn, D-53115 Bonn, Germany.
}
\author{P. Rab\'an}
\email{praban@ucm.es}\affiliation{\ucm}
\author{J.~Ruiz de Elvira}
\email{jacobore@ucm.es}\affiliation{\ucm}
\begin{abstract}
We provide new global parametrizations of \pipi scattering for the S2, P, D, F, and G partial waves up to at least 1.8 GeV, easy to implement for
phenomenological use. With earlier S0-wave parametrizations, slightly updated here, they reproduce previous partial wave dispersion analyses up to the $\pi \omega$ threshold. In addition, these new parametrizations have improved their description of recent P-wave data, the inelasticity in various waves, and their fulfillment of Roy-like and forward dispersion relations. The latter now test very high partial waves and have an improved matching with the Regge regime, extending those with P-wave contributions up to 1.6 GeV. Above 1.6 GeV and up to 1.8 GeV, or sometimes somewhat beyond, the parametrizations are simple unconstrained fits to data.
\end{abstract}

\maketitle

\section{Introduction} 

As the lightest mesons, two or more pions appear in most of the final states in hadronic processes. Their subsequent rescattering makes it particularly important
to have a precise and reliable 
description of pion-pion interactions.
Recently, this has become especially relevant given the
unprecedentedly high statistics attained in hadronic observables measured by experimental collaborations such as ALICE, \textit{BABAR}, Belle, and LHCb or to be carried out in future hadronic facilities like FAIR or the EIC. Of course, $\pi\pi$ scattering is also interesting by itself as a source of data for light meson spectroscopy and as a testing ground for the non-perturbative QCD regime and its spontaneous chiral symmetry breaking.
A renewed interest in pion-pion interactions has arisen from other fronts. Within lattice QCD, \pipi scattering partial waves with relatively low pion masses have been recently obtained~\cite{Wilson:2015dqa,Bali:2015gji,Bulava:2016mks,Fu:2016itp,Alexandrou:2017mpi,Andersen:2018mau,Erben:2019nmx,ExtendedTwistedMass:2019omo,Fischer:2020yvw,Rodas:2023gma,Boyle:2024hvv} and analyzed with various 
methods~\cite{Mai:2019pqr,Niehus:2020gmf,Molina:2020qpw,Rodas:2023nec,Cao:2023ntr}. Pion-pion scattering amplitudes have also become an active ground to test bootstrap techniques confronting them with data or effective theories \cite{Guerrieri:2018uew,Guerrieri:2024jkn,He:2024nwd}.

Experiments on \pipi scattering were mostly performed in the 1970s~\cite{Hyams:1973zf,Durusoy:1973aj,Losty:1973et,Cohen:1973yx,Protopopescu:1973sh,Grayer:1974cr,Hyams:1975mc,Hoogland:1977kt,Kaminski:1996da}.
The \pipi data were extracted as indirect measurements from the $\pi N\rightarrow \pi\pi N'$ process, which led to several data sets. Most often, they are inconsistent among themselves even if they come from the same experiment. Consequently, for many years it was considered enough to have a crude description of these data. The situation was somewhat different at low energies. Below the kaon mass, very precise data could be obtained from $K_{l4}$ decays~\cite{Rosselet:1976pu,Pislak:2001bf}, particularly those from the NA48/2 collaboration~\cite{Batley:2010zza}. In addition, the QCD low-energy effective theory, known as Chiral Perturbation Theory (ChPT)~\cite{Gasser:1983yg,Gasser:1984gg},  provides a systematic low-energy expansion in terms of masses and momenta of pions.
These are the ChPT degrees of freedom since, as the Nambu-Goldstone bosons of the spontaneous chiral symmetry breaking of QCD, there is a gap between their masses and those of other mesons.

However, ChPT alone cannot reproduce heavier resonances, possibly the most interesting for phenomenological applications, although it can be extended to this region by combining it with dispersion relations~\cite{Truong:1988zp,Dobado:1989qm,Dobado:1992ha,Dobado:1996ps} and unitarity constraints. These techniques are generically known as Unitarized ChPT (UChPT), which, in different versions or approximations, generate or reconstruct all \pipi elastic resonances below 1.2 GeV that were not originally present in the ChPT Lagrangian. These are the $\sigma/f_0(500)$ and the $\rho(770)$. This approach can be extended to kaons and etas and successful results exist for $\pi K$ or $\pi\pi\to K \bar K$ scattering, describing the
$\kappa/K_0^*(700)$, $K^*(892)$, $a_0(980)$, and the $f_0(980)$ resonances~\cite{Oller:1997ti,Oller:1998hw,GomezNicola:2001as,Pelaez:2004xp,Nieves:1999bx,Ledwig:2014cla}.  These methods provide analytic expressions, but they become very lengthy when dealing with coupled channels. 
Above 1.2 GeV, other resonances can be explicitly introduced with additional chiral lagrangians, resulting in a successful description of the data~\cite{Oller:1998zr}, albeit with the same caveats as before and with even more elaborate expressions. All in all, these unitarized approaches are interesting, first because they provide a fairly good description of data, including values of resonance poles, and second because they have much better properties (analyticity, chiral symmetry, unitarity, etc.) than simple popular models, such as the superposition of simple resonances, isobar models, different versions of Breit-Wigner shapes, etc.
We refer the reader to~\cite{Pelaez:2015qba,Oller:2019opk,Oller:2020guq,Yao:2020bxx,Pelaez:2021dak} for recent reviews on these topics. Unfortunately, within UChPT, the lack of a systematic expansion does not allow for a precise estimation of uncertainties, and the formalism is generally restricted to two-body or quasi-two-body states. 

Still, modern Hadron Physics demands precise and model-independent meson-meson scattering parametrizations. This requires the use of dispersion relations, which over roughly the last two decades have been successfully applied to describe scattering data on $\pi\pi$~\cite{Ananthanarayan:2000ht,Colangelo:2001df,DescotesGenon:2001tn,Kaminski:2002pe,GarciaMartin:2011cn,Kaminski:2011vj,Moussallam:2011zg,Caprini:2011ky,Albaladejo:2018gif,RuizdeElvira:2018hsv}, but also on $\pi N$~\cite{Ditsche:2012fv,Hoferichter:2015hva}, $\gamma\gamma\to\pi\pi$~\cite{Hoferichter:2011wk,Hoferichter:2019nlq}, $\pi K$~\cite{Buettiker:2003pp,Pelaez:2016tgi} and $\pi\pi\to K\bar K$~\cite{Pelaez:2018qny} 
(see~\cite{Pelaez:2020gnd} for a review of the last two).

Unfortunately, such rigorous dispersive results,
are not always suitable for later practical use. There are several reasons for this. First, they are often obtained numerically from integral equations, which makes them inconvenient for further phenomenological or experimental analyses. Second, they are frequently parametrized in the real axis by piecewise functions. They provide flexibility, but cannot be directly continued to the complex plane in search of poles. Finally, of the usual constraints used in the literature, partial-wave dispersion relations are in practice limited to energies around  $1\,$GeV for $\pi\pi$ and $\pi K$ scattering, and to $1.4\,$
GeV for forward dispersion relations. Above these energies, partial-wave experimental data have not been tested nor described with dispersive constraints.
In this work, we will partially alleviate these caveats.

Recently, two of us with another collaborator \cite{Pelaez:2019eqa} provided a set of relatively simple and ready-to-use ``global'' parametrizations that describe the scalar-isoscalar (S0) and vector-isovector (P1 or just P)
partial waves of $\pipi$ scattering data up to somewhat below 2 GeV. They mimic the central values and uncertainties of the dispersive data analysis in~\cite{GarciaMartin:2011cn}, which was dispersively constrained up to 1.42 GeV. Beyond this energy, these parametrizations were three purely phenomenological fits to three different data sets from the literature.
In addition, being analytic expressions, these parametrizations could be evaluated in the complex plane and they were made to reproduce the dispersive results, including the poles associated with the \sig, $f_0(980)$ and $\rho(770)$ resonances. Later, it was found in~\cite{Pelaez:2022qby} that the poles of the $f_0(1370)$ and $f_0(1500)$ are also present in the dispersive analysis and global parametrization of the S0 wave. Furthermore, these global parametrizations are also consistent with the threshold parameters and the value of the S0-wave Adler zero in~\cite{GarciaMartin:2011cn}.
They have been widely used by both the theoretical and experimental hadron and particle physics communities.

The aim of this work is threefold.
First, extend this ``global" parametrization approach beyond the partial waves S0 and P up to those with angular momentum $\ell=4$, 
trying to avoid piecewise parametrizations whenever possible.
Second, improve the treatment of the P-wave inelasticity, allowing it to start at the $\pi \omega$ threshold.
This was neglected in~\cite{GarciaMartin:2011cn,Pelaez:2019eqa}\footnote{We thank G. Colangelo, M. Hoferichter, and P. Stoffer for pointing this to us.}, which focused more on the S0 wave and its resonances, but 
there is a considerable interest in the precise description of this wave.
Third, extend the forward dispersion relation constraints to higher energies, while improving the matching with the high-energy Regge description.
These improvements will have a relatively small effect on the S0 wave, whose global parameters will be slightly updated, but only because of the indirect effects of the changes in the other waves.

The plan of this work is as follows:  Sec.~\ref{sec:input} presents the data to be described. In Sec.~\ref{Sec:parametrizations} we present the global partial-wave parametrizations. In Sec.~\ref{sec:disprel} we will revisit the dispersive constraints
and obtain constrained Global Fits. We will discuss and summarize our results in Sec.~\ref{sec:discussion}.

\section{The input to be described}\label{sec:input}

There are several data sets on \pipi scattering partial waves $t^{(I)}_\ell(s)$, which almost reach a center-of-mass energy of $\sqrt{s}\sim2\,$ GeV~\cite{Hyams:1973zf,Durusoy:1973aj,Losty:1973et,Cohen:1973yx,Protopopescu:1973sh,Grayer:1974cr,Hyams:1975mc,Hoogland:1977kt,Kaminski:1996da}. 
Here $I=0,1,2$ stands for isospin and $\ell=0,1,2,3,4...$ for angular momentum, although we will follow the
usual spectroscopic notation and refer to them as S, P, D, F, G... waves, followed by their isospin number.

Unfortunately, these are indirect measurements, extracted as a sub-process of $\pi N\to \pi\pi N'$ scattering, which is the observed reaction. These analyses assume the dominance of an almost on-shell one-pion exchange, along with other approximations.
As a consequence of this indirect extraction, $\pi\pi$ data sets have large systematic uncertainties,  and are often incompatible with each other, even when extracted from the same $\pi N\to \pi\pi N'$ experiment. Furthermore, simple fits to individual data sets or to averaged sets do not comply well with dispersion relations~\cite{Pelaez:2004vs,Kaminski:2006yv,Yndurain:2007qm,Kaminski:2006qe,GarciaMartin:2011cn}. 
This is why such relations were used to eliminate inconsistent data sets or as constraints to obtain a Constrained Fit to Data (CFD)~\cite{GarciaMartin:2011cn}. In addition, these CFDs fulfill the normality requirements of the residual distribution~\cite{Perez:2015pea}, thus allowing for standard error-propagation methods. 
Consequently, the CFD parametrizations will be part of our input.

Why, then, not to use the CFD directly? Actually, it is perfectly fine to use them, although they were constructed with piecewise functions and only up to energies around $1.4\,$GeV. Moreover, the focus of~\cite{GarciaMartin:2011cn}
was the low-energy parameters and the region of the  
\sig, $f_0(980)$, and $\rho(770)$ resonances. 
The description above 1.1 GeV, or of higher waves, was provided primarily to serve as input for the integral representation of the S0 and P amplitudes below that energy. 
The need for ``global" parametrizations is justified in that many applications require a wider energy range and waves with higher angular momentum, and because the CFD piecewise expressions cannot be 
straightforwardly continued to the complex plane to describe resonance poles. This is why in~\cite{Pelaez:2019eqa} some of us provided such global parametrizations for the S0 and P waves. Here we extend this approach to six more waves: S2, D0, D2, F, G0, and G2. In addition, we improve some details of the P wave above 0.9 GeV, and expand the dispersive constraints further than 1.4 GeV for all the waves. Moreover, above roughly 1 GeV, we improve the precision of all waves beyond the S0.

Note that in~\cite{GarciaMartin:2011cn}, two complementary kinds of dispersion relations were considered. On the one hand, crossing symmetric partial-wave dispersion relations were imposed on the S0, P, and S2 partial waves, either with two subtractions (Roy equations~\cite{Roy:1971tc}) or one subtraction (GKPY equations~\cite{GarciaMartin:2011cn}). These relations constrain partial waves individually but are, in practice, limited to energies below 1.1 GeV. On the other hand,  Forward Dispersion Relations (FDRs), which do not constrain partial waves separately, were imposed on the forward amplitudes. They were studied up to 1.42 GeV, but could, in principle, be extended to arbitrarily high energies. 

Hence, we will use as input the threshold parameters obtained from the CFD parametrizations in~\cite{GarciaMartin:2011cn} or sum rules \cite{Kaminski:2006qe}. In addition, we will consider the CFD in the real $s$ axis as input for the S2 and D waves, although only below 0.9 GeV, as we plan to improve the treatment of their inelasticities, which in~\cite{GarciaMartin:2011cn} were only taken into account above the $K \bar K$ threshold. 

The only wave for which we use scattering data input different from the CFD~\cite{GarciaMartin:2011cn}
below $0.9\gev$ is the P-wave. The reason is that the initial unconstrained P-wave fit in~\cite{GarciaMartin:2011cn} was not obtained from scattering measurements but from a description of data on the pion vector form factor~\cite{DeTroconiz:2001rip}, which was much more precise. In contrast to $\pi\pi$ scattering, whose experiments were mostly made in the 1970s, many more recent experiments, with better precision, have been carried out for the pion vector form factor. In practice, below the $\pi\omega$ threshold, i.e. $0.922$ GeV, our unconstrained P wave will be built from the recent and accurate analysis of P-wave phase shifts in~\cite{Colangelo:2018mtw}.

The second input is, of course, scattering data. For waves beyond the S0, we first revisit our unconstrained data fits above 0.9 GeV to include the few data points below the $K\bar K$ threshold with a non-vanishing inelasticity. They belong to the P, S2, and D0 waves, and the inelasticity is so small that it was neglected in~\cite{GarciaMartin:2011cn}. However, here, we aim for a better precision. In addition, we will provide Global Fits for the S2, D, F, and G waves reaching up to energies between 1.8 and 2.1 GeV, depending on how far data reach for each wave.
 
Let us first discuss the isospin $I=0,1$ waves, which are attractive in almost the whole energy range of interest.
Above energies around $1.15\,$GeV, almost all data come from the CERN-Munich collaboration, which has several solutions.
Of these, the one published in 1973 (Hyams et al. 73 ~\cite{Hyams:1973zf}) is the most popular. 
For the S0 wave, it was renamed ``Solution b'' in a subsequent collaboration compilation (Grayer et al.~\cite{Grayer:1974cr}) and is very consistent with a later reanalysis using polarized targets~\cite{Kaminski:1996da} and fairly consistent up to 1.43 GeV with ``Solution (- - -)'' in the 1975 collaboration reanalysis (Hyams et al. 75~\cite{Hyams:1975mc}).
In addition, other data, obtained at Berkeley (Protopopescu et al.~\cite{Protopopescu:1973sh}), extend up to 1.15 GeV, although they tend to have somewhat larger uncertainties. We will see that their P and D0 wave phase shifts are quite compatible with those of the CERN-Munich collaboration. However, there are inconsistencies regarding the P-wave inelasticity and the entire F wave.

Concerning the S0 wave for energies above 1.4 GeV, ``Solution b'' and ``Solution (- - -)'' are very incompatible. Other solutions were already disfavored in the very same CERN-Munich 1975 analysis, although the ``(- + -) solution" has recently been resurrected~\cite{Ochs:2013gi}. 
Nevertheless, all these solutions have caveats (see~\cite{Pelaez:2003eh,Pelaez:2004vs,Pelaez:2015qba} for detailed discussions).
In particular, at high energies, the elastic cross section of ``Solution b'' is larger than the inelastic one, which is at odds with the observations in $\pi N$, $KN$, and $NN$ scattering, and there is no obvious reason why $\pi\pi$ scattering should be different. In addition, right above 1 GeV, the S0-wave data of the other CERN-Munich solutions prefer less inelasticity, i.e. the so-called ``non-dip" scenario \cite{GarciaMartin:2011cn}, which is disfavored
by Roy-like dispersive constraints \cite{GarciaMartin:2011cn,Moussallam:2011zg}.
Later on, we will remove such data points for these solutions.
Moreover, if the inelasticity is large, the solution in terms of phase shift and elasticity is not unique~\cite{Atkinson:1969wy,Atkinson:1970pe}. ``Solution b'' is an example of an almost elastic case and ``Solution (- - -)'' is strongly inelastic. 
This, of course, applies to other waves.
Furthermore, a slight modification of the ``(- + -) solution" by one of the members of the CERN-Munich collaboration~\cite{Ochs:2013gi} considering an inelastic S2 wave, seems consistent with the dispersive representation \cite{Pelaez:2019eqa}, finding some qualitative agreement with the GAMS experiment on $\pi^-p\rightarrow\pi^0\pi^0 n$ \cite{Alde:1998mc}.  Such a ``(- + -) solution" differs from the ``b" and ``(- - -)" solutions only above 1.4 GeV and hints at the presence of the $f_0(1500)$ resonance, not evident in other solutions.  Last but not least, the convergence of the partial-wave expansion is questionable, since already at 1.7 GeV the F wave is as large as the P wave, the D0 as the S0, and the D2 is larger than the S2.

Not only the S0 but also the P, D0, and F waves, have three solutions, although 
they are not so different among themselves.
The existence of different data sets leads to three different constrained global fits 
in~\cite{Pelaez:2019eqa}, almost identical up to 1.43 GeV. 
Namely, above 1.43 GeV: i) ``Solution I'' describes  the data of~\cite{Hyams:1973zf,Grayer:1974cr,Kaminski:1996da}, 
ii) ``Solution II'' describes the (- - -) data of~\cite{Hyams:1975mc}, and  iii) ``Solution III" describes the data of the ``(- + -) solution" updated in~\cite{Ochs:2013gi}.
We will follow a similar strategy here, considering the three
sets of data called Solutions I, II, and III again, and their corresponding fits that we will call Global Fits I, II, and III. There are no scattering data for the G0 wave and we will build an educated guess from other information.

Concerning the $I=2$ waves, we also use CFD input below 0.9 GeV for the repulsive S2 and D2, often called ``exotic waves". No CFD was available for the G2 wave, which is also exotic. Above that energy we consider the 1973 data from the Rochester collaboration measured at Brookhaven (Cohen et al.~\cite{Cohen:1973yx}), the Paris-Bari collaboration obtained in 1973 at CERN (Durusoy et al.~\cite{Durusoy:1973aj}), the 1974 CERN-Saclay data (Losty et al.~\cite{Losty:1973et}) and the 1977 Amsterdam-CERN-Munich data (Hoogland et al.~\cite{Hoogland:1977kt}). These experiments are roughly in agreement, partly because these waves are small and the uncertainties become comparatively large. Thus, we do not contemplate alternative data sets for these waves and there will be only one unconstrained fit for each $I=2$ wave. However,  since the dispersive constraints affect all waves simultaneously, 
the alternative data sets from other waves will indirectly affect the
$I=2$ waves, which will also have three different constrained fits, although they will look rather similar in practice. 

Having fits for all these waves up to at least 1.8 GeV, and sometimes even up to 2.1 GeV, will also allow us to increase the matching point with our high-energy Regge description. 
The latter is a semi-local approach, i.e., it describes the amplitude on the average around a given energy~\cite{RuizdeElvira:2010cs} and is only to be used inside integrals. Indeed, in~\cite{GarciaMartin:2011cn} a Regge fit to total cross sections in different $\pi\pi$ channels was used above 1.42 GeV, but no effort was made to match it to the partial-wave series within uncertainties. The reason was that in~\cite{GarciaMartin:2011cn} the high-energy part only contributed as  input for the integrals, and was relatively suppressed when
testing or constraining the low-energy region of interest.
However, since we now want a precise description also up to higher energies, better matching is needed to make sure there are no spurious 
artifacts\footnote{We thank C. Hanhart for expressing interest and calling our attention to this issue.}. 
Then, to improve the precision and consistency of our parametrizations, we will impose the FDR constraints, 
but now with a better matching and up to $1.6\,$GeV for the two FDRs containing P-wave contributions. This is almost $200\,$MeV higher than in~\cite{GarciaMartin:2011cn,Pelaez:2019eqa}. 
All in all, the whole global S0 wave or the other waves below 0.9 GeV will barely change from the CFD and global parametrizations provided in~\cite{GarciaMartin:2011cn,Pelaez:2019eqa}, but there will be a clear improvement for the other waves at higher energies.

Let us now describe the parametrizations used for each partial wave. We will present our parametrizations and illustrate the 
dispersive constraints in terms of Solution I, which is the most popular and, as we will discover later on, seems to be slightly favored by the dispersive checks.

\section{Analytic parametrizations}
\label{Sec:parametrizations}

We will use the following values:
$m_\pi=m_{\pi^\pm}=139.57\,$MeV,
$m_K=\left(m_{K^+}+m_{K^0}\right)/2=495.7\,$MeV, $m_\omega=782.66\,$MeV.

Customarily, the $\pi\pi\to\pi\pi$ partial-wave 
$S^{(I)}_\ell$-matrix element of definite isospin $I=0,1,2$ and angular momentum $\ell$, is parametrized in terms of two real functions:
\begin{equation}\label{eq:Smatrixel}
S^{(I)}_\ell(s)=\eta^{(I)}_\ell(s) e^{2i\delta^{(I)}_\ell(s)}=1+2i\sigma(s)  t^{(I)}_\ell(s),
\end{equation}
where $s$ is the usual Mandelstam variable, the pion center-of-mass (CM) momentum squared
is $k(s)^2=s/4 - m_\pi^2$, and
\begin{equation}
\sigma(s)=\frac{2k(s)}{\sqrt{s}}=\sqrt{1-\frac{4 m_\pi^2}{s}}.
\end{equation}
The functions $\delta^{(I)}_\ell=\arg\left(S^{(I)}_\ell\right)/2$  and
$\eta^{(I)}_\ell=\vert S^{(I)}_\ell\vert$ 
are the phase shift and elasticity, respectively,
which are real for $s\geq 4 m_\pi^2$.
The fact that the $S$-matrix is unitary implies that $0\leq \eta^{(I)}_\ell \leq 1$. It is also convenient to
define the inelasticity as $\sqrt{1-\eta^{(I)\;2}_\ell}$, whose value lies also between zero and one.
Following Eq.~\eqref{eq:Smatrixel}, we define partial-wave amplitudes as 
\begin{equation}
t^{(I)}_\ell(s)=\frac{\eta^{(I)}_\ell(s) e^{2i \delta^{(I)}_\ell(s)}-1}{2 i\sigma(s)}.
\end{equation}

Note that we are not describing the full $S$-matrix, but just its first diagonal element $S_{11}$, where $1=\pi\pi$ (we have suppressed the spin and isospin indices momentarily). On this single element, unitarity above the $\pi\pi$ threshold only implies that $\eta(s)=\vert S_{11}\vert\leq1$.

In contrast, in the popular coupled channel approach,
one describes all the $S_{ij}$-matrix elements. For that, one has to choose a priori how many channels  $i,j=1,...n$ to consider and to identify each of them (and their thresholds). Within that approach the elasticity $\eta(s)\equiv\vert S_{11}\vert=\sqrt{1- \vert{S_{12}}\vert^2 ....  -\vert{S_{1n}}\vert^2  }$,
and thus it is possible to know the partial contribution of each coupled channel to the inelasticity of $S_{11}$. This is impossible in our approach since we only determine the total elasticity. On the positive side, our approach avoids the model dependence due to deciding what channels are open or not and how to describe them, which, for multiple-meson states, is usually done in strongly model-dependent quasi-two-body approximations.

In practice, to fit the data, we have to decide where we allow the inelasticity to set in. In certain channels, the choice is easy because the inelasticity data appears right after a prominent two-body threshold like $K\bar K$, $\omega \pi$. However, multibody thresholds like $4\pi$  or $\pi\pi K\bar K $ or six pions, etc., open up more smoothly (but in coupled channel approaches are frequently modeled as $\rho\rho$, $\sigma\sigma$, $\rho\pi\pi$, ...).
When inelasticity is observed with no obvious two-body threshold right below, we have allowed it to open up at some effective threshold below the inelastic data. But remember that once we have a non-vanishing total inelasticity, it contains all possible inelastic contributions.

This said, we return to our previous notation and define the total amplitude $F^{(I)}(s,t)$ of definite isospin $I$, normalized as follows:
\begin{eqnarray}
F^{(I)}(s,t)&=&\frac{8}{\pi} \sum_{\ell=0}^{\infty}(2\ell+1)P_\ell(z(s,t))t^{(I)}_\ell(s),\label{eq:FI}\\
t^{(I)}_\ell(s)&=&\frac{\pi}{16}\int_{-1}^{1} \text{d}z P_\ell(z) F^{(I)}(s, t(s,z)),\label{eq:pw}
\end{eqnarray}
where $z=\cos\theta$ is the cosine of the scattering angle $\theta$ in the CM frame, and the second Mandelstam variable $t=-2k(s)^2\left(1-\cos\theta\right)$. The third Mandelstam variable is fixed to $u=4m_\pi^2-s-t$ and we omit it for brevity.
Note that in the literature it is also common to use a different normalization for the total amplitude $T^{(I)}(s,t)= 4\pi^2 F^{(I)}(s,t)$.

In the elastic case, when $\eta^{(I)}_\ell=1$, 
the partial-wave amplitude only depends on the phase shift and can be 
written as:
\begin{equation}\label{eq:elastict}
t^{(I)}_\ell(s)=\frac{e^{i\delta^{(I)}_\ell(s)}\sin\delta^{(I)}_\ell(s)}{\sigma(s)}=
\frac{1}{\sigma(s)}\frac{1}{\cot\delta^{(I)}_\ell(s)-i}.
\end{equation}

It is then customary to define the threshold parameters as the coefficients of the partial-wave threshold expansion in powers of the pion momentum, as follows:
\begin{equation}
    \frac{1}{m_\pi k^{2\ell}} \re t^{(I)}_\ell(s)\simeq
    a^{(I)}_\ell+b^{(I)}_\ell(s)\,k^2+ {\cal O}(k^4).
\end{equation}
Their values for all the waves of interest are given in Appendix~\ref{app:Threshold}.

We will use several types of parametrizations
for $t^{(I)}_\ell$, $\delta^{(I)}_\ell$, $\eta^{(I)}_\ell$, or
$\cot \delta^{(I)}_\ell$.
Some parametrizations will have physical features built in, such as factors to describe poles, zeros, peaks, or a specific threshold behavior, described in detail for each wave below.
In this way, we try to mimic the appearance of some new effect at a given energy, like a new channel opening up with small uncertainties that grow at higher energies.
However, other parametrizations will be purely phenomenological, in terms of the powers of some energy variables. They will be used in regions where we want to describe a relatively uniform error band and provide somewhat smaller correlations between their parameters. We will often use Chebyshev polynomials of the first kind,  $p_k(x)$. They satisfy: $p_0(x)=1$, 
$p_1(x)=x$ and the rest are defined recursively in the interval $x\in [-1,1]$ as:
\begin{equation}
p_{n+1}(x)=2 x\,p_n(x)-
p_{n-1}(x).
\end{equation}
We will frequently use that $p_n(-1)=(-1)^n$, and  $p_n^\prime(-1)=n^2(-1)^{n-1}$,
where the prime here means the derivative of the function
with respect to its variable.

Next, let us describe our global parametrizations, starting with the attractive P, D0, F, and G0 waves, all of which present some resonant behavior, followed by the repulsive S2, D2, and G2 waves. The indirect and minor changes of the S0 wave with respect to \cite{Pelaez:2019eqa}, are discussed in Appendix~\ref{app:S0wave}.

\subsection{P-wave parametrization}

A global parametrization for this wave was already provided in~\cite{Pelaez:2019eqa}. Here we will improve it to allow its inelasticity to start at the $\pi \omega $ threshold at $\sqrt{s_{\pi\omega}}\equiv m_\pi+m_\omega=922.23\,$MeV.

\begin{figure}
\centering
\includegraphics[width=0.48\textwidth]{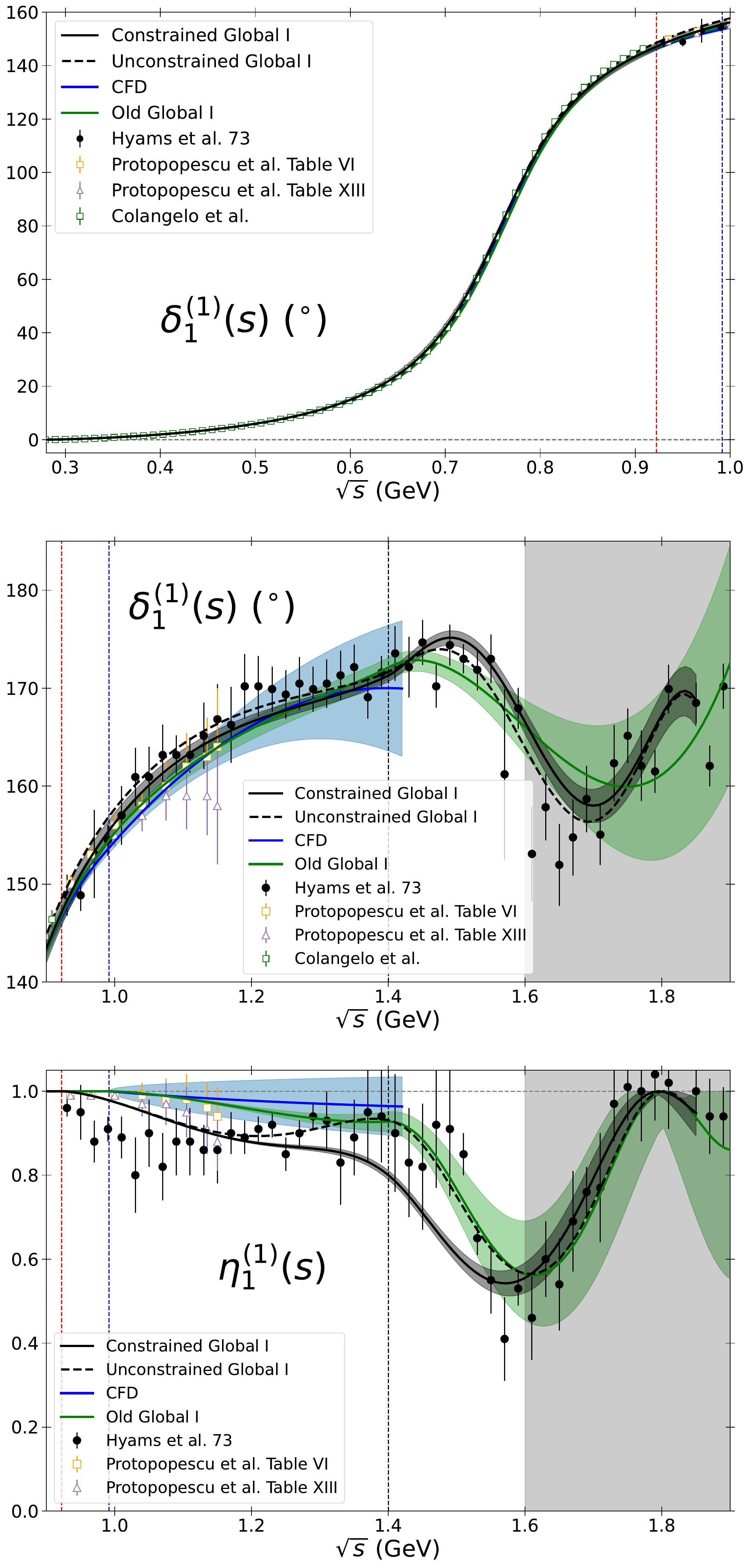}
\caption{We show our unconstrained and dispersively constrained P-wave Global Fit I. Top: phase shift below 1 GeV. Center:  phase shift above 0.9 GeV. Bottom: elasticity from 0.9 GeV. We also show the ``Old Global I" and CFD results from~\cite{Pelaez:2019eqa} and~\cite{GarciaMartin:2011cn}, respectively, whose inelasticity opened at $K\bar K$ threshold (blue vertical line).
We now allow the inelasticity to start from 
 the $\pi\omega$ threshold (red vertical line).
Below that energy, only $a_1^{(1)}$ and $b_1^{(1)}$ from Table~\ref{tab:ThresholdParameters} and the analysis of Colangelo et al.~\cite{Colangelo:2018mtw} 
are used as input for the fit. Above, we show data from Hyams et al.~\cite{Hyams:1973zf} and Protopopescu et al.~\cite{Protopopescu:1973sh} although the latter are not fit. The black vertical line at 1.4 GeV indicates the matching point with the high-energy parametrization. Our Global Fit I should not be extrapolated beyond 1.85 GeV.}\label{fig:P}
\end{figure}

In addition, this is the only wave where we will update the phase shift data in the elastic region below the $\pi \omega $ threshold. As explained before, this is because the dispersive analyses in~\cite{GarciaMartin:2011cn}
were not based on scattering data, but on a 2001 fit to the pion vector form factor~\cite{DeTroconiz:2001rip}. Since then, many other data on the form factor have appeared, and we will therefore use as input the output of the pion vector form factor dispersive analysis by Colangelo et al. in~\cite{Colangelo:2018mtw}, 
which we show as data points with uncertainties in the top panel of Fig.~\ref{fig:P}.\footnote{ We thank C. Colangelo, M. Hoferichter and P. Stoffer for kindly providing us with their results.}  Note that this analysis does not include the recent CMD-3 $e^+e^-\to\pi^+\pi^-$ data~\cite{CMD-3:2023alj}, which show sizable tension with previous results. Nevertheless, the impact of these data on the $\pi\pi$ P-wave phase shift analysis of~\cite{Colangelo:2018mtw} below the $\pi\omega$ threshold has been found to be relatively small~\cite{Stoffer:2023gba,HoferichterCMD3}, and therefore do not affect our results.

As stated above, we aim to provide simple formulae that can be easily implemented later for other purposes, while precisely describing the central value and uncertainty band. This last requirement makes it very difficult to provide a single parametrization covering the whole energy range when a wave has a lot of structure, as seen in the middle and bottom panels of Fig.~\ref{fig:P}. The data there come from \cite{Hyams:1973zf}, which we call Solution I here.

Therefore, our parametrization still has two pieces. The first one below 1.4 GeV, where the inelasticity is small ($\eta^{(1)}_1\geq 0.85$) and
will be constrained by all FDRs (and Roy-like equations below 1.1 GeV). 
Note that this piece will now be a single analytic parametrization, whereas in~\cite{GarciaMartin:2011cn} it also consisted of two pieces, above and below the inelastic threshold.
The second part extends from 1.4 GeV to $\sim$1.8 GeV, where the inelasticity can be rather large at some energies, i.e. $\eta\sim0.5$, and only two FDRs provide constraints up to 1.6 GeV. Above this energy, it is just an unconstrained fit to data.
The applicability of the FDRs is dictated by the energy at which the partial-wave and Regge representations match within uncertainties.

Let us provide the detailed expressions of this parametrization.

\subsubsection{P wave below 1.4 GeV}

Following~\cite{Pelaez:2019eqa}, below  $\sqrt{s_m}=1.4\,$GeV we  build our partial-wave 
$S$-matrix as $S^{(1)}_1=S^{(1)}_{1,\text{conf}}\,\tilde S^{(1)}_1$, where the first factor is always elastic, $\vert S^{(1)}_{1,\text{conf}}\vert=1$, and is given by a simple conformal expansion. The inelasticity is produced by the second factor, which we will modify here. If we now define:
 \begin{align}
      S^{(1)}_1(s)=&1+2i\,\sigma(s)\,t^{(1)}_1(s)\,,\nonumber\\ 
      S^{(1)}_{1,\text{conf}}(s)=&1+2i\,\sigma(s)\,\tau^{(1)}_{1,\text{conf}}(s)\,, \nonumber\\ 
      \tilde S^{(1)}_1(s)=&1+2i\,\sigma(s)\,\tau^{(1)}_1(s)  \,,
 \end{align}
the partial-wave amplitude is
\begin{equation}
t^{(1)}_1(s)=\tau^{(1)}_{1,\text{conf}}(s)+\tau^{(1)}_1(s)+2i\,\sigma(s)\,\tau^{(1)}_{1,\text{conf}}(s)\,\tau^{(1)}_1(s).
\label{eq:finalP}
\end{equation}
Note that both $\tau(s)$ functions are just convenient auxiliary functions to define our parametrization; they are not amplitudes by themselves. The scattering amplitude is the full $t_1^{(1)}(s)$.

As we commented before, we are not describing the full $S$-matrix, just its first diagonal element $S_{11}$, and unitarity just imposes $\eta(s)=\vert S_{11}\vert\leq1$ (for simplicity, spin and isospin indices are suppressed). For this reason, the factorization scheme $S_{11}=\Pi_k S_{11}^{\{k\}}$ with $\vert S^{\{k\}}_{11}\vert\leq1$, is common in the literature~\cite{Zheng:2003rw,Zhou:2004ms,Colemanprocs,Pelaez:2016tgi,Pelaez:2020gnd}. In principle, below a given $k$-th threshold $S_{11}^{\{k\}}=1$ but, with an appropriate momentum dependence, it becomes less than one above. 
This description is enough to describe $S_{11}$ in the real axis as we do here. For other purposes, each $S_{11}^{\{k\}}$ can be given a more detailed analytic structure.

Let us first discuss the elastic regime, which in this work we take 
below the $\pi \omega$ threshold.
Here, $\tau^{(1)}_1$ also fulfills elastic unitarity, i.e.,  $\vert \tilde S^{(1)}_1\vert=1$. 
Consequently, elastic unitarity is satisfied for the whole partial wave, i.e., $\vert S^{(1)}_1\vert=\eta^{(1)}_1(s)=1$.

In this regime, the entire amplitude is constructed such that the conformal part alone provides a reasonable description of the wave, i.e., $t_1^{(1)}\simeq \tau^{(1)}_{1,\text{conf}}$, while $\tau^{(1)}_1$ is comparatively much smaller. In particular, since the elastic regime is dominated by the $\rho(770)$ resonance, a peak is enforced in the conformal part near the $\rho(770)$ mass, with an explicit factor $(\hat m_\rho^2-s)$. Namely
\begin{align}
&\tau^{(1)}_{1,\text{conf}}(s)=
\frac{1}{\sigma(s)}\frac{1}{\Phi^{(1)}_1(s)-i}, \quad s\leq s_m, 
 \nonumber\\
&\Phi^{(1)}_1(s)=\frac{1}{\sigma(s) k(s)^2}(\hat m_\rho^2-s)\left(\frac{2 m_\pi^3}{\hat m_\rho^2 \sqrt{s}}+\sum_{n=0}^4{B_n w(s)^n}\right), \nonumber\\
& w(s)=\frac{\sqrt{s}-\alpha \sqrt{s_0-s}}{\sqrt{s}+\alpha \sqrt{s_0-s}}, \, \, s_0=\left(1.43 \gev\right)^2, \, \, \alpha=0.3. 
\label{eq:generalconformalP}
\end{align} 
Just with five $B_i$ we get good quality fits.
Note that $\hat m_\rho$ is merely a parameter indicating where the phase of the conformal part alone reaches $\pi/2$. However\footnote{The distinction between $\hat m_\rho$ as a parameter and 
the true peak position $m_\rho^{peak}$ was not clearly discussed in~\cite{Pelaez:2019eqa}.}, due to the $1/\sigma(s)$ factor in the conformal part, and the presence of additional non-conformal factors in the full $S$-matrix, the $\rho(770)$ peak position $m_\rho^{peak}$ lies near but not exactly at $\hat m_\rho$. 
The precise value of the peak and the energy at which the phase shift reaches $\pi/2$ must be determined using the full amplitude $t^{(1)}_1$, not just $\tau^{(1)}_{1,\text{conf}}$.

For the full amplitude we also need $\tau^{(1)}_1(s)$, which, following~\cite{Pelaez:2019eqa} once again, we parametrize in terms of a function $\Delta^{(1)}_{1}(s)$ as follows:
\begin{equation}
\tau^{(1)}_1(s)=\frac{e^{2i \Delta^{(1)}_{1}(s)}-1}{2 i\,\sigma(s)}.
\end{equation}
In the elastic regime $\Delta^{(1)}_{1}(s)$ will be real,
so that $\vert \tilde S^{(1)}_1 \vert = \vert e^{2i \Delta^{(1)}_{1}}\vert =1$
and therefore the whole $S^{(1)}_1$ is unitary, $\vert S^{(1)}_1\vert=\eta^{(1)}_1=1$.
This is the same as in~\cite{Pelaez:2019eqa}, but the expression for $\Delta^{(1)}_{1}(s)$, which is purely phenomenological, is adjusted to ensure a non-vanishing inelasticity from the $\pi \omega$ threshold,
\begin{equation}
\Delta^{(1)}_{1}(s)=\bar{J}_{\pi\omega}(s) \frac{k(s)^3}{\sqrt{s} \, m_\pi^2} \frac{k_{\pi \omega}(s)^2}{s_{\pi \omega}}\sum_{n=0}^3 K_n  
\left(\frac{s}{s_{\pi\omega}}-1\right)^n,
\label{eq:ineP}
\end{equation}
where $s_{\pi \omega}=(m_\pi+m_\omega)^2$, the $K_n$ are constant parameters, and 
\begin{align}
& \bar{J}_{\pi\omega}(s)=\frac{1}{\pi}\left[ 1+ \left( \frac{\Delta}{s} - \frac{\Sigma}{\Delta} \right)\log\left(\frac{m_\omega}{m_\pi}\right) \right. \nonumber\\
&\left.+ \frac{\nu(s)}{2s}\left(\log\left(\frac{\nu(s)-s+\Delta}{\nu(s)+s+\Delta}\right)+ \log\left(\frac{\nu(s)-s-\Delta}{\nu(s)+s-\Delta}\right)\right) \right],\nonumber\\
\end{align}
with 
\begin{align}
\nu(s)=&\sqrt{(s-(m_\pi+m_\omega)^2)(s-(m_\pi-m_\omega)^2)},\nonumber\\
k_{\pi \omega}(s)^2=&\frac{\nu(s)^2}{4s},\quad\Delta=m_\pi^2-m_\omega^2, \quad\Sigma=m_\pi^2+m_\omega^2.
\end{align}
Note that the whole $t^{(1)}_1$ amplitude in Eq.~\eqref{eq:finalP} exhibits the appropriate kinematic behavior around the thresholds. Namely, it behaves as $k_{\pi \omega}(s)^2$ near the $\pi \omega$ threshold (since parity conservation requires that the $\pi$ and $\omega$ are produced in a P wave) and as $k(s)^3/\sigma(s)\sim k(s)^2$ near the $\pi \pi$ one.
In addition, since $\bar{J}_{\pi\omega}(s)$ is real below the $\pi \omega$ threshold, this ensures that the entire $t^{(1)}_1$ is elastic for $s<s_{\pi \omega}$. Above the $\pi \omega$ threshold, $\bar{J}_{\pi\omega}(s)$ has an imaginary part, making both $\tau^{(1)}_1 $ and the whole $t^{(1)}_1$ amplitude inelastic. The advantage of using the $\bar J$ function is that this parametrization is analytic over the whole energy range from $s=0$ up to $(1.4 \gev)^2$, allowing for a straightforward continuation into the complex plane---something the usual step function does not provide. 

Once again, we emphasize that this is just a convenient 
parametrization. In practice, we have found that four parameters $K_{i}$ with $i=0, \cdots, 3$, 
together with the conformal parametrization in Eq.~\eqref{eq:generalconformalP}, are sufficient to describe the phase shift and elasticity on the real axis below 1.4 \gev. Notably, $K_3$ is consistently much smaller than the other three parameters, and adding a fifth term does not improve the fits.

Let us clarify that setting the onset of the inelasticity above the $\pi\omega$ threshold does {\it not} imply that all inelasticity arises from the $\pi \omega$ state. We simply allow the elasticity to be $\eta^{(1)}_1\neq1$ above the $\pi \omega$ threshold, as this is roughly where experiments begin to observe such values. Near this threshold, we have imposed a kinematic behavior consistent with $\pi \omega$ being the dominant channel. However, far from this threshold, we neither specify nor assume which states contribute to the inelasticity. 

\subsubsection{P wave above 1.4 GeV}

Above $s_m=(1.4 \gev)^2$, the P-wave phase shift is described using Chebyshev polynomials in terms of the following variable 
\begin{align}
    x(s)&=2\frac{\sqrt{s}-\sqrt{s_m}}{2 \gev-\sqrt{s_m}}-1, \nonumber 
\end{align}
which maps the $\sqrt{s}\in[1.4, 2]\gev$ region into the $[-1,1]$ segment, where Chebyshev polynomials are defined.

The phase shift in the region above 1.4 \gev reads:
\begin{eqnarray}\label{eq:P_phase-high}
   \delta^{(1)}_1(s)=\sum_{n=0}^6 d_n \,p_{n}\left(x(s)\right),
\end{eqnarray}
where the five $d_n$ for ${n\geq 2}$ will be the fit parameters.
To ensure a continuous and differentiable matching at $s_m$, 
 $d_0$ and $d_1$ are fixed as follows: 
\begin{align} 
    d_1&=\frac{\delta^{{(1)}\,\prime}_1(s_m)}{x'(s_m)}+\sum_{n=2}^6 (-1)^n n^2d_n, \nonumber\\
    d_0& = \delta^{(1)}_1(s_m)-\sum_{n=1}^6 (-1)^nd_n,
\end{align}
where the values of $\delta^{(1)}_1(s_m)$ and  $\delta^{{(1)}\,\prime}_1(s_m)$ in the previous two equations are determined from the parametrization below 1.4 GeV, as outlined in the previous subsection. 

Seven Chebyshev polynomials are required to achieve an acceptable $\chi^2/d.o.f$ in describing the P-wave phase shift data above 1.4 GeV in~\cite{Hyams:1973zf}, which we refer to as Solution I. This means three more parameters than in~\cite{Pelaez:2019eqa}. Other data sets or ``Solutions" will be discussed below.

For the P-wave elasticity, we employ an exponential with a negative exponent to ensure $0\leq \eta^{(1)}_1\leq1$. In this case, we found that using up to five Chebyshev polynomials is good enough to describe this exponent for all data solutions without introducing spurious oscillations. Thus,
\begin{equation}\label{eq:P_ine-high}
    \eta^{(1)}_1(s)=\exp\left[-\left(\sum_{n=0}^4 \epsilon_n p_n(x(s))\right)^2\right].
\end{equation}
Note that only $\epsilon_2, \epsilon_3$ and $\epsilon_4$ are free parameters, since the continuity of the derivative fixes $\epsilon_1$ to
\begin{align}
    \epsilon_1&=\frac{-\eta^{{(1)}\,\prime}_1(s_m)}{2 \eta^{(1)}_1(s_m)x'(s_m)\sqrt{-\log(\eta^{(1)}_1(s_m))}}+\sum_{n=2}^4  (-1)^n n^2 \epsilon_n,
\end{align}
which, in turn, imposing continuity, fixes $\epsilon_0$ to
\begin{equation}
    \epsilon_0=\sqrt{-\log(\eta^{(1)}_1(s_m))}-\sum_{n=1}^4 (-1)^n\epsilon_n.
   \label{eq:eta11e0}
\end{equation}
Both $\eta^{(1)}_1(s_m)$ and its derivative are obtained from the 
parametrization below 1.4 GeV provided in the previous subsection. Thus, at most three free parameters are required.

\subsubsection{P-wave Global Fit and parameters}

To begin with, for our P-wave Global Fit, shown in Fig.~\ref{fig:P}, we have used as input the 
values of the two first threshold parameters, $a^{(1)}_1$ and $b^{(1)}_1$, obtained from the CFD in~\cite{GarciaMartin:2011cn} and listed in Table~\ref{tab:ThresholdParameters} of Appendix~\ref{app:Threshold}.
In addition, we have considered the P-wave phase shift and elasticity data collected in Fig.~\ref{fig:P}.  The parameters used to plot these curves for the P wave are given in Table~\ref{tab:Pparameters}. They were obtained by fitting the data discussed above while imposing the fulfillment of forward and Roy-like dispersion relations, as detailed in the next section.

Let us first discuss the $\delta_1^{(1)}$ phase shift, shown from the $\pi\pi$ threshold up to 1 GeV in the top panel and from 0.9 GeV to 1.9 GeV in the central one. These two panels overlap to make it clear that the phase-shift function is always differentiable.
In the figure, we show the phase shift used as input below the $\pi\omega$ threshold, obtained by Colangelo et al.~\cite{Colangelo:2018mtw} from the pion vector form factor.  As explained before, our input above the $\pi\omega$ threshold is the CERN-Munich data of Hyams et al.~\cite{Hyams:1973zf}, which has significantly larger uncertainties than the input below. As usual, we use Global Fit I for illustration.

In Fig.~\ref{fig:P}, we also include two data sets from Protopopescu et al.~\cite{Protopopescu:1973sh} for comparison. These data sets are not included in the fit for three reasons.  First, they do not extend to energies beyond 1.2 GeV. Second, above 1.1 GeV they consist of only a few data points whose central values differ visibly from those in~\cite{Hyams:1973zf}. Nevertheless, they are still consistent with~\cite{Hyams:1973zf} due to their much larger uncertainties.
This means that even if we included them in the fit, they would be dominated by the data from~\cite{Hyams:1973zf}. Third, as we will see below, for the F wave, the sets from \cite{Protopopescu:1973sh} contain some artifacts clearly inconsistent with other data.

In addition, Fig.~\ref{fig:P} shows our Global Fit I, which incorporates constraints from dispersion relations that will be explained in detail in the next section.  
While the uncertainties are fairly uniform across the entire energy range, reflecting the size of the experimental error bars, they are significantly smaller in the elastic region due to the high-precision vector form-factor data. Similarly, the uncertainties are reduced in the region where the inelasticity opens up, as its growth from zero is determined by the momentum dependence imposed by the angular momentum barrier. 
Note that the phase shift of the constrained fit does not 
separate from the unconstrained one by more than two deviations.

The phase-shift data set from Table VI in~\cite{Protopopescu:1973sh} is remarkably consistent with our constrained Global Fit, as it is already compatible with~\cite{Hyams:1973zf}. 
The data from Table XIII exhibits somewhat different central values, but its uncertainties are so large that our constrained Global Fit I only lies slightly more than one standard deviation away from them. As already commented, including these data from~\cite{Protopopescu:1973sh} would have a negligible impact on our analysis.

Let us observe that we have excluded the last two phase-shift data points of Solution I from our fit. The reason is that they are highly incompatible with each other and, given the size of our uncertainties, attempting to fit both would result in a significant oscillation that we believe would be 
non-physical. Thus, our Global Fit I should only be used up to 1.85 GeV.

Concerning the elasticity $\eta_1^{(1)}$, it is shown in the bottom panel of Fig.~\ref{fig:P}, starting from 0.9 GeV, slightly below the $\pi\omega$ threshold and the first
datum we show of Hyams et al.~\cite{Hyams:1973zf}. Once again, we show the two data sets from~\cite{Protopopescu:1973sh}, although, as explained above, they are not included in the fit. Nevertheless, they are roughly consistent within uncertainties with the data from~\cite{Hyams:1973zf} above 1.05 GeV, but clearly incompatible below, as they do not show any inelasticity before the opening of the $K \bar K$ threshold.

The Global Fit I elasticity below 1.15 GeV is fairly compatible with the data from~\cite{Protopopescu:1973sh}, but even with the inelasticity opening up at the $\pi\omega$ threshold there are four data points from Hyams et al.~\cite{Hyams:1973zf} that
lie about two standard deviations below the curve. These points cannot be reconciled with the data from Protopopescu et al.~\cite{Protopopescu:1973sh}, and our Global Fit I lies somewhere between the two experiments.
Overall, allowing the inelasticity to start at the $\pi\omega$ threshold yields a smaller elasticity, consistently with the whole data set from~\cite{Protopopescu:1973sh} and the Hyams et al. data~\cite{Hyams:1973zf} above 1.05 GeV. 
It is worth noticing that the constrained Global Fit I elasticity deviates significantly from the unconstrained one between 1.2 and 1.6 GeV. 
This is the starting energy of the purely phenomenological fit, without any dispersive constraint. Finally, the Global Fit I uncertainties are now more uniform across the whole energy region and do not grow exaggeratedly with the energy as they did before in \cite{GarciaMartin:2011cn,Pelaez:2019eqa}.

\newpage

However, the most relevant feature of the Global Fits is that the elasticity changes curvature between 0.9 and 1.4 GeV. 
In the unconstrained fits there was indeed a maximum, which is softened in the constrained Global Fit.
We believe this is not an artifact, but a consequence of the well-established $\rho(1450)$ resonance, although it has not been explicitly included in our parametrizations.

\begin{widetext}
\begin{center}
\begin{table}[h]
\renewcommand{\arraystretch}{1.3}
  \centering
  \begin{tabular}{lcccc}
 P wave    & Parameters& Global I values& Global II values& Global III values\\
\toprule
\multirow{ 6}{*}{$S^{(1)}_{1,\text{conf}}$}& $ B_0$ &$1.11\pm0.05$ &$1.20\pm0.06$ &$1.18\pm0.06$ \\
 &  $B_1$&$-0.833\pm0.013$&$-1.193\pm0.016$&$-1.144\pm0.017$\\
  & $ B_2$ &$0.80\pm0.04$ &$1.47\pm0.05$&$1.30\pm0.05$\\
   &$B_3$&$2.74\pm0.08$&$3.17\pm0.09$&$3.26\pm0.10$\\
   &$ B_4$ &$-1.82\pm0.10$ &$-2.71\pm0.11$&$-2.72\pm0.14$ \\
    &  $ \hat m_{\rho}$ &$(770.7\pm1.2)\, \mev$ &$(769.7\pm1.2)\, \mev$&$(769.6\pm1.2)\, \mev$ \\
\colrule
 \multirow{ 4}{*}{$\tilde S^{(1)}_1$}&$K_0$ & $(39.6\pm0.3)\, 10^{-2}$  &  $(44.75\pm0.15)\, 10^{-2}$ & $(46.38\pm0.19)\, 10^{-2}$ \\
 & $K_1$ & $(-58.25\pm0.08)\, 10^{-2}$ & $(-69.36\pm0.08)\, 10^{-2}$& $(-68.50\pm0.11)\, 10^{-2}$ \\
  & $K_2$ & $(26.16\pm0.05)\, 10^{-2}$ & $(35.29\pm0.05)\, 10^{-2}$& $(31.51\pm0.07)\, 10^{-2}$ \\
   & $K_3$ & $(-1.20\pm0.03)\, 10^{-2}$ & $(-4.59\pm0.03)\, 10^{-2}$ & $(-2.23\pm0.05)\, 10^{-2}$ \\
\colrule
\multirow{ 5}{*}{$ \delta^{(1)}_1\big\vert_{s>s_m}$}&$ d_2$ &$(-64.64\pm0.26) \, ^{\circ}$  &$(0.3\pm0.4) \, ^{\circ}$  &$(0.6\pm0.5) \, ^{\circ}$ \\
&$ d_3$ &$(-46.95\pm0.08) \, ^{\circ}$ &$(0.61\pm0.10) \, ^{\circ}$&$(0.22\pm0.13) \, ^{\circ}$ \\
&$ d_4$ &$(-31.14\pm0.04) \, ^{\circ}$ &$\equiv 0 \, ^{\circ}$&$\equiv 0 \, ^{\circ}$ \\
&$ d_5$ &$(-10.82\pm0.03) \, ^{\circ}$ &$\equiv 0 \, ^{\circ}$ &$\equiv 0 \, ^{\circ}$ \\
&$ d_6$ &$(-0.928\pm0.018) \, ^{\circ}$ &$\equiv 0 \, ^{\circ}$&$\equiv 0 \, ^{\circ}$ \\
\colrule
\multirow{ 3}{*}{$\eta^{(1)}_1\big\vert_{s>s_m}$}&$\epsilon_2$&$(-2.6\pm0.9)\, 10^{-2}$&$(-19.0\pm1.2)\, 10^{-2}$&$(-28.2\pm1.3)\, 10^{-2}$\\
&$\epsilon_3$&$(32.1\pm0.4)\, 10^{-2}$&$(-2.4\pm0.3)\, 10^{-2}$&$(-6.7\pm0.4)\, 10^{-2}$\\
&$\epsilon_4$&$(13.1\pm0.3)\, 10^{-2}$&$(0.33\pm0.16)\, 10^{-2}$&$(-1.71\pm0.18)\, 10^{-2}$\\
\botrule
  \end{tabular}
  \caption{P-wave parameters of the constrained Global Fits I, II, and III. Recall that $s_m=(1.4 \gev )^2$ for this wave .
}
  \label{tab:Pparameters}
\end{table}   
\end{center}
\vspace{-.3 cm}
\end{widetext}

In Fig.~\ref{fig:P}, we also compare our results with the dispersively constrained CFD~\cite{GarciaMartin:2011cn} and with the ``Old" Global I parametrization~\cite{Pelaez:2019eqa},
whose curves below 1 GeV are indistinguishable. 
Since we now use as input an updated pion vector form factor, our Global Fit I is slightly different but still quite consistent with both of them up to the inelastic threshold. From that energy up to 1.2 GeV, the new phase shift is slightly higher than before, and from 1.2 to 1.42 GeV, they are all compatible again. From 1.42 GeV we can only compare our new Global Fit I to the old one, whose qualitative behavior is similar.
Finally, the elasticity of Global Fit I also departs notably from the CFD and Old Global descriptions for all energies below 1.6 GeV.

The most striking feature when comparing 
our new Global Fit I with the old one or the CFD is 
the drastic reduction of the uncertainty bands in the new P-wave, which are now much more uniform and do not grow artificially fast. As we will see below, this improved accuracy will make the compliance with FDRs much more demanding. Hence, having them fulfilled will be a more remarkable accomplishment.

\subsubsection{The three P-wave Global Fits}

\begin{figure}
\centering
\includegraphics[width=0.48\textwidth]{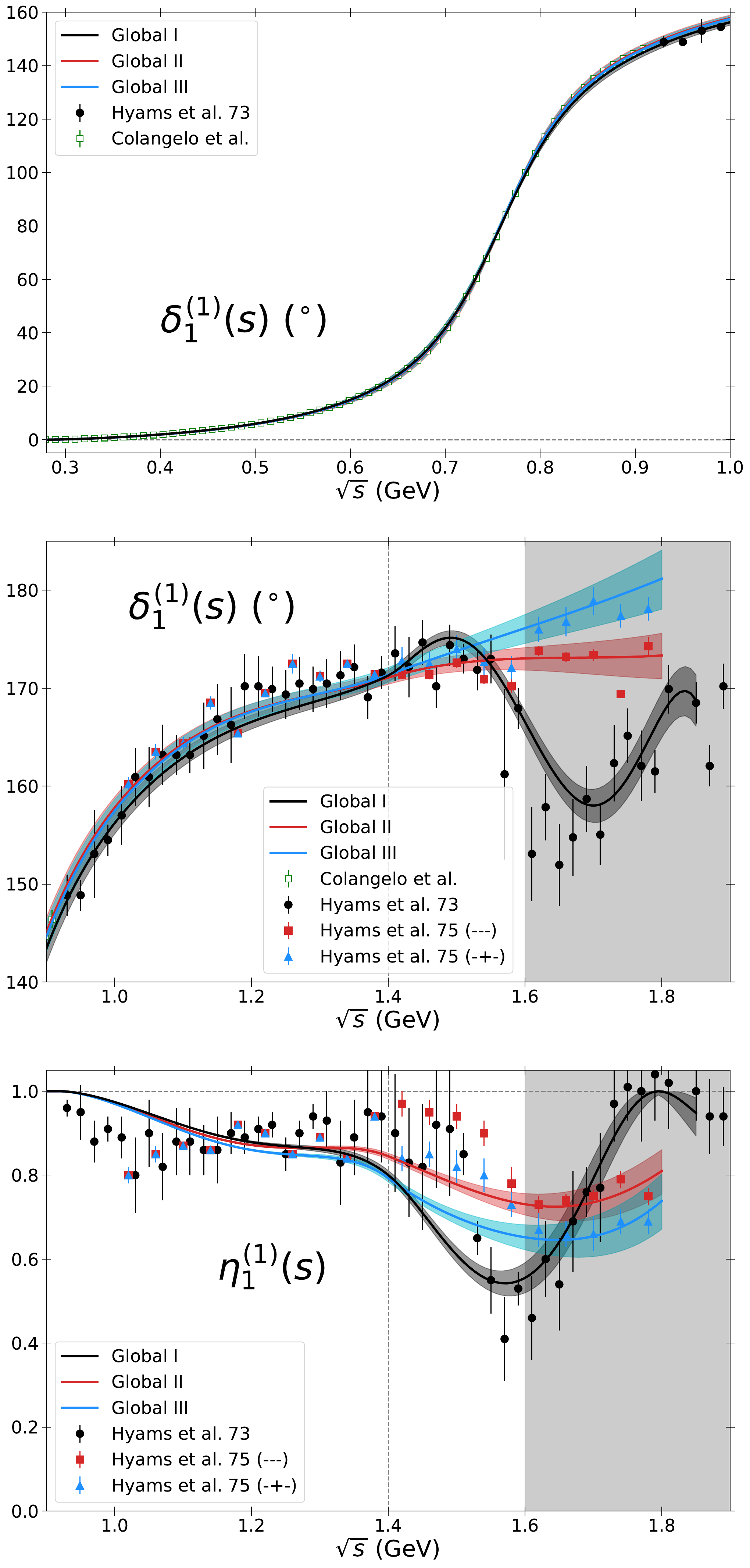}
\caption{Comparison among the P-wave Global Fits I, II, and III.
They are quite similar below the matching point at 1.4 GeV, although, strictly speaking, their elasticities are incompatible within uncertainties. Above 1.4 GeV, the three Global Fits differ qualitatively in the phase shift and elasticity. Moreover, the dispersive constraints drag them all slightly below their respective elasticity data, softening the maximum around 1.4 GeV present in the unconstrained fits. Recall we have only imposed dispersive constraints up to 1.6 GeV. Above that energy (shaded region) they are mere unconstrained data fits. Note that we only plot our Global Fits up to the round energy value closest to the last data point in the fit.
Experimental data are taken from  \cite{Hyams:1973zf} (Hyams et al. 73)
and~\cite{Hyams:1975mc} (Hyams et al. 75) .}\label{fig:P_sols}
\end{figure}

In Fig.~\ref{fig:P_sols}, we show the P-wave phase shift and elasticity for the final Global Fits I, II, and III, obtained following the procedure described in previous subsections but applied to data Solutions I, II, and III, respectively. 
We provide their parameters in Table~\ref{tab:Pparameters}.

These fits have a very similar phase shift up to 1.4 GeV. Their elasticities are, strictly speaking, incompatible among themselves due to their small uncertainty. However, the qualitative behavior of the three elasticities is very similar, with a smooth decrease up to 1.2 GeV, where they flatten out to decrease again around 1.35 GeV.
However, from 1.4 to 1.8 GeV there is a clear qualitative difference in both the phase shift and elasticity.
In particular, the elasticity of Solution I decreases much faster and has a clear minimum slightly above 1.55 GeV. In addition, the fact that the phase shift data of Solutions II and III have less structure, allows us to describe these fits with lower-degree Chebyshev polynomials. Namely, for the Global Fits II and III, we have set to zero the $d_4, d_5$ and $d_6$ coefficients in Table~\ref{tab:Pparameters}.

Despite these differences, we have been able to make the three Global Fits satisfy FDRs by deforming slightly their phase shift but visibly their elasticity. Namely, in all cases, the elasticity bump around 1.4 GeV is less pronounced than in the data, and in the 1.4 to 1.6 GeV region the dispersive representation prefers a lower $\eta(s)$ than suggested by the data. However, the three of them still present a clear change in the behavior of the elasticity from convexity to concavity and back, absent in the CFD and old global fits, and most likely related to the presence of the $\rho(1450)$ resonance.

It is important to recall that, although our global parametrizations extend up to 1.8 GeV (1.85 GeV for Global Fit I), they are only dispersively constrained up to 1.6 GeV with the $F^{0+}$
and $F^{I_t=1}$ FDRs. In the region beyond that energy, marked with a gray background in Fig.~\ref{fig:P_sols}, they are just mere fits to data with the parametrization used above 1.4 GeV. Note also that we only plot our Global Fits up to the round value of the energy after the last data point included in each fit (for this wave, 1.85 GeV for Global Fit I and 1.8 GeV for Global Fits II and III). Extrapolating our fits beyond that energy is meaningless and may produce oscillations that are mere artifacts.

\subsection{D0-wave parametrization}

This is an attractive wave, largely dominated by the well-established
$f_2(1270)$ resonance. For Solution I, a few data points in~\cite{Hyams:1973zf} between 0.9 and 1 GeV suggest a very small inelasticity. Thus, contrary to what was done in~\cite{GarciaMartin:2011cn}, we now allow the inelasticity of the Global Fit I to open at 0.9 GeV, which is just an effective phenomenological threshold. 
A non-vanishing inelasticity below the $K\bar K$ threshold is suggested not only by Solution I but also by the fact that the $f_2(1270)$ resonance, which dominates this channel, lies only one and a half widths away from the $K\bar K$ threshold, and its branching ratio to $4\pi$ is twice as large as that to $K\bar K$.
 However, for Global Fits II and III, 
we will keep the parametrization elastic up to the $K\bar K$ threshold, $\sqrt s\sim 0.992\,$~GeV, since the original data analysis~\cite{Hyams:1975mc} explicitly  {\it imposed} the onset of inelasticity at this point. In that reference, a K-matrix formalism was employed with only the $\pi\pi$ and $K\bar K$ channels, where the $K\bar K$
channel was used to effectively parametrize ``all the inelastic channels"~\cite{Hyams:1975mc}. 

As usual, we will illustrate our procedure using the Global Fit I. Below 1.4 GeV, we will adopt a parametrization very similar to that used in~\cite{GarciaMartin:2011cn}, which describes the phase shift in two parts matched at $2m_K$. However, now we will ensure that the matching is also differentiable. Additionally, we will extend the fit up to 1.9 GeV.

\subsubsection{D0 wave below 1.4 GeV} 

When $s^{1/2}\leq 2m_K$, 
we parametrize the phase shift by:
\begin{eqnarray}
&&\cot\delta_2^{(0)}(s)=\frac{m_{f_2}^2-s }{\sigma(s) k(s)^4}m_\pi^2\big(
B_0+B_1 w_l(s)\big),
\nonumber\\
&&w_l(s)=\frac{\sqrt{s}-\sqrt{s_l-s}}{\sqrt{s}+\sqrt{s_l-s}},\quad
s_l^{1/2}=1.05 \,\gev.
  \label{eq:D0lowparam} 
\end{eqnarray}
In~\cite{GarciaMartin:2011cn}, the $f_2(1270)$ resonance ``peak-mass" was fixed 
at $m_{f_2}=1275.4 \,\mev$, the central value in the Review of Particle Physics~\cite{ParticleDataGroup:2022pth}.
However, here we will also consider its uncertainty $\Delta m_{f_2}=\pm 0.8 \mev$.

In the intermediate energy region, $2m_K  \leq s^{1/2}\leq 1.4 \,\gev$,
we use a rather similar parametrization:
\begin{eqnarray}
&&\cot\delta_2^{(0)}(s)=\frac{m_{f_2}^2-s }{\sigma(s) k(s)^4}m_\pi^2 
\sum_{n=0}^2 B_{h n} \,w_h(s)^n, \nonumber\\
&&w_h(s)=\frac{\sqrt{s}-\sqrt{s_h-s}}{\sqrt{s}+\sqrt{s_h-s}},\quad
s_h^{1/2}=1.45 \,\gev.
  \label{eq:D0highparam} 
\end{eqnarray}
By imposing continuity and, now, also differentiability at the matching point $s_K=4m_K^2$, the parameters $B_{h0}$ and $B_{h1}$ are determined as follows:
\begin{align}
B_{h1}=&B_1 \frac{w_l'(s_K)}{w_h'(s_K)}-2B_{h2}w_h(s_K), \nonumber\\
B_{h0}=&B_0+B_1 w_l(s_K)-B_{h1} w_h(s_K)-B_{h2} w_h(s_K)^2, \nonumber\\
   \label{eq:D0match_mk} 
\end{align}
where the prime denotes differentiation with respect to $s$, such that:
\begin{equation}
  \frac{w_l'(s_K)}{w_h'(s_K)}=
\frac{s_l}{s_h}\frac{\sqrt{s_h-s_K}}{\sqrt{s_l-s_K}}
\left( \frac{\sqrt{s_K}+\sqrt{s_h-s_K}}{\sqrt{s_K}+\sqrt{s_l-s_K}}\right)^2.
\end{equation}

For Global Fit I, we allow the elasticity to differ from 1 at $s>\hat s=(0.9 \, \gev)^2$. In contrast, 
for Global Fits II and III, we set $\hat s=4m_K^2$.
For $\hat s\leq s \leq (1.4 \, \gev)^2$ we write:
\begin{equation}
\eta_2^{(0)}=
1-\epsilon\left( \frac{1-\hat s/s}{1-\hat s/m_{f_2}^2} \right)^{5/2}
\!\left[1+r\left(1-\frac{s-\hat s}{m_{f_2}^2-\hat s}\right)\right]\!\!.
\label{eq:D0etalow}
\end{equation}

This parametrization is built up so that at the resonance peak $\eta^{(0)}_2=1-\epsilon$.
Moreover, when the branching ratio of a naive Breit-Wigner shape to a given channel is bigger than 0.5, as it is the case here for the $\pi\pi$ decay,  that branching ratio can be written as $1-\epsilon/2$  (otherwise it would be $\epsilon/2$).
With this very naive approximation, we find an $f_2(1270)$ branching ratio to $\pi\pi$ of
$0.871\pm0.008$. This is consistent with the value obtained from the RPP\cite{ParticleDataGroup:2022pth}, namely $\Gamma_{\pi\pi}/\Gamma_\text{tot}= 0.843^{+0.028}_{-0.010}$, with $\Gamma_{\pi\pi}$ and $\Gamma_\text{tot}$ the partial width to $\pi\pi$ and the total width of the $f_2(1270)$, respectively.

\subsubsection{D0 wave above 1.4 GeV} 

As in other cases, we describe the partial-wave phase shift above $s_m^{1/2}=1.4 \, \gev$ employing Chebyshev polynomials of the first kind. For this, we once again map the region of interest onto the $[-1,1]$ segment using the following variable
\begin{equation}
x(s)=2\frac{\sqrt{s}-\sqrt{s_m}}{2 \gev-\sqrt{s_m}}-1. 
\end{equation}
Then, we write:
\begin{eqnarray}
   \delta^{(0)}_2(s)=\sum_{n=0}^4 \d_n\, p_{n}(x(s)), 
\end{eqnarray}
where to ensure a continuous and differentiable matching, we set
\begin{align} 
    d_1&=\frac{\delta^{(0)\,\prime}_2(s_m)}{x'(s_m)}+\sum_{n=2}^4 (-1)^nn^2d_n, \nonumber\\
    d_0& = \delta^{(0)}_2(s_m)-\sum_{n=1}^4 (-1)^nd_n.
   \label{eq:matchD014} 
\end{align}

Concerning the elasticity, we use: 

\begin{equation}
\eta^{(0)}_2(s)=\exp\left[-\left(\sum_{n=0}^{4} \epsilon_n \left(\frac{s}{s_m}-1\right)^n\right)^2\right],
    \label{eq:highinelD0} 
\end{equation}
and the continuous and differentiable matching requires:
\begin{equation}
    \epsilon_0=\sqrt{-\log(\eta^{(0)}_2(s_m))}, \quad \epsilon_1=-\frac{s_m}{2\epsilon_0}\frac{{\eta^{(0)\,\prime}_2}(s_m)}{{\eta^{(0)}_2}(s_m)}.
\end{equation}

The quantities $\delta^{(0)}_2(s_m)$, $\delta^{(0)}_2\,^\prime(s_m)$, $\eta^{(0)}_2(s_m)$ and $\eta^{(0)}_2\,^\prime(s_m)$ are calculated from the parametrization below 1.4 GeV.

\subsubsection{D0-wave Global Fit and parameters}

\begin{figure}
\centering
\includegraphics[width=0.48\textwidth]{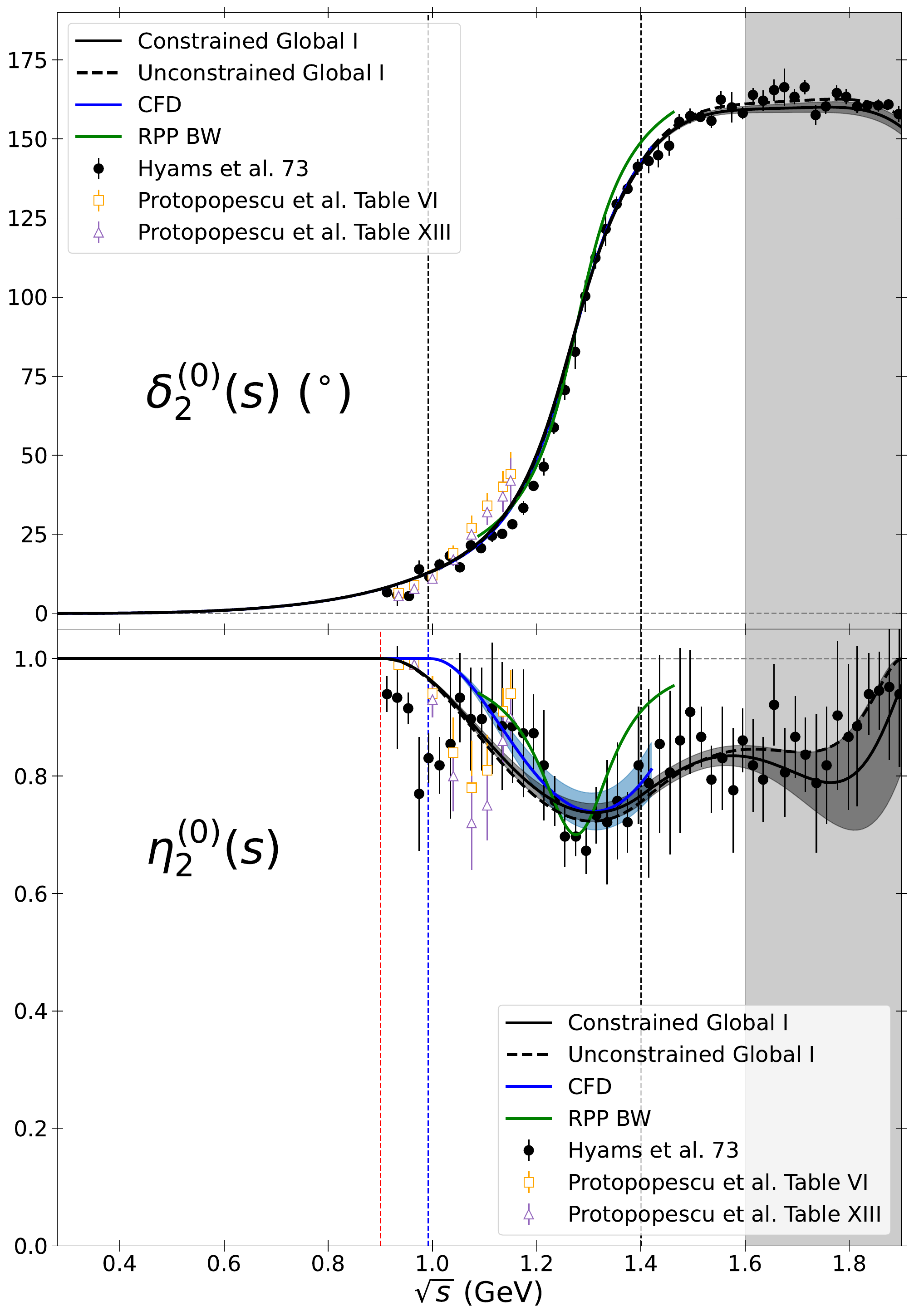}
\caption{D0-wave phase shift (top) and elasticity (bottom).  We show our unconstrained and dispersively constrained Global Fit I, together with the CFD from~\cite{GarciaMartin:2011cn}. Experimental data, only plotted and fit above 0.9 GeV, are taken from Hyams et al.~\cite{Hyams:1973zf} and Protopopescu et al.~\cite{Protopopescu:1973sh}, although the last ones are not included in the fit. Below 0.9 the input is the CFD and the $a_2^{(0)}$ and $b_2^{(0)}$ CFD threshold parameters in Table~\ref{tab:ThresholdParameters} of the
Appendix~\ref{app:Threshold}.
The black vertical lines stand at matching points described in the text. The red vertical line at 0.9~GeV marks the new onset of inelasticity, which was at $2m_K\simeq0.992\gev$ for the CFD (blue vertical line).}\label{fig:D0}
\end{figure}

In Fig.~\ref{fig:D0}, we show the scattering data used for our fits above 0.9 GeV for Global Fit I, whose parameters are given in Table~\ref{tab:D0parameters}. Below 0.9 GeV the input consists of the dispersively constrained CFD set of~\cite{GarciaMartin:2011cn}, as well as the  $a^{(0)}_2$ and $b^{(0)}_2$ threshold parameters, obtained from the CFD in~\cite{GarciaMartin:2011cn}, and listed in Table~\ref{tab:ThresholdParameters} of Appendix~\ref{app:Threshold}. The data of Protopopescu et al.~\cite{Protopopescu:1973sh} only reaches up to 1.15 GeV, and typically have much larger uncertainties than those of Hyams et al.~\cite{Hyams:1973zf}, which would dominate any fit. For this reason and other inconsistencies in the F wave, to be discussed below, we only include the set from \cite{Hyams:1973zf} in the fit.

The phase shift, shown in the upper panel of Fig.~\ref{fig:D0}, exhibits a distinct resonant shape characterized by a sharp $\sim160^\circ$ rise, due to the $f_2(1270)$ resonance. The elasticity, depicted in the lower panel, features a dip in the same energy region. 

In the 0.9-1.1 GeV range, the Global Fit I phase shift describes well the Hyams et al.~\cite{Hyams:1973zf} data, but lies between them and the Protopopescu et al.~\cite{Protopopescu:1973sh} sets in the
1.1 to 1.25 GeV region. 
Below 1.6 GeV, there are no significant differences between our dispersively constrained Global Fit I and the unconstrained one. Above that energy, the constrained
elasticity is clearly lower but remains within one deviation from the unconstrained one.

Our new phase shift is indistinguishable from the CFD result from~\cite{GarciaMartin:2011cn} up to where it was defined, i.e., 1.42 GeV. In contrast, there are two important differences above that energy.
First, our Global Fit I now reaches 1.9 GeV (and should not be used beyond). Second,
the inelasticity is now allowed to open at 0.9 GeV instead of the $K \bar K$ threshold. This value should be regarded as an effective scale since some experiments---though not all---observe a non-zero inelasticity slightly above this point. Consequently, the elasticity is lower than that of the CFD in the 0.9-1.2 GeV range, although it becomes consistent with the CFD again beyond this region. 

The lowest six data points from Hyams et al.~\cite{Hyams:1973zf} systematically lie below our curve, which, nevertheless, is very compatible with the data from Protopopescu et al.~\cite{Protopopescu:1973sh}.
Note that both sets of data exhibit a dip in the elasticity, although at energies about 100 MeV apart. The data from~\cite{Hyams:1973zf} reach the minimum of its dip ($\eta^{(0)}_2\simeq 0.8$), slightly before 1 GeV, while the data from~\cite{Protopopescu:1973sh} reach their local minimum ($\eta^{(0)}_2\simeq 0.75-0.8$) slightly before 1.1 GeV. In neither case there is a significant associated change in the phase shift. hence, we suspect these different dips might be just artifacts (see the discussion on the F wave below).
Above 1.05 GeV, both data sets are roughly consistent within uncertainties.

Note we have studied the possibility of maintaining the opening of the inelasticity at the $K \bar K$ threshold, as in the CFD~\cite{GarciaMartin:2011cn}. However, 
as we will see in Sec.~\ref{sec:disprel}, forward dispersion relations clearly favor its onset at 0.9 GeV. 
This preference is not only evident in Global Fit I, but also in Global Fits II and III, which, as discussed below, are based on solutions whose inelasticity was set to begin at the $K \bar K$ threshold.

Finally, for illustration, we show in Fig.~\ref{fig:D0} how a naive Breit-Wigner (BW) formula compares to our global parametrization. This is represented by a continuous green line that covers the energy range of plus or minus one width around the peak.  The BW has $\sim85\%$ elasticity at the peak, and for its parameters we have taken the central values of the $f_2(1270)$ average mass and width provided at the RPP~\cite{ParticleDataGroup:2022pth}. 
We see that for the phase shift, this shape is almost identical to our parametrization between 1.1 and 1.3 GeV. In particular, in the 1.1 to 1.2 GeV region where the two data sets are
in clear tension, it is very close to our parametrization, that lies in between both sets.
However, starting around 1.3 GeV, there is a growing deviation of the BW with respect to our global parametrization, which is more pronounced in the elasticity. This asymmetry between the two sides of the Breit-Wigner may be due to the presence of the $K\bar K$ threshold on the left but also to the well-established $f_2^\prime(1525)$ resonance on the right, whose estimated width at the RPP is $86\pm5\mev$. Nevertheless, the $f_2^\prime(1525)$ coupling to two pions is very small, and thus, its shape might go almost unnoticed in $\pi\pi$ scattering, while still producing a small deformation in the naively expected $f_2(1270)$ shape.

\begin{widetext}
    \begin{center}
\begin{table}[h]
\renewcommand{\arraystretch}{1.3}
  \centering
  \begin{tabular}{lcccc}
 D0 wave    & Parameters & Global I values& Global II values& Global III values\\
 \toprule
 &$m_{f_2}$& $1274.6\pm0.8$ MeV& $1275.6\pm0.8$ MeV& $1275.4\pm0.8$ MeV\\
 &$\hat s$& $(0.9\,\gev)^2$& $4m_K^2$ & $4m_K^2$\\
\colrule
\multirow{2}{*}{$\delta^{(0)}_2\big\vert_{s<4m_K^2}$}& $ B_0$ &$12.34\pm0.13$ &$12.42\pm0.13$&$12.48\pm0.13$  \\
 &  $B_1$&$10.12\pm0.15$&$10.00\pm0.15$&$10.02\pm0.15$\\
\colrule
$\delta^{(0)}_2\big\vert_{4m_K^2<s<s_m}$ & $B_{h2}$ & $4.5\pm1.8$& $33\pm4$& $34\pm5$\\[3pt]
\colrule
 \multirow{ 2}{*}{$\eta^{(0)}_2\big\vert_{\hat s<s<s_m}$}&$\epsilon$&$0.258\pm0.015$ &$0.317\pm0.026$&$0.322\pm0.024$ \\
 & $r$&$0.94\pm0.03$&$1.10\pm0.04$&$1.16\pm0.03$ \\
\colrule
\multirow{ 3}{*}{$ \delta^{(0)}_2\big\vert_{s>s_m}$}&$ d_2$ &$(-12.18\pm0.17)\, ^{\circ}$ &$(-39.7\pm0.5)\, ^{\circ}$&$(-82.4\pm0.5)\, ^{\circ}$\\
&$ d_3$ &$(-1.94\pm0.05)\, ^{\circ}$&$(-19.28\pm0.14)\, ^{\circ}$&$(-42.65\pm0.14)\, ^{\circ}$ \\
&$ d_4$ &$(-3.01\pm0.03)\, ^{\circ} $ &$(-8.86\pm0.07)\, ^{\circ} $&$(-14.38\pm0.07)\, ^{\circ} $\\
\colrule
\multirow{ 3}{*}{$\eta^{(0)}_2\big\vert_{s>s_m}$}&$\epsilon_2$&$0.22\pm0.13$&$11.99\pm0.11$&$10.20\pm0.20$\\
&$\epsilon_3$&$3.71\pm0.22$&$-25.49\pm0.23$&$-18.8\pm0.4$\\
&$\epsilon_4$&$-4.3\pm0.4$&$16.3\pm0.4$&$14.2\pm0.7$\\
\botrule
  \end{tabular}
  \caption{D0-wave parameters of the constrained Global Fits I, II, and III.
  Recall that $s_m=(1.4 \gev)^2$ for this wave.}
  \label{tab:D0parameters}
\end{table}
    \end{center}
\end{widetext}

\subsubsection{The three D0-wave Global Fits}

\begin{figure}
\centering
\includegraphics[width=0.46\textwidth]{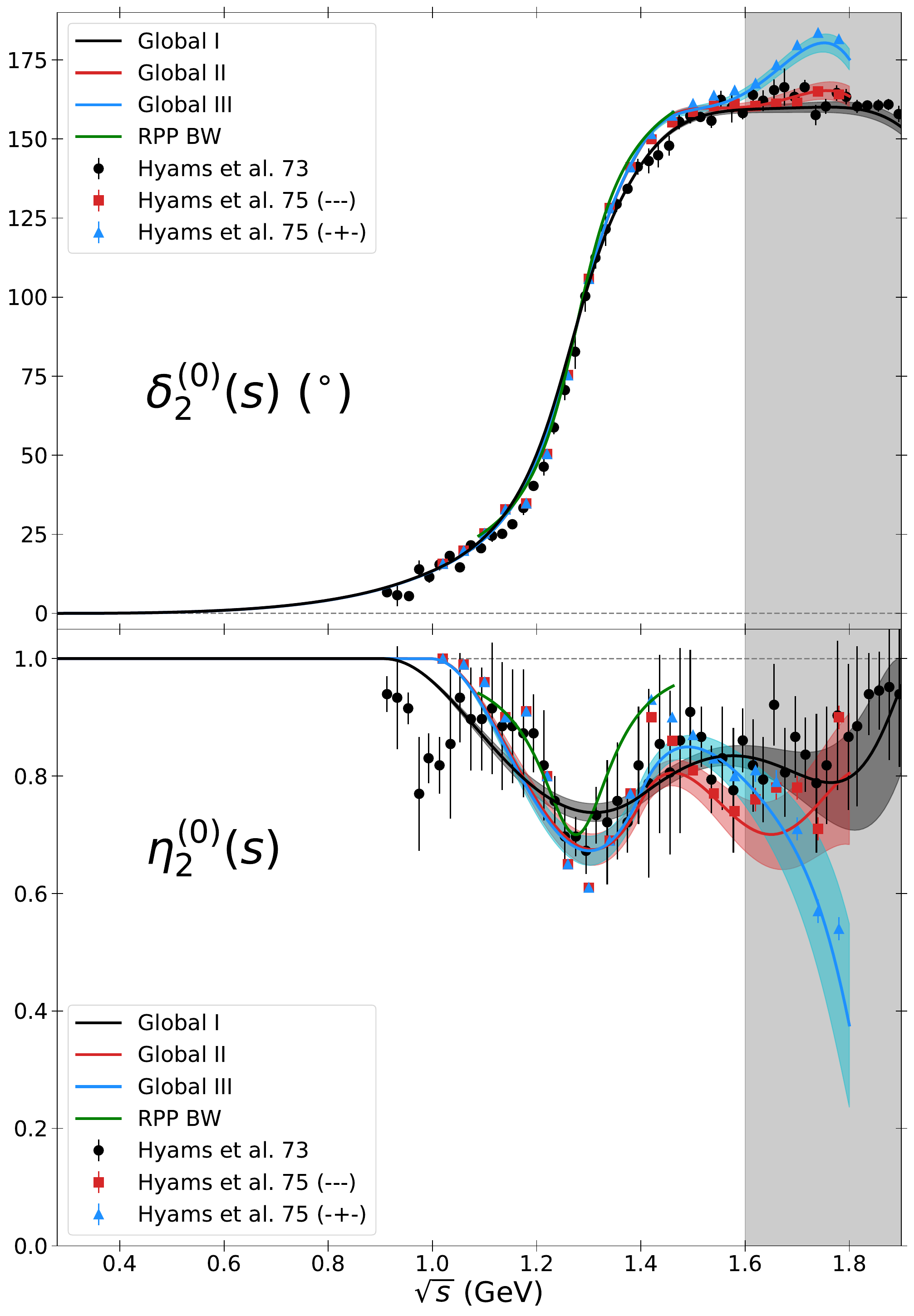}
\caption{Comparison among Global Fits I, II, and III for the D0 wave. Their phase shifts are identical up to 1.3 GeV and deviate above that energy. The elasticities are rather different in the whole energy range since the inelasticity opens up at different energies. The behavior of all three fits is similar, although they are incompatible in several regions.  Above 1.6 GeV (shaded region) they are pure data fits since we have only imposed dispersive constraints up to 1.6 GeV. Note that we only plot our Global Fits up to the round energy value closest to the last data point included in the fit. 
The data comes from~\cite{Hyams:1973zf} (Hyams et al. 73)
and from \cite{Hyams:1975mc} (Hyams et al. 75).}\label{fig:D0_sols}
\end{figure}

In Fig.~\ref{fig:D0_sols}, we now show the final result for Global Fits I, II, and III after imposing the dispersive constraints.
Their phase shifts are almost indistinguishable below 1.3 GeV. However, at this energy, Fits II and III begin to increase slightly faster than Fit I, which explains why their $B_{h2}$ parameter is larger than that of Global Fit I.
In contrast, their elasticities show notable discrepancies below 1.4 GeV, primarily due to their differing onset energies.
Beyond 1.4 GeV, the three Global Fits exhibit significant differences either in the phase shift, elasticity, or both. 
For this reason, the elasticity parameters $\epsilon, r,\epsilon_2,\epsilon_3$, and $\epsilon_4$ in Table~\ref{tab:D0parameters} are quite different among the three Global Fits.

Here, it is worth recalling our pending comment that the
 inelasticity of Solutions II and III was set to begin at the $K \bar K$ threshold. Actually, they were obtained using a D0 wave parametrized as a two-channel K-matrix only above the $K \bar K$ threshold. In Fig.~\ref{fig:D0_sols} we see
that the data of Solutions II and III below 1.6 GeV do not have error bars because in~\cite{Hyams:1975mc} uncertainties were only provided above that energy. Consequently, to obtain reasonable error bands, we have fit these data assuming an uncertainty similar to that of Solution I.
In addition, it is worth noticing that 
the data from Solutions II and III overlap below 1.4 GeV, making only one set visible in Fig.~\ref{fig:D0_sols}.
Finally, in these two solutions, the elasticity was set to $\eta_2^{(0)}(s)=1$ below $K \bar K$ threshold. As explained in~\cite{Hyams:1975mc} ``in fact, all inelastic effects" are parametrized in the $K\bar K$ channel of their two-channel K-matrix.

 Imposing the dispersive representation yields the uncertainties reported for the Global II and III fit parameters in Table~\ref{tab:D0parameters}. 
While the fulfillment of forward dispersion relations improves when the inelasticity onset is set at 0.9 GeV, we have adhered to the data-analysis assumptions in~\cite{Hyams:1975mc} and kept it at the $K\bar K$ threshold for these two fits.

Nevertheless, we emphasize that, within our formalism, we cannot specify the origin of the inelasticity. We just set a phenomenological threshold based on where the data suggests the process is no longer elastic.
For instance, our inelasticity may well be due in part to 
the $4\pi$ channel, known to play a significant role in the $f_2(1270)$ decays. Most likely such a contribution dominates the inelasticity below the $K \bar K$ threshold.  However, how much of the inelasticity arises from a particular channel cannot be discerned within our approach. 
We only deal with the total $\eta_\ell^{(I)}$.
On the positive side, our formalism avoids the model-dependence
required to identify how many channels are open and what they are made of.

Note also that the central values of the final Global Fits II and III deviate significantly more from their original data sets than the deviation observed between Global Fit I and its original data.  In addition, while the overall fulfillment of FDRs is comparably good on average across the entire energy region where they are applied as constraints, the fulfillment of the $I_t=1$ FDR in the 0.93-1.06 GeV region is somewhat worse for Global Fits II and III than for Global Fit I. Actually, in that region, the average squared distance between the input and the dispersive output of Global Fits II and III is larger than one, as will be discussed quantitatively in section~\ref{sec:disprel}. 
This poorer fulfillment is primarily due to the choice of energy where the 
D0-wave inelasticity opens; we have explicitly checked that the FDR fulfillment improves when the inelasticity onset is set to 0.9 GeV.  
We believe that this original choice made for these two solutions, setting the D0-wave inelasticity onset at the $K\bar K$ threshold, that we have followed here, not only worsens the FDR fulfillment in that region but also influences the shape of the partial wave at higher energies. This caveat and the previously mentioned lack of uncertainties in the original Solutions II and III data, make the D0-wave Global Fits II and III slightly disfavored compared to Global Fit I.

It is also worth noticing that the phase shift around the $f_2(1270)$ of the three Global Fits has a rather similar shape to a BW resonance,
but that is not the case for the elasticity, 
since it is very asymmetric. For illustration, we have also represented in Fig.~\ref{fig:D0_sols} a BW shape as a green line using the present $f_2(1270)$ RPP parameters~\cite{ParticleDataGroup:2022pth}.
Recall that, as explained right after Eq.~\eqref{eq:D0etalow},  the branching ratio is $1-\epsilon/2$ when naively using a BW. Thus, from Table~\ref{tab:D0parameters}, we see that the $f_2(1270)$ branching ratio to two pions is smaller for Global Fits II and III than for Fit I:
respectively $0.842\pm0.013$, $0.839\pm0.012$ and $0.871\pm0.008$.

Once again, we recall that our Global Fits are only dispersively constrained up to 1.6 GeV. Beyond that energy (the shaded region in Fig.~\ref{fig:D0_sols}), they are just simple fits to data. Moreover, Global Fit I should not be used beyond 1.9 GeV, nor Global Fits II and III beyond 1.8 GeV.

\subsection{F-wave parametrization}

This wave is attractive and, above 1.2 GeV, is dominated by the $\rho_3(1690)$, which has a width of $161\pm10$ MeV \cite{ParticleDataGroup:2022pth}. However, its shape differs significantly from the P and D0 waves, since this is a highly inelastic wave in the resonant region, with a branching ratio to $\pi\pi$ of only $\sim 23\%$. As a result, its phase shift is small up to 1.9 GeV, where data exist. The RPP also lists a second F-wave resonance at 1.99 GeV, but this state is currently omitted from the summary tables. Given this situation--- and considering that we will fit scattering data up to 1.9 GeV and use this parametrization inside dispersive integrals only below 1.6 GeV---we do not find it necessary to try to describe this second non-confirmed state.

In~\cite{GarciaMartin:2011cn}, the dispersive representation was studied only up to 1.4 GeV, where the role of the F wave was almost insignificant.
A very naive, non-resonant parametrization was used to describe the data below 1.4 GeV, along with the scattering length obtained from sum rules. The inelasticity was neglected below 1.4 GeV.

Thus, we propose here a completely new parametrization. It will describe the phase shift data, starting somewhat above 0.9 GeV, as well as the elasticity data that starts around 1.3 GeV. We will also fit the scattering length, $a^{(1)}_3$, and effective range, $b^{(1)}_3$, obtained from the CFD in~\cite{GarciaMartin:2011cn} and listed in Table~\ref{tab:ThresholdParameters} of Appendix~\ref{app:Threshold}. In this region, the effect of the resonance at 1.69 GeV is negligible.
For this reason, our parametrization will have an elastic low-energy part, described by a conformal expansion, and a high-energy part, an inelastic Breit-Wigner parametrization with a Blatt-Weisskopf barrier, as used in~\cite{Hyams:1973zf}.

In particular, for the phase shift at $s^{1/2}\leq s_m^{1/2}=1.2\, \gev$ we use:
\begin{eqnarray}
&&\cot\delta_3^{(1)}(s)=\frac{m_\pi^{6} }{\sigma(s)\,k(s)^6}
\sum_{n=0}^3 B_{n} \,w(s)^n, \nonumber 
\\
&&w(s)=\frac{\sqrt{s}-\alpha \sqrt{s_0-s}}{\sqrt{s}+\alpha \sqrt{s_0-s}},\quad
s_0^{1/2}=1.5 \,\gev, \quad \alpha=\frac{1}{2}.\nonumber\\
\label{eq:Flowparam}
\end{eqnarray}

No inelasticity is observed below 1.3 GeV, however, we have conservatively allowed the wave to be formally inelastic from the $\pi\omega$ threshold.  Nevertheless, after constraining the fit, the inelasticity is only visible above 1.2 GeV. Hence, we set $\eta_3^{(1)}(s)=1$ for $s$ below
$\hat s=(m_{\pi} + m_{\omega})^2$. From that energy to $s_m^{1/2}=1.2\, \gev$, we use,
\begin{eqnarray}
\eta_3^{(1)}(s)=1-\epsilon\left(1-\frac{\hat s}{s}\right)^{7/2}\left(1+r\left(1-\frac{s_m}{s}\right)\right).
\label{eq:Flowparamine} 
\end{eqnarray}

The requirement of a continuous and differentiable matching at $s_m^{1/2}=1.2$ \gev 
fixes the $B_0$ and $B_1$ parameters in Eq.~\eqref{eq:Flowparam}, and the $\epsilon$ and $r$ in Eq.~\eqref{eq:Flowparamine} as follows:
\begin{align}
   \epsilon  =& \Big(1-\eta_3^{(1)}(s_m)\Big)\left(1-\frac{\hat s}{s_m}\right)^{-7/2}, \nonumber\\
    r =& -\frac{s_m\,\eta_3^{ (1)\,\prime}(s_m)}{1-\eta_3^{(1)}(s_m)} + \frac{7}{2}\frac{\hat s}{s_m-\hat s},\nonumber \\
    B_1 =& \frac{1}{m_{\pi}^6w'_m}\left(\sigma(s) k(s)^6 \cot \delta_3^{ (1)}(s)\right)^{\prime}\bigg\vert_{s=s_m}\nonumber \\
    &-2B_2w_m-3B_3w_m^2, \nonumber\\
    B_0 =& \frac{\sigma(s_m) k(s_m)^6}{m_{\pi}^6} \cot \delta_3^{ (1)}(s_m) -\sum_{n=1}^3 B_n w_m^n, 
\end{align}
where $w_m=w(s_m)$, $w'_m=w'(s_m)$ and the phase shift, elasticity, and their first derivatives 
at $s_m$ in the previous expressions must be
calculated using Eqs.~\eqref{eq:BWBW1} and Eq.~\eqref{eq:BWBW2} below, which provide the expression for the high-energy region, to be explained next.
Note that,  unlike the waves discussed so far, in this case, we fix the parameters of the low-energy part using values from the high-energy parametrization (provided below). We have adopted this matching scheme because the F wave is much better known in the high-energy region. Consequently, the uncertainty of the threshold parameters from the Global Fit will be larger than from their input (see Table~\ref{tab:ThresholdParameters}).

The region $1.2\, \gev\leq s^{1/2}\leq 1.9 \,\gev$ lies within three widths of the  $\rho_3(1690)$ resonance. Thus, our parametrization in this range employs an inelastic Breit-Wigner shape with a potential barrier, similar to the approach used by the experimentalists in~\cite{Hyams:1973zf}:
\begin{eqnarray}
&&  t^{(1)}_3(s)=\frac{1}{\sigma (s)}\frac{x_{\rho_3} m_{\rho_3}\Gamma_{\rho_3}(s)}{m_{\rho_3}^2-s-im_{\rho_3}\Gamma_{\rho_3}(s)},
\label{eq:BWBW1}
\end{eqnarray}
where
\begin{align}
\Gamma_{\rho_3}(s) =& \Gamma_{\rho_3}\left(\frac{k(s)}{k_{\rho_3}}\right)^7\frac{D_3(k_{\rho_3}R_{\rho_3})}{D_3(k(s)R_{\rho_3})}, 
\nonumber\\
 D_3(x)=&225 + 45x^2 + 6x^4 + x^6, \nonumber\\
k_{\rho_3}=&k(m_{\rho_3}^2), \label{eq:BWBW2}
\end{align}
are the familiar Blatt-Weisskopf~\cite{Blatt:1952ije} angular momentum barrier factors.
For brevity, we will refer to this functional form as BWBW. This equation only differs significantly from the naive BW formula, i.e., with a constant width $\Gamma_{\rho_3}$ in Eq.~\eqref{eq:BWBW1}, for energies far from $m_{\rho_3}$.

Why not use this simple BWBW shape in the whole range?
There are several reasons. First, even though the $k(s)^7$ factor provides the expected threshold leading behavior, it does not account for the subleading contributions. In addition, Eq.~\eqref{eq:BWBW1} corresponds to an inelastic partial wave, although at threshold, it should be elastic. Thus, Eq.~\eqref{eq:BWBW1} alone yields threshold parameters that are very inconsistent with those obtained from sum rules with the dispersive representation in~\cite{GarciaMartin:2011cn} or with ChPT.
Although it is possible to adjust the numerator of the BW shape---e.g., by promoting the $x_{\rho_3}$ parameter to a function of $s$---to reproduce these threshold parameters with a shape similar to our Global Fits, the resulting expression would fail to remain elastic in the elastic region.
In particular, elastic unitarity implies $\im t^{(1)}_3/\sigma \vert \,t^{(1)}_3 \vert^2=1$, and with such a naively modified inelastic BWBW formula, this ratio is off by an order of magnitude, even though $\eta^{(1)}_3$ may still be very close to one. This has enormous consequences for the FDRs.

For these reasons, our global parametrization has a strictly elastic piece in the elastic region
matched to the BWBW shape in the resonance region.
These observations will also drive our choice of parametrization for the G0 wave.

\begin{figure}
\centering
\vspace{.2cm}
\includegraphics[width=0.48\textwidth]{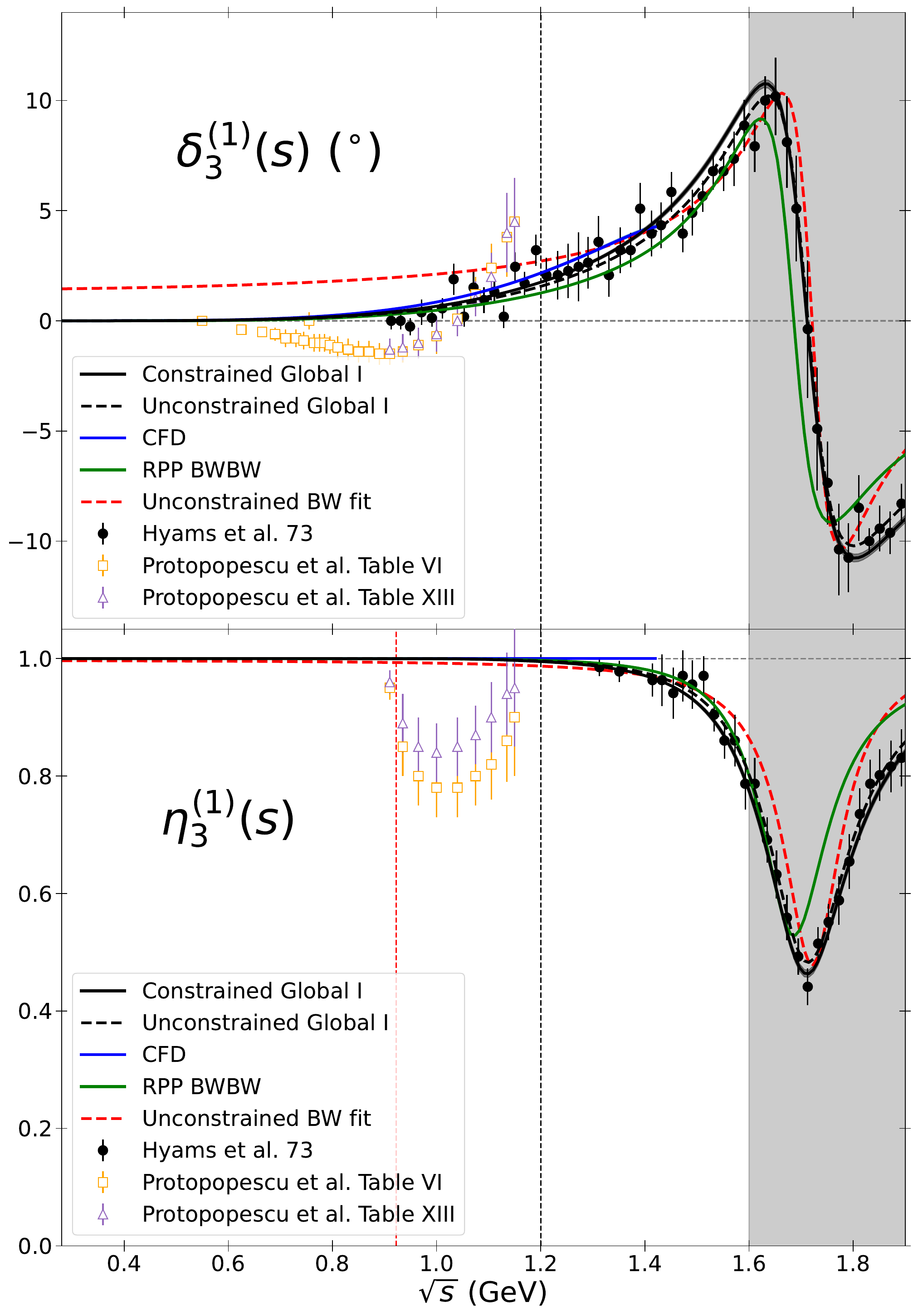}
\caption{F-wave phase shift (top) and elasticity (bottom). We show our unconstrained and dispersively constrained Global Fit I, together with the CFD  from~\cite{GarciaMartin:2011cn}. Our new parametrization fits the Hyams et al.~data from~\cite{Hyams:1973zf}.
The data from Protopopescu et al.~\cite{Protopopescu:1973sh} are only included for completeness.
The red vertical line stands at the allowed onset of the inelasticity, whereas the black one marks the energy of the matching with the BWBW parametrization Eqs.~\eqref{eq:BWBW1} and \eqref{eq:BWBW2}. For illustration, we also show a BWBW curve with the RPP~\cite{ParticleDataGroup:2022pth} estimates for the mass and width. We also illustrate the inability of a simple BW parametrization to describe the data and provide a good low-energy description.
}\label{fig:F}
\end{figure}

\subsubsection{F-wave Global Fit and parameters}

This parametrization is entirely new. Therefore, we do not use CFD as input at low energies.
Instead, we use the threshold parameter values $a^{(1)}_3$ and $b^{(1)}_3$ provided in Appendix~\ref{app:Threshold}. In addition, we fit the CERN-Munich scattering data
of Hyams et al.~\cite{Hyams:1973zf}, shown with solid symbols in Fig.~\ref{fig:F}.
This wave is very small, and its phase shift remains below $2^\circ$ up to 1.1 GeV, where the influence of the $\rho_3(1690)$ starts to be felt.
Our dispersively constrained Global Fit I is shown in the same figure
and its parameters are listed in Table~\ref{tab:Fparameters}.
We also show the unconstrained Global Fit I. Although it looks qualitatively very similar to the constrained one, it is quite apart in terms of deviations. In addition, we show an unconstrained BW fit to illustrate the need for a more elaborate description, like the BWBW form. 
The naive BW formula roughly reproduces the minimum in the elasticity and the zero in the phase shift. However, it fails to describe the shape of the 
phase shift, the width of the elasticity dip---which comes out too narrow---and it also has a disastrous low-energy behavior, impossible to match smoothly to the conformal parametrization.

Since we fit the scattering data, our $\rho_3(1690)$ resonance mass and width are compatible with those
 obtained from the phase shift analysis by the CERN-Munich collaboration~\cite{Hyams:1973zf}, i.e., $m_{\rho_3}=1713\pm4\,$MeV, $\Gamma_{\rho_3}=228\pm10\,$MeV, and $x_{\rho_3}=0.26\pm0.02$.
Nevertheless, our Blatt-Weisskopf phenomenological radius
is smaller than their value: $R_{\rho_3}=6.38\pm0.44\,\gev^{-1}$.
Overall, our uncertainties are smaller because we impose consistency with the elastic region and the threshold parameters we use as input
are very precise. The most extreme case is the uncertainty of $R_{\rho_3}$, which comes out about twenty times smaller than 
the CERN-Munich result. 
This could be expected since, as explained above,  well below the resonance nominal mass, this parameter controls the BWBW behavior and dominates its uncertainty. 
A $R_{\rho_3}$ variation from its central value as large as that of the CERN-Munich one standard deviation would result in a scattering length more than ten deviations away from the input value.

Note, however, that the mass and width differ by several MeV from the RPP values, dominated by the CERN-Munich BWBW analysis of their $\pi N\to \pi\pi N'$  momentum distributions. To illustrate this difference, Fig.~\ref{fig:F} shows a naive BWBW shape using the mass and width values estimated by the RPP. The resulting shape is qualitatively similar to our Global Fit, but the phase shift is slightly displaced, and the data are described worse. The width is also smaller and misses all the elasticity data above 1.65 GeV. Since we are analyzing scattering data, we adopt our parametrization and do not fit the RPP estimates. 

For completeness, in Fig.~\ref{fig:F}, we also show 
the F-wave data of Protopopescu et al.~\cite{Protopopescu:1973sh}. In contrast to the Hyams et al.~data~\cite{Hyams:1973zf}, these phase shifts are small and negative between 0.6 and approximately 1.05 GeV.  Moreover, the elasticity data from~\cite{Protopopescu:1973sh} exhibit an unusual dip around 1.05 GeV, which is difficult to reconcile with any physical interpretation. 
We believe that this behavior is largely an artifact, particularly since the closest known resonance in this channel, the $\rho_3(1690)$, lies well above this range.  No other resonances are expected or claimed around 1 GeV in this wave, and, in their absence, the inelasticity in this channel should be relatively smooth due to the angular momentum barrier. 
Together with other caveats, this argument has already been used to discard the 
data from~\cite{Protopopescu:1973sh} in this region for other waves.

\begin{widetext}
    \begin{center}
\begin{table}[h]
\renewcommand{\arraystretch}{1.3}
  \centering
  \begin{tabular}{lcccc}
 F wave    &Parameters& Global I values & Global II values& Global III values\\
\toprule
\multirow{ 2}{*}{$\delta^{(1)}_3\big\vert_{s<s_m}$}&$ B_2$ &$(57.5\pm2.4)\,10^3$&$(55.2\pm1.6)\,10^3$&$(129.0\pm1.6)\,10^3$ \\
  & $B_3$&$(-103\pm6)\,10^3$&$(-123\pm8)\,10^3$&$(10\pm8)\,10^3$\\
 \colrule
\multirow{ 4}{*}{$t^{(1)}_3\big\vert_{s>s_m}$}&$m_{\rho_3}$&$(1711\pm2)\,\mev$&$(1724\pm2)\,\mev$&$(1719\pm2)\,\mev$\\
&$\Gamma_{\rho_3}$&$(252\pm6)\,\mev$&$(278\pm7)\,\mev$&$(242\pm5)\,\mev$\\
&$x_{\rho_3}$&$0.269\pm0.003$&$0.284\pm0.009$&$0.235\pm0.002$\\
&$R_{\rho_3}$&$(5.06\pm0.02)\,\gev^{-1}$&$(4.92\pm0.02)\,\gev^{-1}$&$(5.86\pm0.02)\,\gev^{-1}$\\\botrule
  \end{tabular}
  \caption{F-wave parameters of the constrained Global Fits I, II, and III.
  Recall that $s_m=(1.2 \gev)^2$ for this wave.
  }
  \label{tab:Fparameters}
\end{table}
    \end{center}
   \vspace*{-.5cm}
\end{widetext}

\subsubsection{The three F-wave Global Fits}

\begin{figure}[h!]
\centering
\includegraphics[width=0.475\textwidth]{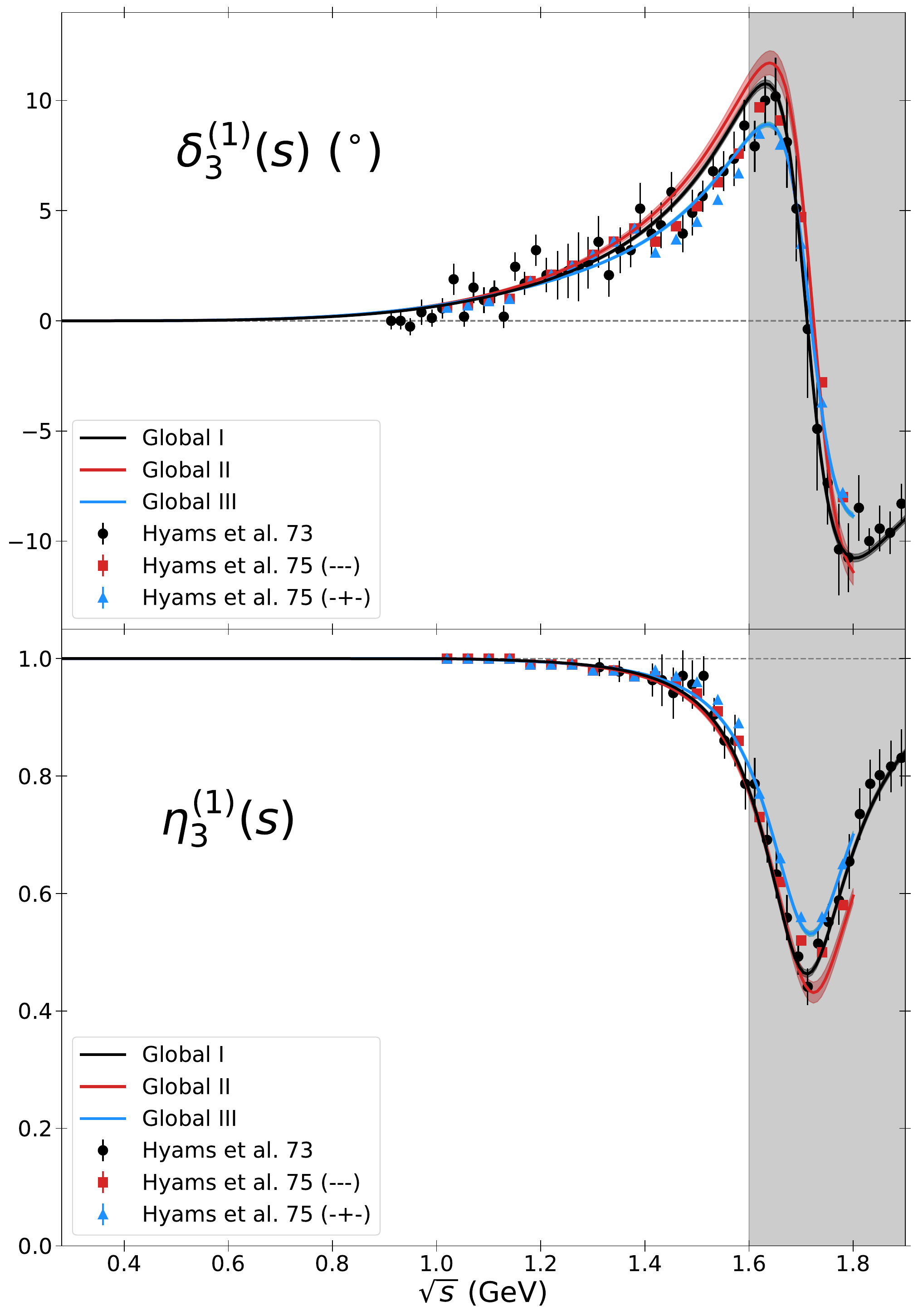}
\vspace{-.3cm}
\caption{Comparison among Global Fits I, II, and III for the F wave. 
Qualitatively, they all display a clear resonance shape, although their parameters are incompatible within their uncertainties. Note that the dispersively constrained Global Fit II lies far away from its corresponding data. Above 1.6 GeV (shaded region), the Global Fits are pure data fits since dispersive constraints are only imposed up to that energy. Note that we only plot our Global Fits up to the round energy value closest to the last data point included in the fit. The data come from~\cite{Hyams:1973zf} (Hyams et al.~73) and
\cite{Hyams:1975mc} (Hyams et al.~75).}\label{fig:F_sols}
\end{figure}

In Fig.~\ref{fig:F_sols}, we show the resulting F wave for Global Fits I, II, and III, after imposing the dispersive constraints on the three data Solutions I, II, and III, respectively. They are qualitatively very similar, always displaying a clear resonance with a BW-like shape with a large inelasticity. It should be noted that, as with other waves, the data from Solutions II and III have no uncertainties. For their fits, we add an uncertainty, similar to that of Solution I, to get realistic and comparable error bands.

We provide the parameters of all three Global Fits in Table~\ref{tab:Fparameters}. They are fairly incompatible with one another. Fits II and III prefer a heavier $\rho_3(1690)$, but that is the only common feature they share compared to Fit I.

It is worth noticing that, contrary to Global Fits I and III, Global Fit II lies far away from its data set. We consider that this large displacement, required to fulfill dispersion relations, makes Global Fit II somewhat disfavored compared to the other two.

As a final remark, our Global Fit I should only be used up to 1.9 GeV, whereas Global Fits II and III only up to 1.8 GeV.

\subsection{G0-wave parametrization}

This wave should be very small at low energies due to the potential barrier and dominated at high energies by the $f_4(2050)$, whose mass and width in the RPP~\cite{ParticleDataGroup:2022pth} are $M_{f_4}=2018\pm11\,$MeV and $\Gamma_{f_4}=237\pm18\,$MeV.
The RPP also lists another possible G0 resonance at 2.3 GeV, but it is currently omitted from the summary tables. Given that we will use this wave as input inside our integrals only up to 1.6 GeV, it seems more than enough to describe only the first resonance.

Unfortunately, no scattering data are available for this wave. Previous dispersive analyses, which focused on the dispersive representation up to 1.4 GeV, relied on a very simple estimate of its imaginary part~\cite{Kaminski:2006qe}.
Still, based on our previous experience with the F wave, we expect to achieve a fair description of the $f_4(2050)$ resonance by describing it with a Breit-Wigner-like form, using its RPP mass and width as input. This will then be matched at low energies to a conformal expansion, where we will impose the scattering length obtained from sum rules using previous dispersive analyses~\cite{Kaminski:2006qe}. 

We will choose the matching point at 1.4 GeV and will allow a non-vanishing inelasticity from 1.05 GeV, which crudely corresponds to the $\pi\pi\rho$ threshold. This is purely phenomenological since, in practice, the inelasticity will be imperceptible below 1.6 GeV.

Thus, for the phase shift at energies $s^{1/2}\leq s_m^{1/2}=1.4\, \gev$ we use:
\begin{eqnarray}
&&\cot\delta_4^{(0)}(s)=\frac{m_\pi^{8} }{\sigma(s) k(s)^8}
\left( B_0+B_1 \,w(s) \right),\nonumber\\
&&w(s)=\frac{\sqrt{s}-\alpha \sqrt{s_0-s}}{\sqrt{s}+\alpha \sqrt{s_0-s}},\quad
s_0^{1/2}=1.75 \,\gev, \quad \alpha=\frac{1}{2}.
\nonumber\\
 \label{eq:G0lowparam} 
\end{eqnarray}

For the elasticity, we set $\eta_4^{(0)}(s)=1$ for $s$ below
$\hat s=(1.05 \gev)^2$. From that energy to $s_m^{1/2}=1.4\, \gev$, we use:
\begin{eqnarray}
\eta_4^{(0)}(s)=1-\epsilon\left(1-\frac{\hat s}{s}\right)^{9/2}\left(1+r\left(1-\frac{s_m}{s}\right)\right).
\label{eq:G0lowparamine} 
\end{eqnarray}
Above 1.4 GeV, we use a Breit-Wigner with a potential barrier for the $f_4(2050)$ resonance:
\begin{eqnarray}
&&  t^{(0)}_4(s)=\frac{1}{\sigma(s)}\frac{x_{f_4} m_{f_4}\Gamma_{f_4}(s)}{m_{f_4}^2-s-im_{f_4}\Gamma_{f_4}(s)},
\label{eq:BWBWG0a}
\end{eqnarray}
where
\begin{eqnarray}
&&\Gamma_{f_4}(s) = \Gamma_{f_4}\left(\frac{k(s)}{k_{f_4}}\right)^9\frac{D_4(k_{f_4}R_{f_4})}{D_4(k(s)R_{f_4})},
\nonumber\\
&& D_4(x)=11025 + 1575x^2 + 135x^4 + 10x^6 + x^8 \,, \nonumber\\
\label{eq:BWBWG0b}
\end{eqnarray}
and $k_{f_4}=k(m_{f_4}^2)$.
The parameters $B_0$ and $B_1$ from Eq.~\eqref{eq:G0lowparam}, and $\epsilon$ and $r$ from Eq.~\eqref{eq:G0lowparamine}, are obtained by imposing continuity and differentiability at the matching point $s_m^{1/2}=1.4 \gev $
\begin{align}
   \epsilon & = \Big(1-\eta_4^{(0)}(s_m)\Big)\left(1-\frac{\hat s}{s_m}\right)^{-9/2}, \nonumber \\
    r& = -\frac{s_m \eta_4^{(0)\,\prime}(s_m)}{1-\eta_4^{(0)}(s_m)} + \frac{9}{2}\frac{\hat s}{s_m-\hat s},\nonumber \\
    B_1& = \frac{1}{m_{\pi}^8w'_m}\left(\sigma(s) k(s)^8 \cot \delta_4^{ (0)}(s)\right)^{\prime}\bigg\vert_{s=s_m},
    \nonumber\\
    B_0& = \frac{\sigma(s_m) k(s_m)^8}{m_{\pi}^8} \cot \delta_4^{ (0)}(s_m) 
    -B_1 w_m, 
\end{align}
where $w_m=w(s_m)$, $w'_m=w'(s_m)$ and the phase shift, elasticity and their first derivatives 
at $s_m$ in the previous expressions must be evaluated using Eqs.~\eqref{eq:BWBWG0a} and \eqref{eq:BWBWG0b}.

\subsubsection{G0-wave Global Fit and parameters}

As commented before, there are no scattering data for this wave. Thus our input consists of the $a^{(0)}_4$ scattering length provided in Table~\ref{tab:ThresholdParameters} in Appendix~\ref{app:Threshold}, along with the mass and width of the $f_4(2050)$ resonance taken from the weighted average of their determinations listed in the RPP~\cite{ParticleDataGroup:2022pth} with two pions in the final state. This yields $m_{f_4}=2030\pm23\mev$ and $\Gamma_{f_4}=248\pm80\mev$. The central values are consistent within uncertainties with the central RPP averages but have uncertainties roughly two to four times larger, which we consider a very conservative estimate.  The input value for $x_{f_4}=0.17\pm0.03$ was also taken from the RPP but conservatively doubling again its estimated uncertainty. Finally, the value of $R_{f_4}$ does not affect the resonant shape much, and given the simplicity of the parametrization, it gets completely fixed from the high precision of the scattering length and the differentiable matching conditions.

\begin{figure}[h]
\centering
\includegraphics[width=0.48\textwidth]{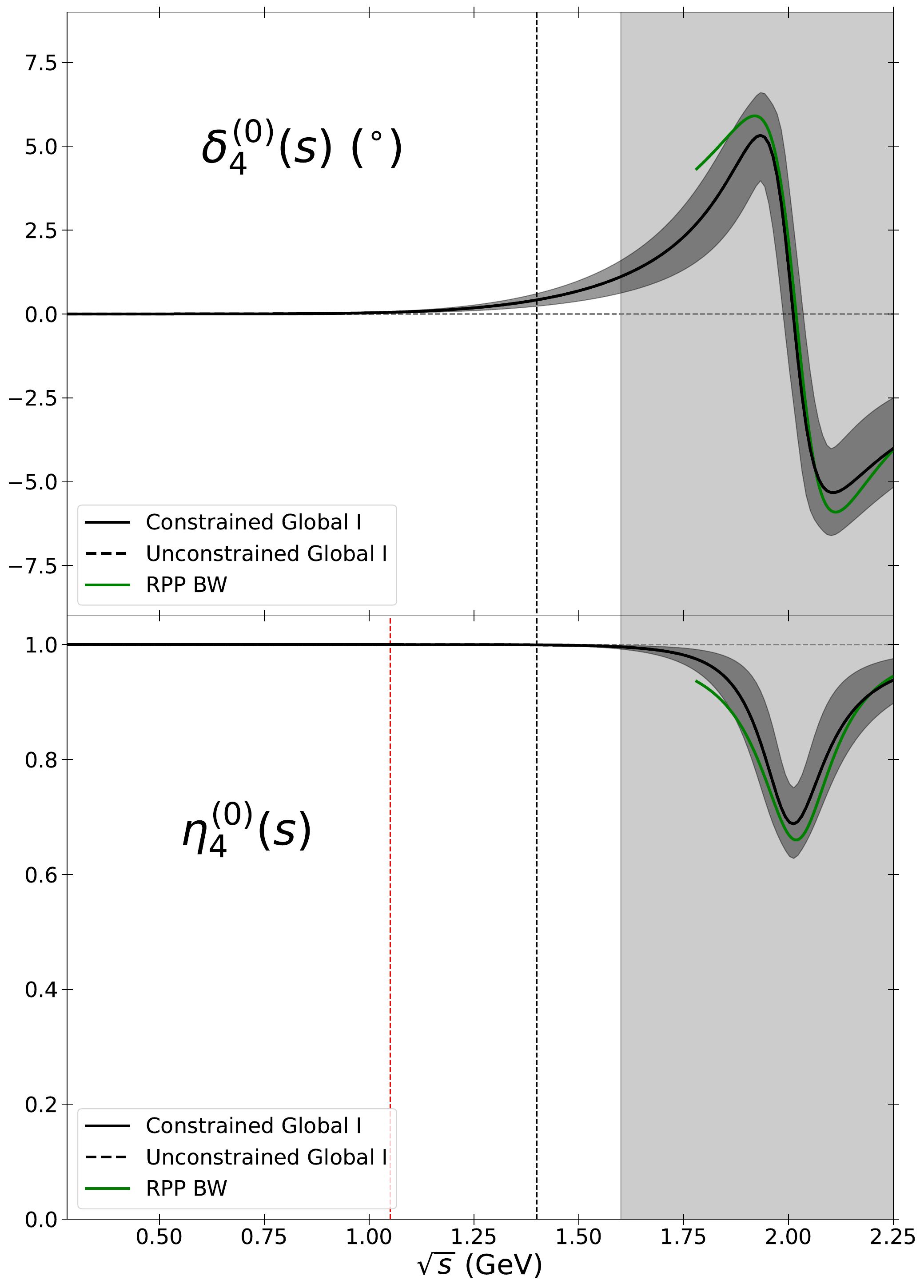}
 \vspace*{-.7cm}
\caption{G0-wave phase shift (top) and elasticity (bottom). The red vertical line stands at the onset of the inelasticity, while the black one marks the energy of the matching between the low-energy and the BWBW parametrizations, Eqs.~\eqref{eq:BWBWG0a} and \eqref{eq:BWBWG0b}. 
The unconstrained fit is not visible because it is identical to the constrained one.
For illustration, we show a naive Breit-Wigner with the RPP parameters in the region within one width on either side of its mass.  Above 1.6 GeV (shaded region), there are no dispersive constraints.}\label{fig:G0}
\end{figure}

In Fig.~\ref{fig:G0}, we show the resulting phase shift (top panel) and elasticity (bottom panel) from the Global Fit I. As already commented, there are no scattering data available for comparison. For illustration, we show in the figure a naive Breit-Wigner shape with the mass and width averages of the RPP. The difference in the central curve arises because we consider only width determinations from two-pion final states, we include Blatt-Weisskopf factors and impose a matching to the scattering length.
Still, as far as one width away from its peak, this simple shape is pretty consistent with our Global Fit within uncertainties.

The G0-wave parameters, obtained
after matching with the low energy parametrizations and using the dispersion relations as constraints, are given in Table~\ref{tab:G0parameters}.
They are very compatible with the input values and those in the RPP.
The value of $R_{f_4}$ falls within the bulk of values used in the literature for different resonances, which is of the order of a few  $\gev^{-1}$. It is indeed very close to the one we found for the F wave and also in very good agreement with the values found for the $\rho(770)$ in the literature, typically
$4-5\gev^{-1}$~\cite{VonHippel:1972fg,Hyams:1973zf,Kaminski:1996da,CERN-Cracow-Munich:1982jjg,Nebreda:2011di}. 
However, given that with this parameter we are basically fitting just one high-precision datum, which is the scattering length, The resulting uncertainty is smaller than 0.5 MeV, and, in practice, $R_{f_4}$ becomes a fixed constant in our phenomenological parametrization. We have nevertheless allowed it to vary when imposing the dispersive constraints.

\subsubsection{The three G0-wave Global Fits}

In practice, as seen in Table~\ref{tab:G0parameters}, our three G0-wave Global Fits share the same parameters up to the precision that we give them. The fact that the three of them share as input the same value for the scattering length with its small uncertainty, fixes the parameters within that precision, even after imposing 
the dispersive constraints. Hence, contrary to other waves, there is no need to provide a figure comparing the three Global Fits as they overlap almost exactly with the curve and band in Fig.~\ref{fig:G0}.
Indeed, let us recall that in Fig.~\ref{fig:G0}, the unconstrained and constrained Global Fits I also overlap completely.

\begin{table}[h]
\renewcommand{\arraystretch}{1.3}
  \centering
  \begin{tabular}{cc}
G0 Parameters & All Global Fits \\
\toprule
$m_{f_4}$&$(2011\pm23)\,\mev$\\
$\Gamma_{f_4}$&$(206\pm80)\,\mev$\\
$x_{f_4}$&$0.16\pm0.03$\\
$R_{f_4}$&$4.671\,\gev^{-1}$\\\botrule
  \end{tabular}
  \caption{G0-wave common parameters for the three constrained Global Fits I, II, and III.
  In practice, $R_{f_4}$ has no uncertainty after the fit, and we just keep its first four significant figures.
  Recall that $s_m=(1.4 \gev)^2$ for this wave.}
  \label{tab:G0parameters}
\end{table}

\subsection{S2-wave parametrization}

The S2 wave does not have much structure as no resonances appear there since it is a repulsive channel. Its CFD description in~\cite{GarciaMartin:2011cn} was fairly good up to 1.42 GeV, but here we want to obtain a global fit up to the last available data point, almost at 2.1 GeV. In addition, we will allow its inelasticity to start at $\hat s=(0.915\, \gev)^2$,
which is just a convenient phenomenological value, with no other physical meaning than the approximate energy 
where the inelasticity seems to open in Solution B of~\cite{Losty:1973et}. This adds more flexibility
compared to \cite{GarciaMartin:2011cn}, where the inelasticity opened up at $1.05\, \gev$. 
Lowering the inelasticity effective threshold and extending the parametrization to higher energies require some minor changes and one more parameter compared to the CFD parametrization in   \cite{GarciaMartin:2011cn}, but our global parametrization remains rather simple and easy to implement.

In particular, for $s^{1/2}\leq s_m^{1/2}=0.85\, \gev$ we use:
\begin{eqnarray}
&&  \cot\delta^{(2)}_0(s)=\frac{1}{\sigma(s)}\frac{m_\pi^2}{s-2z_2^2}\big(
B_0+B_1 w_l(s)\big),
\nonumber\\
&&w_l(s)=\frac{\sqrt{s}-\sqrt{s_l-s}}{\sqrt{s}+\sqrt{s_l-s}},\quad s_l^{1/2}=1.05 \,\gev,
  \label{eq:S2lowparam}
\end{eqnarray}
whereas at intermediate energies, $0.85\, \gev=s_m^{1/2}\leq s^{1/2}\leq 2.1 \,\gev$, we use:
\begin{align}
\cot\delta^{(2)}_0(s)=&\frac{1}{\sigma(s)}
\frac{m_\pi^2}{s-2z_2^2}\sum_{n=0}^3 B_{hn}
 \big(w_h(s)-w_h(s_m)\big)^n ,
 \nonumber\\
 w_h(s)=&\frac{\sqrt{s}-\sqrt{s_h-s}}{\sqrt{s}+\sqrt{s_h-s}},
\quad s_h^{1/2}=2.3 \,\gev.
\end{align}
Here we have added one more parameter than in~\cite{GarciaMartin:2011cn}, $B_{h3}$. In addition, since the data points in~\cite{Cohen:1973yx} extend to 2.1 \gev, we have also increased the value of $s_h$.

Imposing continuity and differentiability at the
matching point $s=s_m$, the $B_{h0}$ and $B_{h1}$ parameters are fixed as
\begin{equation}
B_{h0}=B_0+B_1 w_l(s_m),\quad
B_{h1}=B_1 \frac{w_l'(s_m)}{w_h'(s_m)},
\label{eq:S2highparam}
\end{equation}
where, once again, the prime denotes the derivative with respect to $s$ and, therefore,
\begin{equation}
\frac{w_l'(s_m)}{w_h'(s_m)}=\frac{s_l}{s_h}\frac{\sqrt{s_h-s_m}}{\sqrt{s_l-s_m}}
\left( \frac{\sqrt{s_m}+\sqrt{s_h-s_m}}{\sqrt{s_m}+\sqrt{s_l-s_m}}
\right)^2.
\end{equation}

For the elasticity, we set $\eta_0^{(2)}(s)=1$ for $s$ below
$\hat s=(0.915\, \gev)^2$ and above we use the empirical fit
\begin{equation}
\eta_0^{(2)}(s)=1-\epsilon\left(1-\frac{\hat s}{s}\right)^{3/2}.
\label{eq:S2inel}
\end{equation}

The elasticity data is so poor and scarce that one parameter is enough to describe them. Since no resonances are known below 2.1 GeV, above the last data point at 1.4 GeV, we will just show an educated extrapolation of the elasticity up to 2.1 GeV because that is the energy up to where data on the phase shift exists.
Note, however, that we will only use this wave up to 1.62 GeV
as input in our dispersive representation. 
\begin{figure}
\centering
\includegraphics[width=0.48\textwidth]{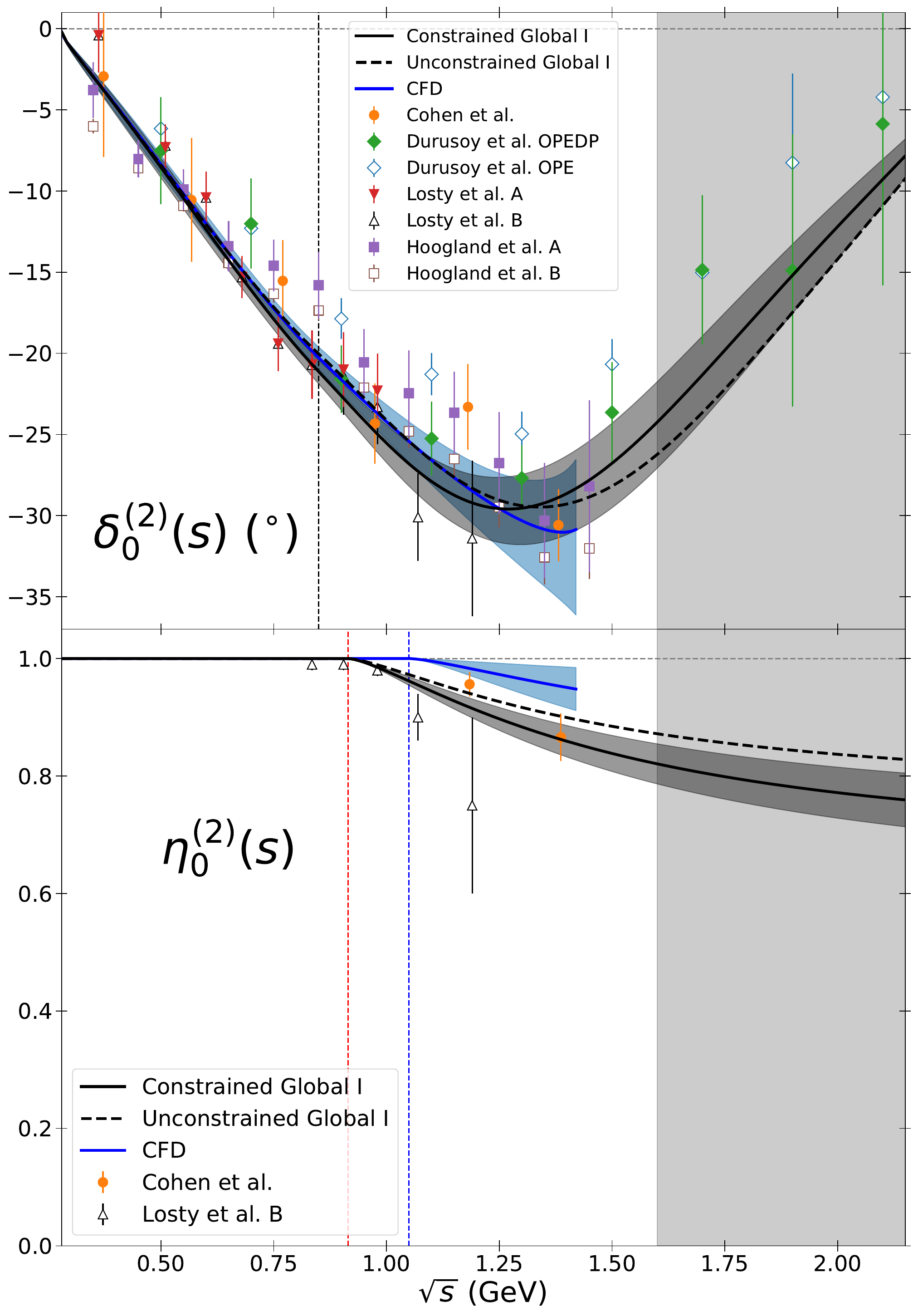}
\caption{ 
S2-wave phase shift (top) and elasticity (bottom). We show our unconstrained and dispersively constrained Global Fit I 
together with the CFD from \cite{GarciaMartin:2011cn}. 
As explained in the text, the data used in the fit above 0.9 GeV are those with solid symbols from Cohen et al.~\cite{Cohen:1973yx}, OPEDP from Durusoy et al.~\cite{Durusoy:1973aj} and the A solutions from Losty et al.~\cite{Losty:1973et} and Hoogland et al.~\cite{Hoogland:1977kt}.
The other data are shown for completeness.
In the top panel, the black vertical line at 0.85~GeV marks the matching point with the high-energy parametrization. In the bottom one, the red vertical line stands at 0.915~GeV, where we allow the inelasticity to open up instead of 1.05~GeV (blue vertical line) used in the CFD.}\label{fig:S2}
\end{figure}

\subsubsection{S2-wave Global Fit and parameters}

In Fig.~\ref{fig:S2}, we show the phase shift and elasticity of our constrained
Global Fit I, together with all available scattering data. 
However, the input we use in our fit is the CFD below 0.9 GeV, the data represented by solid symbols above 0.9 GeV and the $a_0^{(2)}$ and $b_0^{(2)}$ threshold parameters in Table~\ref{tab:ThresholdParameters} in Appendix~\ref{app:Threshold}.

Regarding the phase shifts, Cohen et al. \cite{Cohen:1973yx} only provide one set of data, whereas Durusoy et al.~\cite{Durusoy:1973aj}, Losty et al.~\cite{Losty:1973et}, and Hoogland et al.~\cite{Hoogland:1977kt} each provide two sets of data, which correspond to slightly different methods of analyses that overlap within uncertainties.
We show all sets for completeness but fit only one per experiment (the solid symbols in Fig.~\ref{fig:S2}). 
For \cite{Durusoy:1973aj} and \cite{Hoogland:1977kt} we have chosen to fit only the set that includes a D\"urr-Pilkuhn~\cite{Durr:1965} form factor for consistency with the data 
from \cite{Cohen:1973yx}. This is the most conservative choice since it has somewhat larger uncertainties.
Concerning \cite{Losty:1973et}, we fit Solution A, discarding Solution B for two reasons.
First, it has larger uncertainties than the rest of the data, and including them or not in the S2 fit is almost irrelevant.
The fit is still compatible with Solution B as long as we allow the inelasticity to open up at 0.915 GeV.
Second, we will see that for the G2 wave, and 
despite these huge uncertainties, Solution B is incompatible with the rest of the data.
We will keep these choices of data input for all the $I=2$ waves.

The Global Fit I phase shift is compatible with the CFD parametrization in~\cite{GarciaMartin:2011cn}, valid only up to 1.42 GeV.
However, the new phase shift, which fits data up to 2.1 GeV, reaches a minimum near 1.3 GeV and grows toward zero at higher energies.

A significant difference with the CFD is seen in the inelasticity, which we now allow to start from 0.915 GeV instead of 1.05 GeV, as for the CFD. We reiterate that this value has no other physical meaning than the energy where the data starts showing some non-vanishing inelasticity. 
We prefer our new global parametrization in Eq.~\eqref{eq:S2inel} since it is continuous, differentiable,
and more consistent with the two data points from Cohen et al.~\cite{Cohen:1973yx}.
As explained above, the Solution B data from Losty et al.~\cite{Losty:1973et}, even if not included in the fit, are quite compatible with it due to their large uncertainties.
Beyond 1.5 GeV, our elasticity is just a naive extrapolation since no data exist beyond that point. Durusoy et al.~\cite{Durusoy:1973aj} used a linear extrapolation in $\sqrt{s}$ of the two data points of Cohen et al.~\cite{Cohen:1973yx}. Nevertheless, they say that their phase shifts are not very sensitive to different elasticity guesses. 
The S2-wave Global Fit I parameters, after using the dispersion relations as constraints are given in Table~\ref{tab:S2parameters}.

\begin{widetext}
    \begin{center}
\begin{table}[h]
\renewcommand{\arraystretch}{1.3}
  \centering
  \begin{tabular}{lcccc}
S2 wave    & Parameter &Global I values&Global II values&Global III values\\
\toprule
\multirow{ 3}{*}{$\delta^{(2)}_0\big\vert_{s<s_m}$}&$ B_0$ &$-76.5\pm2.8$&$-80.2\pm2.8$&$-78.1\pm2.8$ \\
&$B_1$& $-57\pm11$& $-64\pm11$& $-58\pm11$\\
 &$z_2$ & $142 \pm 4 \, \mev$& $147 \pm 4 \, \mev$& $145 \pm 4 \, \mev$\\
\colrule
\multirow{ 2}{*}{$\delta^{(2)}_0\big\vert_{s>s_m}$}&$B_{h2}$&$290\pm109$&$340\pm109$&$327\pm109$\\
&$B_{h3}$&$-2466\pm355$&$-2489\pm355$&$-2615\pm355$\\
\colrule
$\eta^{(2)}_0\big\vert_{s>\hat s}$&$\epsilon$&$0.32\pm0.06$&$0.33\pm0.06$&$0.30\pm0.06$\\[1pt]
\botrule
  \end{tabular}
  \caption{S2-wave parameters of the constrained Global Fits I, II, and III.  Recall that $s_m=(0.85 \gev)^2$ for this wave.}
  \label{tab:S2parameters}
\end{table}
    \end{center}
     \vspace*{-.7cm}
\end{widetext}

\subsubsection{The three S2-wave Global Fits}

In contrast to the S0, P, D0, and F waves,  each $I=2$ wave has only one data set to fit from the start. Therefore, we expect the three constrained Global Fits to the $I=2$ waves to differ very little among themselves since their separation is an effect induced indirectly from the other waves used as input in the dispersion relations.

\begin{figure}[ht!]
\centering
\includegraphics[width=0.48\textwidth]{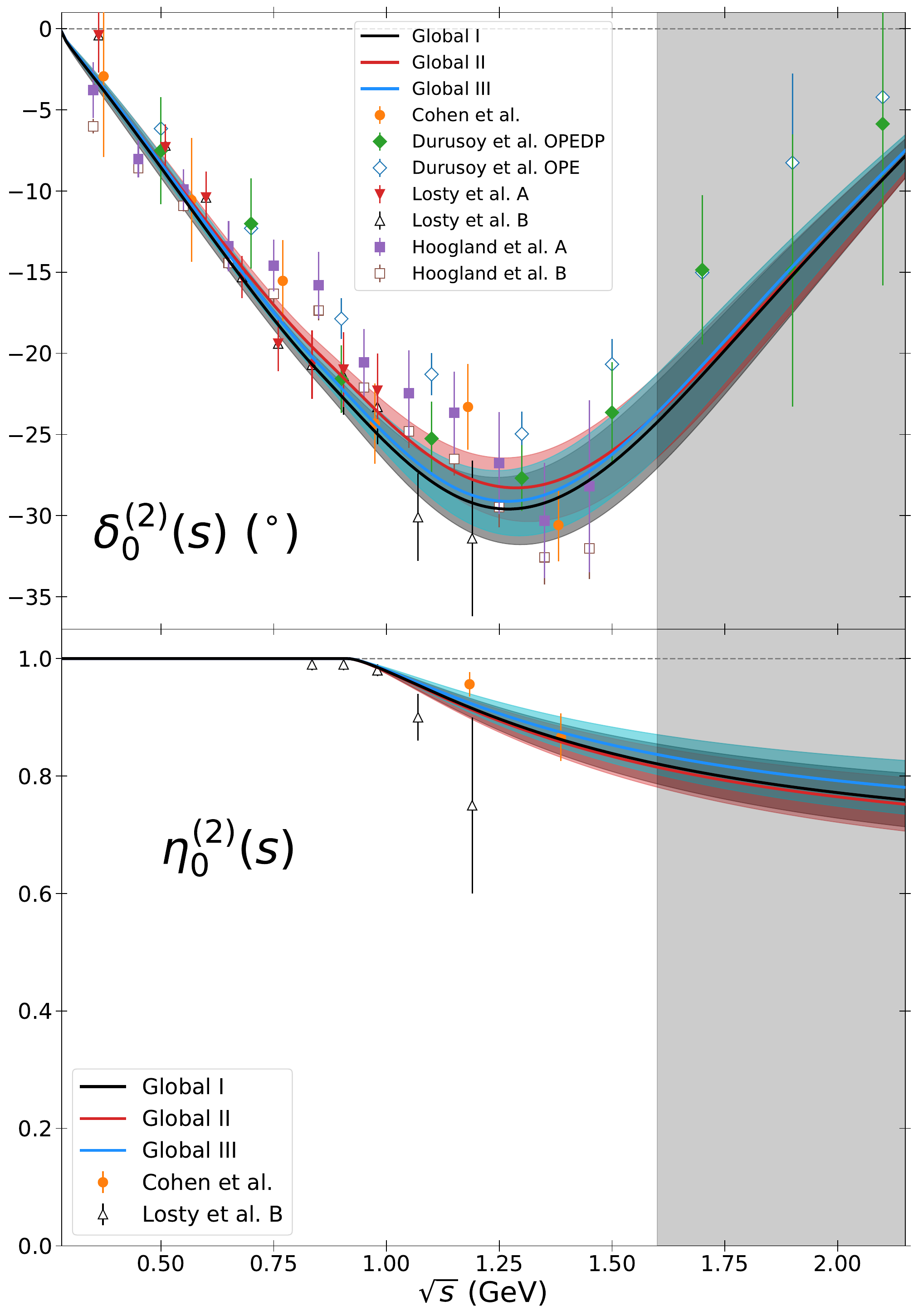}
\vspace*{-.7cm}
\caption{ 
Comparison among Global Fits I, II, and III for the S2 wave.
Within uncertainties, the three fits are compatible in their phase shift and elasticity. Above 1.6 GeV (shaded region) there are no dispersive constraints.
Data references as in Fig.~\ref{fig:S2}.}\label{fig:S2_sols}
\end{figure}

Figure~\ref{fig:S2_sols} shows the three constrained Global Fits of the S2 wave. As expected, they are remarkably compatible among themselves.
The Fit I and III phase shifts are almost indistinguishable. 
This compatibility is also reflected in the values of the parameters for the three fits, listed in Table~\ref{tab:S2parameters}. 
Nevertheless, the Fit II phase shift is slightly less negative than the other two below 1.4 GeV. In addition, Fit II is slightly more inelastic, and Fit III is more elastic than Fit I.

\subsection{D2-wave parametrization}

Except close to the $\pi\pi$ threshold, this is a repulsive wave, with no resonances and little structure, even smaller than the S2. The CFD parametrization in~\cite{GarciaMartin:2011cn}
only reached $\sim1.4\,$GeV. Here, we extend the parametrization up to 
2.1 GeV, where a few data on its phase shift exist. It will have two pieces with continuous and differentiable 
matching at 1.4 GeV.

The phase shift parametrization will be very similar to that of the S2, except for the appropriate angular momentum factors 
and another factor to ensure a positive scattering length (confirmed by sum rules and Chiral Perturbation Theory calculations), despite the wave being mostly repulsive.

In particular, for $s^{1/2}\leq s_m^{1/2}=0.85\, \gev$ we use:

\begin{align}
\cot\delta_2^{(2)}(s)=&\frac{s }{\sigma(s) k(s)^4}\frac{m_\pi^4\big(B_0+B_1 w_l(s)+B_2 w_l(s)^2\big)}{4(m_\pi^2+\Delta^2)-s},
\nonumber\\
w_l(s)=&\frac{\sqrt{s}-\sqrt{s_l-s}}{\sqrt{s}+\sqrt{s_l-s}},\quad
s_l^{1/2}=1.45 \,\gev,
  \label{eq:D2lowparam} 
\end{align}
whereas at intermediate energies, $0.85\, \gev\leq s^{1/2}\leq 2.1 \,\gev$, we use 
\begin{eqnarray}
 &&\cot\delta^{(2)}_2(s)=\frac{s }{\sigma(s) k(s)^4}\frac{m_\pi^4}{4(m_\pi^2+\Delta^2)-s}\
\nonumber\\
&&\hspace{20mm}\times\sum_{n=0}^3 B_{hn}
 \big(w_h(s)-w_h(s_m)\big)^n ,
 \nonumber\\
 &&w_h(s)=\frac{\sqrt{s}-\sqrt{s_h-s}}{\sqrt{s}+\sqrt{s_h-s}},
\quad s_h^{1/2}=2.4 \,\gev,
\end{eqnarray}
where
\begin{eqnarray}
 B_{h0}&=&B_0+B_1 w_l(s_m)+B_2 w_l(s_m)^2,\nonumber\\
B_{h1}&=&\big(B_1+2 B_2 w_l(s_m)\big)\frac{w_l'(s_m)}{w_h'(s_m)}.
\label{eq:D2highparam}
\end{eqnarray}
Here, the prime denotes the derivative with respect to $s$, so that:
\begin{equation}
\frac{w_l'(s_m)}{w_h'(s_m)}=\frac{s_l}{s_h}\frac{\sqrt{s_h-s_m}}{\sqrt{s_l-s_m}}
\left( \frac{\sqrt{s_m}+\sqrt{s_h-s_m}}{\sqrt{s_m}+\sqrt{s_l-s_m}}
\right)^2.
\end{equation}
With these definitions,
both the parametrization and its derivative are continuous at $s_m$.

\begin{figure}[h]
\centering
\includegraphics[width=0.48\textwidth]{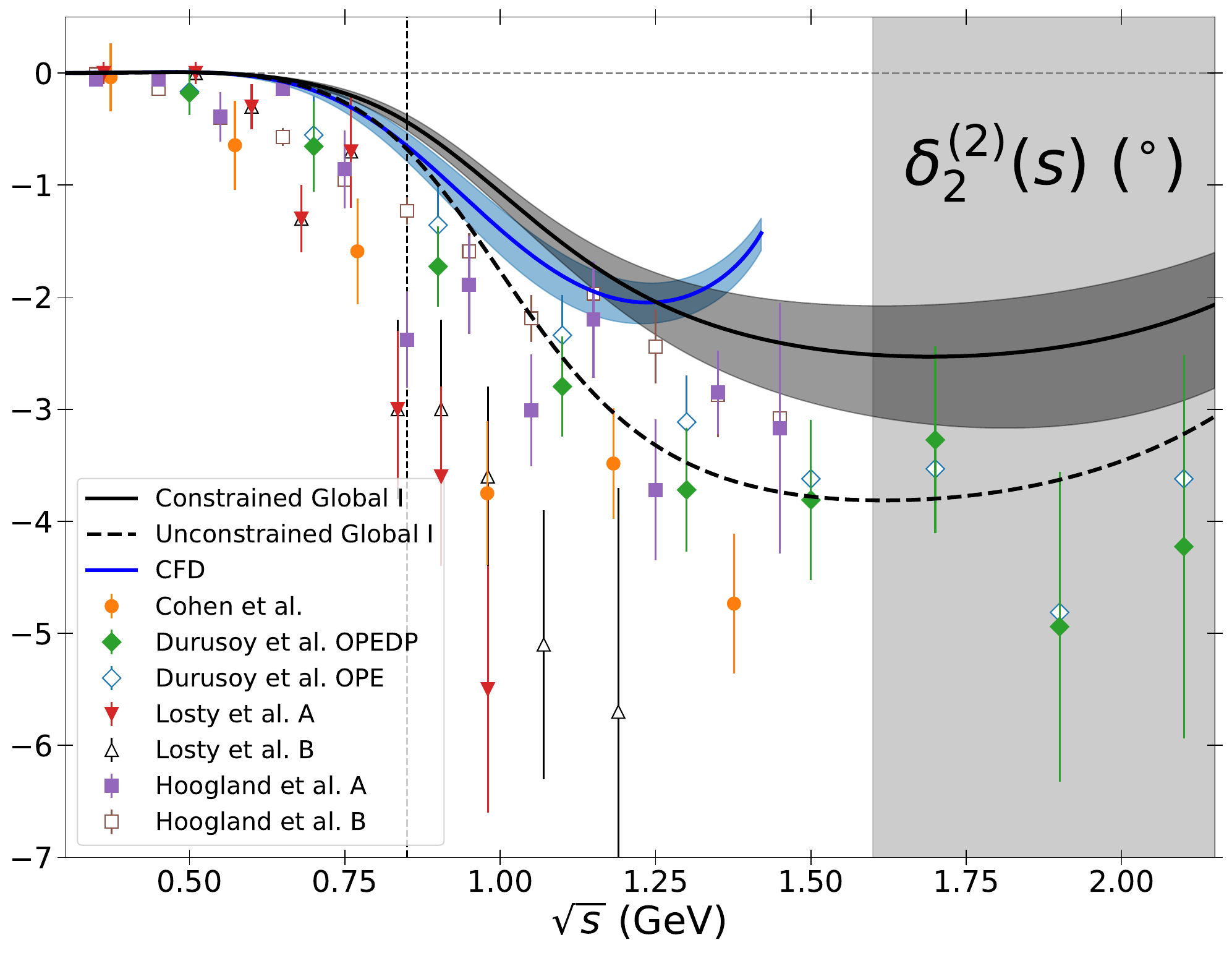}
\vspace*{-.8cm}
\caption{D2-wave phase shift. 
We show our unconstrained and dispersively constrained Global Fit I, together with the dispersively constrained CFD parametrization from~\cite{GarciaMartin:2011cn}. As explained in the text, the data used in the fit above 0.9 GeV are those with solid symbols. The other data are shown for completeness. The black vertical line at 0.85~GeV marks the matching point with the high-energy part.
All experiments either do not observe or assume that there is no inelasticity. Data references as in Fig.~\ref{fig:S2}.}\label{fig:D2}
\end{figure}

No hint of any inelasticity for the D2 and G2 waves has been observed in any of the experiments studying $\pi\pi$ scattering in the $I=2$ channel.
In  particular, Cohen et al.~\cite{Cohen:1973yx} explicitly state
that\footnote{Their mass range is from threshold to 1.4 GeV and they omit the $I=2$ superindex.}: {\it ``We found the d and g waves to be totally elastic throughout our mass range ($\eta_2=\eta_4=1)$, while the s wave became inelastic at about 1.1 GeV"}. 
In addition, Losty et al.~\cite{Losty:1973et}, who also study $I=2$, do find inelasticity for the S2 wave in their Solution B, but show no inelasticity for the D2 and G2 waves.
Finally, Durusoy et al.~\cite{Durusoy:1973aj}, whose phase-shift data reaches 2.1 GeV, consider only inelasticity for the S2 wave. For this reason, we will set $\eta_2^{(2)}(s)=1$ in our whole energy range.

\subsubsection{D2-wave Global Fit and parameters}

For this wave, we include in the fit the 
values of the threshold parameters $a^{(2)}_2$ and 
$b^{(2)}_2$ obtained from the CFD parametrization in~\cite{GarciaMartin:2011cn}, given in Table~\ref{tab:ThresholdParameters} here.
In addition, we fit the CFD results below 0.915 GeV, as well as the data on the phase shift that we show in Fig.~\ref{fig:D2}. 
It should be noted that this was the wave where the dispersive constraints demanded the largest change between the Unconstrained and Constrained Fit to Data (UFD versus CFD in~\cite{GarciaMartin:2011cn}). This is a strong reason to use the CFD as input, and we see that, up to energies around 0.95 GeV, the curves lie above most of the data.

\begin{widetext}
\vspace*{-.7cm}
    \begin{center}
\begin{table}[h]
\renewcommand{\arraystretch}{1.3}
  \centering
  \begin{tabular}{lcccc}
D2 wave    & Parameter &Global I values&Global II values&Global III values\\
\toprule
\multirow{ 4}{*}{$\delta^{(2)}_2\big\vert_{s<s_m}$}&$ B_0$ &$(4.6\pm0.5)\,10^3$&$(4.4\pm0.5)\,10^3$&$(4.6\pm0.5)\,10^3$ \\
  & $B_1$&$(-1.4\pm3.2)\,10^3$&$(0.5\pm3.2)\,10^3$&$(-0.5\pm3.2)\,10^3$\\
    & $B_2$&$(7\pm3)\,10^3$&$(10\pm3)\,10^3$&$(10\pm3)\,10^3$\\
 &$\Delta$ & $235\pm14\,\mev$& $236\pm14\,\mev$&  $240\pm14\,\mev$  \\
 \colrule
\multirow{ 2}{*}{$\delta^{(2)}_2\big\vert_{s>s_m}$}&$B_{h2}$&$(83\pm13)\,10^3$&$(80\pm13)\,10^3$&$(91\pm13)\,10^3$\\
&$B_{h3}$&$(73\pm46)\,10^3$&$(110\pm46)\,10^3$&$(80\pm46)\,10^3$\\\botrule
  \end{tabular}
  \caption{D2-wave parameters of the constrained Global Fits I, II, and III.  Recall that $s_m=(0.85 \gev)^2$ for this wave.}
  \label{tab:D2parameters}
\end{table}
    \end{center}
    \vspace*{-.9cm}
\end{widetext}

Compared to the CFD in~\cite{GarciaMartin:2011cn}, we now extend our parametrization to 2.1 GeV, since there is still a data point at this energy. When two consistent data sets from the same experiment are available, as in Durusoy et al.~\cite{Durusoy:1973aj} and Hoogland et al.~\cite{Hoogland:1977kt}, we only fit one set, chosen with the same criteria explained for the S2 wave.
Once again, the data of Losty et al.~\cite{Losty:1973et} lie systematically below and tend to have much larger uncertainties than other experiments in the 0.8 to 1.2 GeV region.

The D2-wave  parameters are given in Table~\ref{tab:D2parameters}.

\subsubsection{The three D2-wave Global Fits}

\begin{figure}[H]
\centering

\vspace*{-.3cm}

\includegraphics[width=0.48\textwidth]{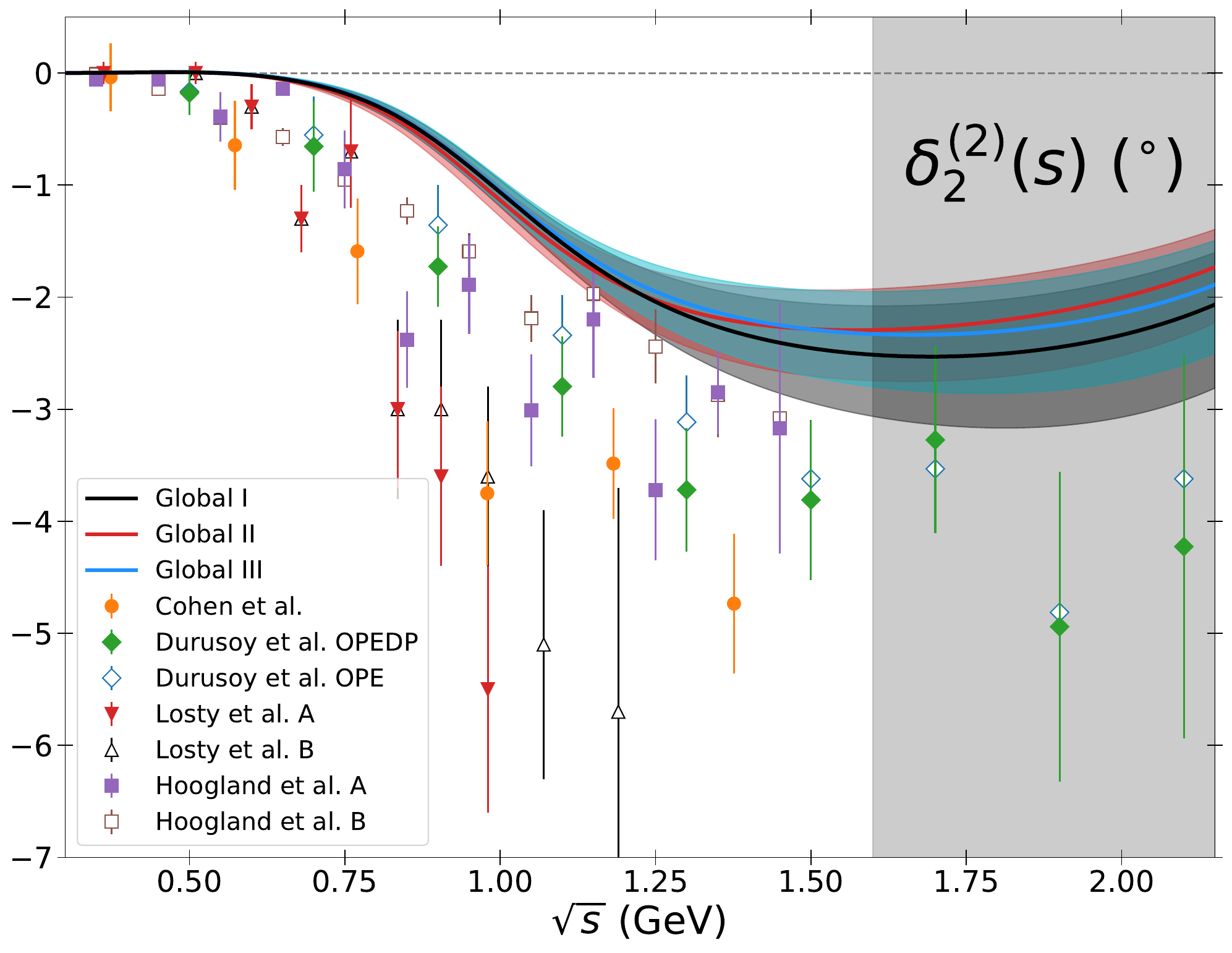}
\vspace*{-.8cm}
\caption{Comparison among Global Fits I, II, and III for the D2 wave, which are very compatible\label{fig:D2_sols}. Above 1.6 GeV (shaded region) there are no dispersive constraints.
Data references as in Fig.~\ref{fig:S2}.}
\end{figure}

The three Global Fits are very compatible, almost identical up to 1.25 GeV, as shown in Fig.~\ref{fig:D2_sols}. 
Their parameters, provided in Table~\ref{tab:D2parameters}
are very compatible too.

As we have done for Global Fit I, we have fixed to one the elasticity of the other two Global Fits in the whole energy region.

\subsection{G2-wave parametrization}

Once again, this is a repulsive wave, very
small in the whole energy range of interest.
In the dispersive analysis of~\cite{GarciaMartin:2011cn}, it was neglected below 1.42 GeV, based on 
previous estimations~\cite{Kaminski:2006qe} that 
 studied its negligible impact on dispersive calculations within this range. However, there are 
a few data points available up to almost 2.1 GeV, and hence, we now provide a global parametrization. In addition, its role is not so small in the two FDRs we study above 1.4 GeV here.

In particular, the parametrization in~\cite{Kaminski:2006qe} was little more than an order of magnitude estimate above 1 GeV, assuming its phase shift to be zero below. No attempt was made to describe the threshold parameters.

We will use a parametrization of the phase shift rather similar to that of the D2 wave. Namely, for $s^{1/2}\leq s_m^{1/2}=0.85\, \gev$ we use a conformal expansion:
\begin{align}
\cot\delta_4^{(2)}(s)=&\frac{s }{\sigma(s) k(s)^8}\frac{m_\pi^{8}\big(B_0+B_1 w_l(s)+B_2 w_l(s)^2\big)}{4(m_\pi^2+\Delta^2)-s},
\nonumber\\
w_l(s)=&\frac{\sqrt{s}-\sqrt{s_l-s}}{\sqrt{s}+\sqrt{s_l-s}},\quad
s_l^{1/2}=1.65 \,\gev,
  \label{eq:G2lowparam} 
\end{align}
whereas at higher energies, $0.85\, \gev\leq s^{1/2}\leq 2.1 \,\gev$, we write:
\begin{eqnarray}
&&\cot\delta^{(2)}_4(s)=\frac{s }{\sigma(s) k(s)^8}\frac{m_\pi^{8}}{4(m_\pi^2+\Delta^2)-s}\
\nonumber\\
&&\hspace{20mm}\times\sum_{n=0}^3 B_{hn}
 \big(w_h(s)-w_h(s_m)\big)^n,\nonumber\\
 &&w_h(s)=\frac{\sqrt{s}-\sqrt{s_h-s}}{\sqrt{s}+\sqrt{s_h-s}},
\quad s_h^{1/2}=2.4 \,\gev,
\end{eqnarray}
where
\begin{eqnarray}
 B_{h0}&=&B_0+B_1 w_l(s_m)+B_2 w_l(s_m)^2,\nonumber\\
B_{h1}&=&\big(B_1+2 B_2 w_l(s_m)\big)\frac{w_l'(s_m)}{w_h'(s_m)}.
\label{eq:G2highparam}
\end{eqnarray}
Once again, the prime denotes the derivative with respect to $s$, so that:
\begin{equation}
\frac{w_l'(s_m)}{w_h'(s_m)}=\frac{s_l}{s_h}\frac{\sqrt{s_h-s_m}}{\sqrt{s_l-s_m}}
\left( \frac{\sqrt{s_m}+\sqrt{s_h-s_m}}{\sqrt{s_m}+\sqrt{s_l-s_m}}
\right)^2,
\end{equation}
and both the parametrization and its derivative with respect to $s$ are continuous at the matching point.

As already commented for the D2 wave, all experiments 
on $I=2$ scattering either do not observe inelasticity in the G2 wave
or assume there is none. For that reason, we keep this wave 
elastic, $\eta^{(2)}_4=1$, in the whole energy range.

\begin{figure}[ht]
\centering
\vspace*{.2cm}
\includegraphics[width=0.48\textwidth]{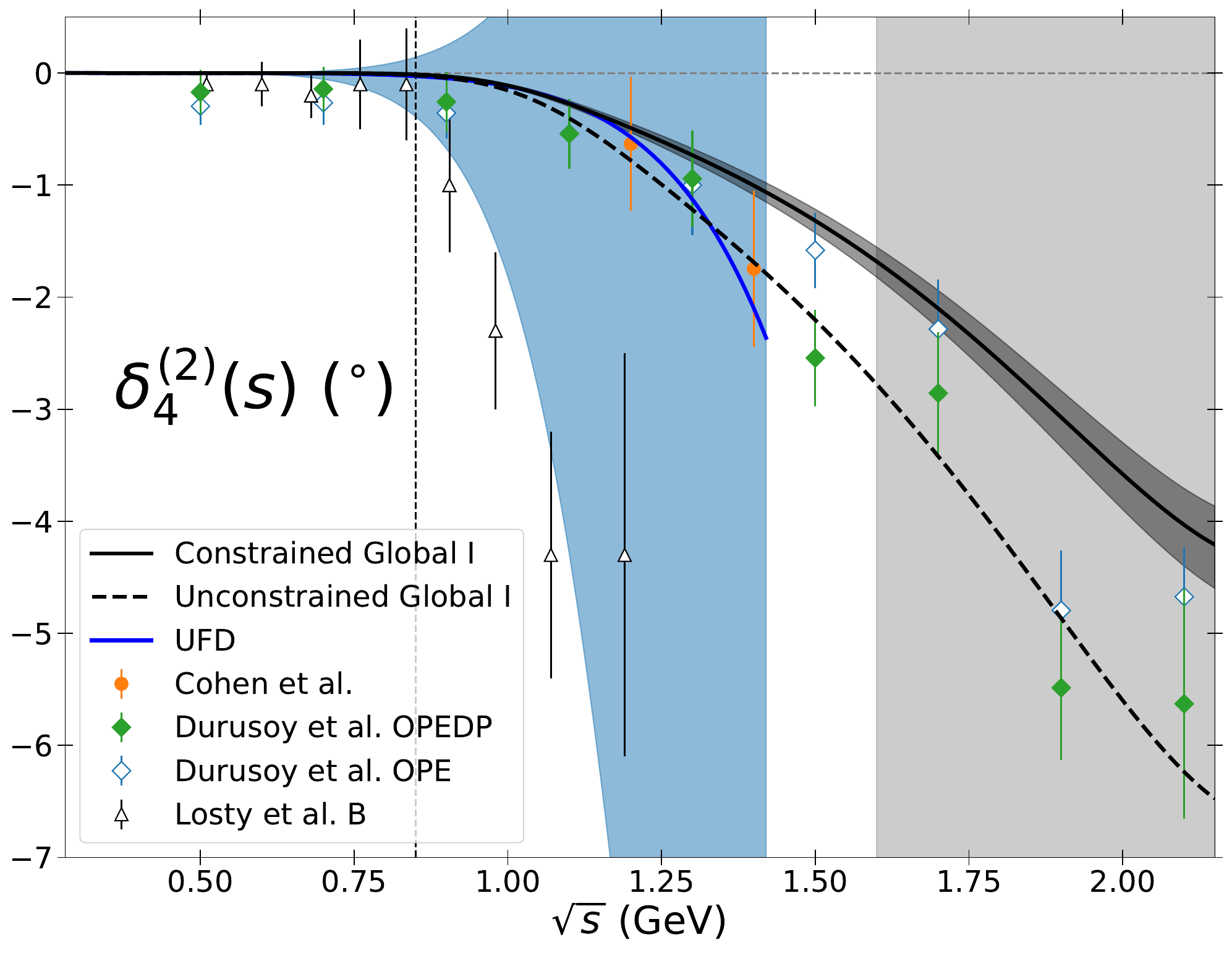}
\vspace*{-.7cm}
\caption{G2-wave 
phase shift.
We show our unconstrained and dispersively constrained Global Fit I, together with the crude UFD estimation in~\cite{GarciaMartin:2011cn} only used there to estimate uncertainties.
The data included in the fit above 0.9 GeV are those with solid symbols. The other data are shown for completeness.
 The black vertical line at 0.85~GeV marks the matching point with the high-energy part.
All experiments either do not observe or assume no inelasticity. Data references as in Fig.~\ref{fig:S2}.
}\label{fig:G2} 
\vspace*{-.2cm}
\end{figure}

\subsubsection{G2-wave Global Fit and parameters}

Three different experimental collaborations, Cohen et al.~\cite{Cohen:1973yx}, Durusoy et al.~\cite{Durusoy:1973aj}, and Losty et al.~\cite{Losty:1973et} provide measurements
of the phase shift for this wave, shown in Fig.~\ref{fig:G2}.
The set from Cohen et al.~\cite{Cohen:1973yx} only has two data points at 1.2 and 1.4 GeV, but they are remarkably compatible with the data from Durusoy et al.~\cite{Durusoy:1973aj}, which has nine points spanning from 0.5 to 2.1 GeV, and hence, dominates our Global Fit.
Following the same criteria we have already used for other $I=2$ waves, we have fit the data from ~\cite{Cohen:1973yx} and only one of the two analyses from Durusoy et al. (OPEDP)~\cite{Durusoy:1973aj}. Moreover, although only Solution B of Losty et al.~\cite{Losty:1973et} considers the G-wave, it has not been included in the fit because it is incompatible with the other two sets above 0.9 GeV, despite its much larger uncertainties.

Since there is no experimental information below 0.5 GeV, but the phase shift seems compatible with zero below 0.75 GeV, we have just fit the scattering length result in~\cite{Kaminski:2006qe}, obtained from sum rules with the CFD input. Its value is provided in Table~\ref{tab:ThresholdParameters} of Appendix~\ref{app:Threshold}.

In Fig.~\ref{fig:G2}. we also show the very conservative estimate provided in~\cite{GarciaMartin:2011cn}, valid only up to 1.4 GeV. In that work, it was only used to confirm that the G2 wave contribution to the dispersive representation below 1.4 GeV could be considered part of the uncertainties.

\begin{widetext}
\vspace*{-.7cm}
    \begin{center}
\begin{table}[h]
\renewcommand{\arraystretch}{1.3}
  \centering
  \begin{tabular}{lcccc}
G2 wave    & Parameter & Global I values& Global II values& Global III values\\
\toprule
\multirow{ 3}{*}{$\delta^{(2)}_4\big\vert_{s<s_m}$}&$ B_0$ &$(-4.57\pm0.09)\,10^6$&$(-4.56\pm0.09)\,10^6$&$(-4.57\pm0.09)\,10^6$ \\
&$B_1$&$(-62.15\pm0.13)\,10^6$&$(-62.18\pm0.13)\,10^6$&$(-62.11\pm0.13)\,10^6$\\
&$B_2$&$(-77.05\pm0.19)\,10^6$&$(-76.98\pm0.19)\,10^6$&$(-77.00\pm0.19)\,10^6$\\
& $\Delta$ & $329\pm23\,\mev$& $340\pm23\,\mev$& $319\pm23\,\mev$  \\
 \colrule
\multirow{2}{*}{$\delta^{(2)}_4\big\vert_{s>s_m}$}&$B_{h2}$&$(188\pm10)\,10^6$&$(187\pm10)\,10^6$&$(194\pm10)\,10^6$\\
&$B_{h3}$&$(86\pm21)\,10^6$&$(110\pm21)\,10^6$&$(120\pm21 )\,10^6$\\\botrule
  \end{tabular}
  \caption{G2-wave parameters of the constrained Global Fits I, II, and III.
  Recall that $s_m=(0.85 \gev)^2$ for this wave.}\label{tab:G2parameters}
\end{table}
\end{center}
\end{widetext}

The G2-wave Global Fit parameters, after imposing the dispersive constraints, are given in Table~\ref{tab:G2parameters}.

\vspace*{.3cm}
\subsubsection{The three G2-wave Global Fits}

\begin{figure}[h]
\centering
\includegraphics[width=0.48\textwidth]{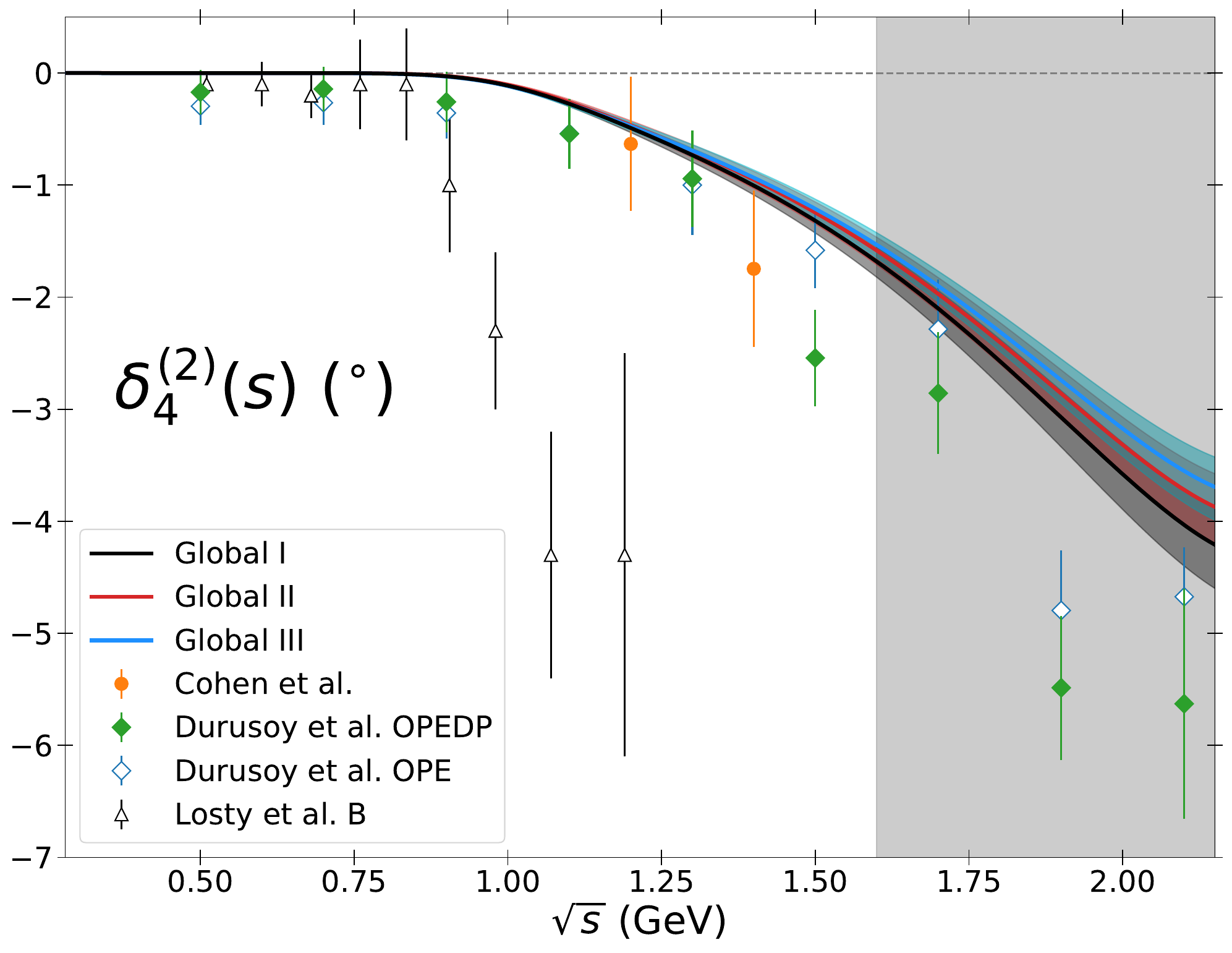}
\vspace*{-.7cm}
\caption{Comparison among Global Fits I, II, and III for the G2 wave, which are very compatible. Above 1.6 GeV (shaded region) there are no dispersive constraints. Data references as in Fig.~\ref{fig:S2}.}\label{fig:G2_sols}
\end{figure}

As seen in Fig.~\ref{fig:G2_sols} and in Table~\ref{tab:G2parameters}, the three Global Fits I, II, and III are remarkably compatible and fully consistent within uncertainties. 

\section{Dispersion relations}
\label{sec:disprel}

Dispersion relations are a consequence of causality, translating into strong analyticity constraints on scattering amplitudes when extended to the complex plane of the Mandelstam variables. For our case of interest, $\pi\pi$ scattering,  both fixed-$t$ and partial-wave amplitudes are analytic except for branch-cut singularities due to thresholds in the $s$-channel (right-hand cut) or crossed channels (left-hand cut). This structure allows us to use Cauchy's Integral Theorem to write integral equations, known as dispersion relations,
relating amplitudes in the complex plane to their imaginary parts along the cuts.  For pedagogical introductions, we refer the reader to \cite{Martin:1970xx,Itzykson:1980rh} and for $\pi\pi$ scattering in particular, to \cite{Martin:1976mb,Pelaez:2015qba}.

Since a two-body scattering amplitude $F(s,t)$ depends on two independent variables, in order to apply Cauchy's integral theorem, it is convenient to either fix $t$ or integrate it to define partial waves.
The first option gives rise to fixed-$t$ dispersion relations and the second to partial-wave dispersion relations. 
Next, we will discuss both the fulfillment and the use as constraints of the two most common instances of both types of dispersion relations: forward dispersion relations and Roy-like equations.

\subsection{Forward Dispersion Relations}

\subsubsection{Definitions}
Let us first discuss the fixed $t=0$ dispersion relations, also known as forward dispersion relations. They are very relevant for three reasons.
First, their applicability can be extended, in principle, to any value of $s$. Second, the optical theorem relates the imaginary part of the forward amplitude to the total cross section, for which data are easier to obtain. Third, when the combination of amplitudes is chosen conveniently, the numerators of the integrands are all positive, and the resulting uncertainties are generically small. 

Actually, from the $F^{(I)}(s,t)$ amplitudes in Eq.~\eqref{eq:FI} we can define these other three $F^i$combinations: 
\begin{eqnarray}
F^{00}=\frac{1}{3}\left( F^{(0)}+2F^{(2)} \right),\;
F^{0+}=\frac{1}{2}\left( F^{(1)}+F^{(2)}\right), \label{eq:Fsym}\\
F^{I_t=1}=\frac{1}{6}\left(2F^{(0)}+3F^{(1)}-5F^{(2)}\right). \label{eq:Fanti}
\hspace{1cm}
\end{eqnarray}
The first two correspond to $\pi^0\pi^0$ and $\pi^0\pi^+$ scattering, respectively, and are symmetric under the $s\leftrightarrow u$ crossing symmetry. 
This allows us to write a once-subtracted Forward Dispersion Relation for the $i=00,0+$ cases, which implies the vanishing of
\begin{eqnarray}
\Delta^i(s)&\equiv&\Re F^i(s,0)-F^i(4M^2_\pi,0)-\frac{s(s-4M^2_\pi)}{\pi} \label{eq:FDRsym}\nonumber\\
&\times& P.P.\int_{4M^2_\pi}^\infty \frac{(2s'-4M^2_\pi) \Im F^i(s',0)\,ds'}{s'(s'-s)(s'-4M^2_\pi)(s'+s-4M^2_\pi)},\nonumber\\
\end{eqnarray}
where $P.P.$ is the principal part of the integral.
These two FDRs are very precise because, from Eq.~\eqref{eq:Fsym},
and recalling that the imaginary parts of $F^{(I)}$ are always positive, all the contributions to the numerator in their integrals are positive.

The amplitude in Eq.~\eqref{eq:Fanti} is antisymmetric under $s\leftrightarrow u$ crossing and corresponds to the exchange
of isospin $I=1$ in the $t$-channel. For it, we can write an unsubtracted FDR, which implies the vanishing of
\begin{eqnarray}
\Delta^{I_t=1}(s)&\equiv& F^{I_t=1}(s,0)-\frac{2s-4M^2_\pi}{\pi}\nonumber
\\
&\times& P.P.\int_{4M^2_\pi}^\infty \frac{\Im F^{I_t=1}(s',0)\,ds'}{(s'-s)(s'+s-4M^2_\pi)}. \label{eq:FDRanti}
\end{eqnarray}
This one is not definite positive, and its uncertainties will be generically larger.

Note that the integral of dispersion relations extends up to infinity. At high energies, above a certain matching energy, we use the Regge parametrizations of $\pi\pi$ amplitudes obtained from $\pi\pi$, $\pi N$, and $NN$ cross sections in~\cite{Pelaez:2003ky} and upgraded and updated in~\cite{GarciaMartin:2011cn} and the review~\cite{Pelaez:2015qba}.
In a later subsection, we will discuss the matching with these Regge parametrizations. 

In what follows, we will first check how well the FDRs are satisfied with the partial-wave fits to the data discussed above, and then we will use them as constraints on the fits.

\begin{figure*}
\centering
\includegraphics[width=0.49\textwidth]{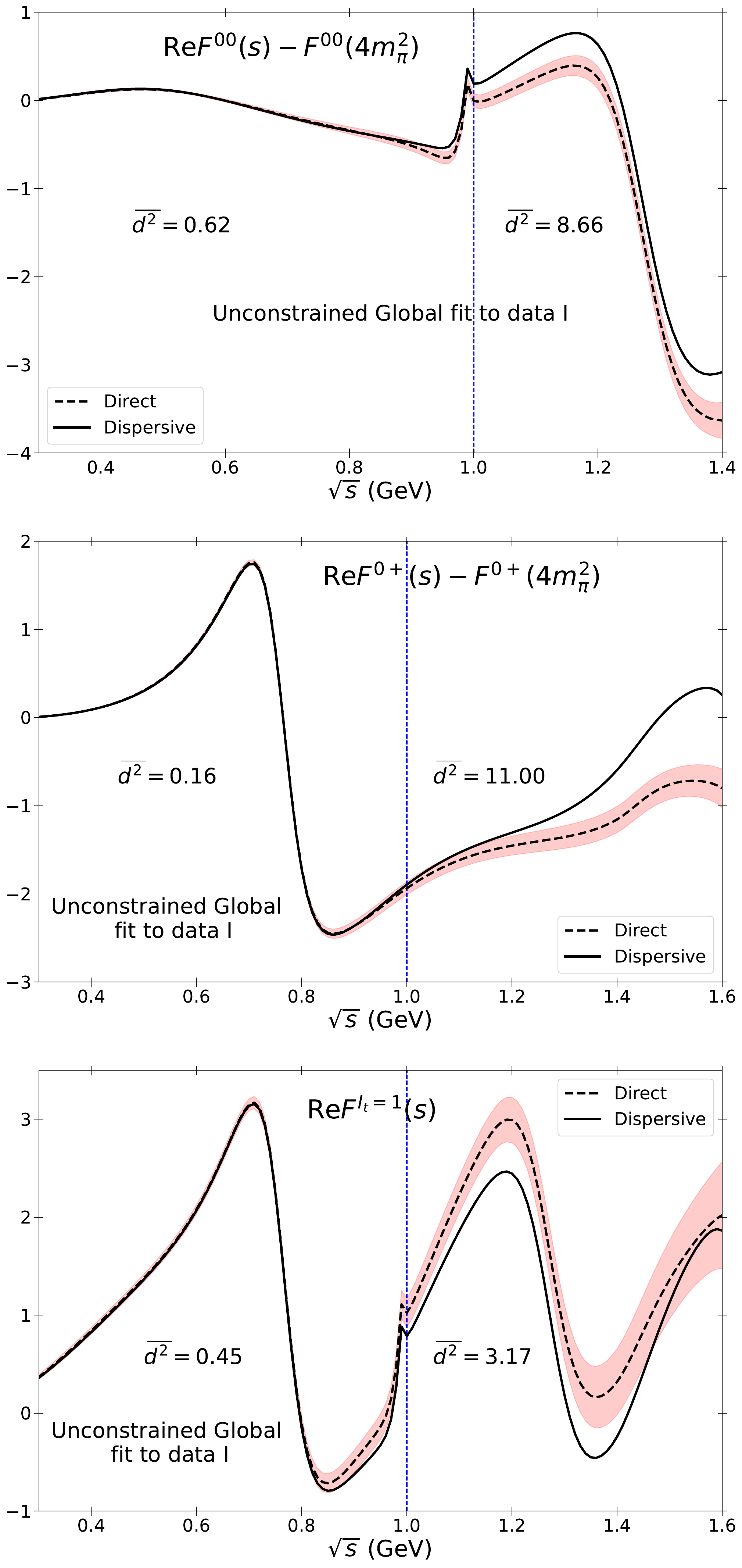}
\includegraphics[width=0.49\textwidth]{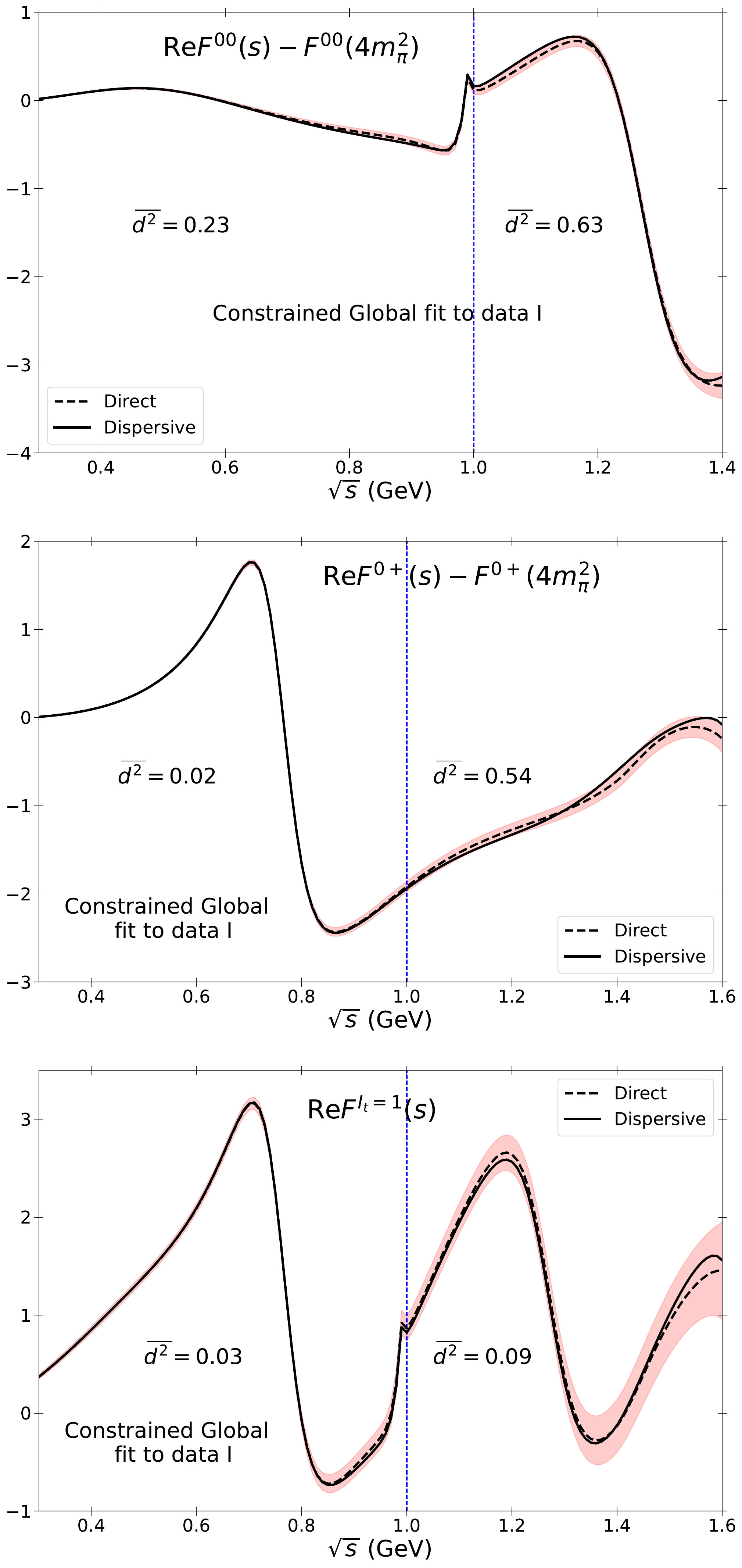}
\caption{Fulfillment of forward dispersion relations before (left column)
and after (right column), they have been imposed as constraints on the Global Fit I. The continuous line corresponds to the evaluation of 
$\Re F^{i}$ with the dispersive integral and using the global parametrizations as input, whereas the dashed line is calculated directly from the global parametrizations. The red band is the error in their difference, that we attach to the ``direct" one for illustration. The vertical blue line at 1 GeV separates the two energy regions, for which we provide separated averaged quadratic distances $\bar d^2$. 
The unconstrained Global Fit I only satisfies well the FDRs below 1 GeV ($\bar d^2<1$), but is rather inconsistent above ($\bar d^2>>1$). In contrast, the constrained Global Fit I satisfies very well the FDRs in the two energy regions.}\label{fig:FDRs} 
\end{figure*}

\subsubsection{FDRs as checks}

In~\cite{Pelaez:2004vs,Kaminski:2006qe,Kaminski:2006yv,Yndurain:2007qm, GarciaMartin:2011cn} the three FDRs in Eqs.~\eqref{eq:FDRsym} and~\eqref{eq:FDRanti} were imposed as constraints of phenomenological fits to data, to be satisfied within uncertainties to obtain the CFD parametrizations.
Moreover, in~\cite{GarciaMartin:2011cn} Roy and GKPY equations---to be explained below---were imposed on the CFD too.
Since our global parametrizations mimic the CFD sets up to 0.9 GeV, it is no surprise that the FDRs are well satisfied by our Global Fits in that region, even without using them as constraints. This can be seen in the left panels of Fig.~\ref{fig:FDRs}, where we show the fulfillment of FDRs {\it before} we impose them as constraints in our Global Fits to data. Note that in Fig.~\ref{fig:FDRs} we show as dashed lines the parts of the FDRs that are calculated directly from the parametrizations and as continuous lines the same quantities obtained from the integral representation, which we call ``dispersive".
In theory (in the isospin limit, without $4\pi$ inelasticity, etc), these two lines should agree, and their difference should be zero. However, since ours is a data analysis, we only expect them to agree within uncertainties. The uncertainties of the ``direct" and ``dispersive" parts are correlated, as they are both calculated from the same parametrizations. Such correlations cancel out to a large degree in their difference.
Thus, we calculate the uncertainty band of their difference and, for illustrative purposes,  we attach it to the direct part. Thus, we aim at a fit such that the continuous line would fall within that red uncertainty band. From the left panels of Fig.~\ref{fig:FDRs} it is clear that the three FDRs are not well satisfied above the 1 GeV region before we impose them as constraints into our Global Fits.

To quantify this disagreement we have defined, following~\cite{GarciaMartin:2011cn}, a $\chi^2$-like quantity as follows: for each FDR $i=00,\,0+,\,I_t=1$, we calculate $\Delta^i(s_k)$ and its uncertainty $\delta\Delta^i(s_k)$, at a collection of points $s_k$, $k=1, ...n$ that cover a given energy region. In addition, as in \cite{GarciaMartin:2011cn}, we also consider a subthreshold point at $2m_\pi^2$, for stability. We then define an average discrepancy for each FDR as
\begin{equation}
 \bar d_i^2 \equiv \frac{1}{n}\sum_{k=1}^{n}\left(\frac{\Delta^i(s_k)}{\delta\Delta^i(s_k)}\right)^2,
 \label{eq:distances}
\end{equation}
which can be interpreted as a quadratic distance weighted by the uncertainty, or an averaged $\chi^2$ of the difference between the curves. We consider the FDR $i$ to be well satisfied in a given region when $\bar d_i^2\leq1$ there.
Thus, in Fig.~\ref{fig:FDRs} we provide these averaged discrepancies in two representative regions. On the one hand, from the subthreshold point to
1 GeV, and on the other, from 1 GeV to the maximum energy we consider for each FDR. Namely, 1.4 GeV for $F^{00}$ and 1.6 GeV for the other two. 
As discussed below, these maximum energies have been chosen somewhat below the round number (1.42 and 1.62 GeV, respectively) closest to the energy where the partial wave series, used below these energies, and the Regge representation, used above them, match within uncertainties. 

Thus, making now quantitative our previous qualitative discussion, in the left panels of Fig.~\ref{fig:FDRs} we see that our Global Fits {\it before} imposing the FDRs as constraints satisfy very well the FDRs below 1 GeV, with all $\bar d_i^2\leq1$, but they do not satisfy them above 1 GeV, with $\bar d_{I_t=1}^2\sim3.2$, $\bar d_{00}^2\sim8.7$ and $\bar d_{0+}^2\sim11$. Since we have fit data, it is clear, as already known from previous works, that the data are inconsistent with the dispersive representation. 

\subsubsection{FDRs as constraints: The constrained Global Fit.}

For the above reasons, we impose the FDRs as constraints of our Global Fits.
Note that we first impose the FDRs, because our global parametrizations here mimic the CFD below 0.9 GeV and only deviate significantly from them above that energy. The applicability range of Roy and GKPY equations extends up to approximately $1.1\,$GeV and were already imposed in the CFD in \cite{Pelaez:2004vs,Kaminski:2006qe,Kaminski:2006yv,Yndurain:2007qm}. Since they have relatively large uncertainties above 0.9 GeV compared to FDRs, they are still fairly well satisfied with only a minor deviation around 1.1 GeV for the S2 GKPY equation. Thus, we first impose the FDRs and, on a second step
discussed in a subsection below, we also impose the Roy and GKPY equations, which only produce a minor modification to reach our final Global Fits. As in previous sections, we only illustrate this procedure in detail for Global Fit I, and we just discuss the results for Global Fit II and III later on.


Thus, following the procedure in~\cite{Pelaez:2004vs,Kaminski:2006qe,Kaminski:2006yv,Yndurain:2007qm}, we minimize the combination
\begin{equation}
    \sum_i W_i^2 \bar d^2_i+\sum_k\left(\frac{p_k-p_k^{U}}{\delta p^U_k}\right)^2+\sum_m\left(\frac{q_m-q_m^{exp}}{\delta q_m^{exp}}\right)^2,
    \label{eq:chisqpenalty}
\end{equation}
where $i$ runs over the three FDRs. Here, we denote by $p_k^{U}$ the parameters of the unconstrained Global Fits described above and their uncertainties
by $\delta p_k^U$.   The $q_m^{exp}$ are just a collection of data points or threshold parameters detailed below.

In principle, when imposing the dispersion relations as penalty functions, it is less computationally costly to perturb all the parameters around their unconstrained values $p_k^{U}$, than to refit all the data $q_m^{exp}$. This is what was done in \cite{GarciaMartin:2011cn}. However, we have now found that 
with this method, some parameter changes are not small. 
Thus, we do not vary those parameters but refit the data in the wave or the region they affect.

In particular, in Eq.~\eqref{eq:chisqpenalty} we do not include in the $k$ sum the $p_k=K_i, d_i$, and $\epsilon_i$ parameters in Eqs.~\eqref{eq:ineP},~\eqref{eq:P_phase-high}, and~\eqref{eq:P_ine-high}, respectively, but instead they are constrained indirectly by refitting the $q_m^{exp}$ P-wave experimental data above the $\pi\omega$ threshold.
The same happens with the S0-wave parameters above 1.4 GeV,
the S2-wave elasticity parameter, and the F-wave Global Fit II parameters.  Instead, they are constrained by including their respective data input in the $m$ sum.
In addition, we also include in that sum all the input threshold parameters from Table~\ref{tab:ThresholdParameters} in the Appendix~\ref{app:Threshold}.

The $W_i$ are weights chosen so that the resulting fit has $\bar d^2_i\lesssim1$ for all FDRs, and not just globally in their entire applicability range, but also in a relatively uniform way throughout the whole region. For this purpose, we divide the calculation of the $\bar d^2_i$ into different energy regions,
assigning them a different weight to ensure a $\bar d^2_i\leq1$ in all of them.
The $\bar d^2_i$ are often called penalty functions in the literature.
As a final remark, let us note that we minimize this pseudo-$\chi^2$ to obtain the central values of the new parameters, but 
then we keep their uncertainties $\delta p_k=\delta p_k^{U}$.

The results of this procedure are the constrained Global Fits already discussed in the previous section and shown in Figs.~\ref{fig:P} to~\ref{fig:G2_sols}.
We have illustrated with the constrained Global Fit I that they still describe the data reasonably well, and we now show that they also satisfy FDRs within uncertainties up to their maximum applicability region. This can be seen in the right panels of Fig.~\ref{fig:FDRs}, where, in contrast to the left panels, the ``direct'' and ``dispersive'' calculations agree within uncertainties. In particular, the $\bar d_i^2\leq1$ for the three FDRs in both regions, below and above 1 GeV. The improvement above 1 GeV is huge and there is even a slight improvement below 1 GeV.
Therefore, Global Fit I satisfies remarkably well the FDRs.

\begin{figure}[ht]
\centering
\vspace*{-0.3cm}
\includegraphics[width=0.46\textwidth]{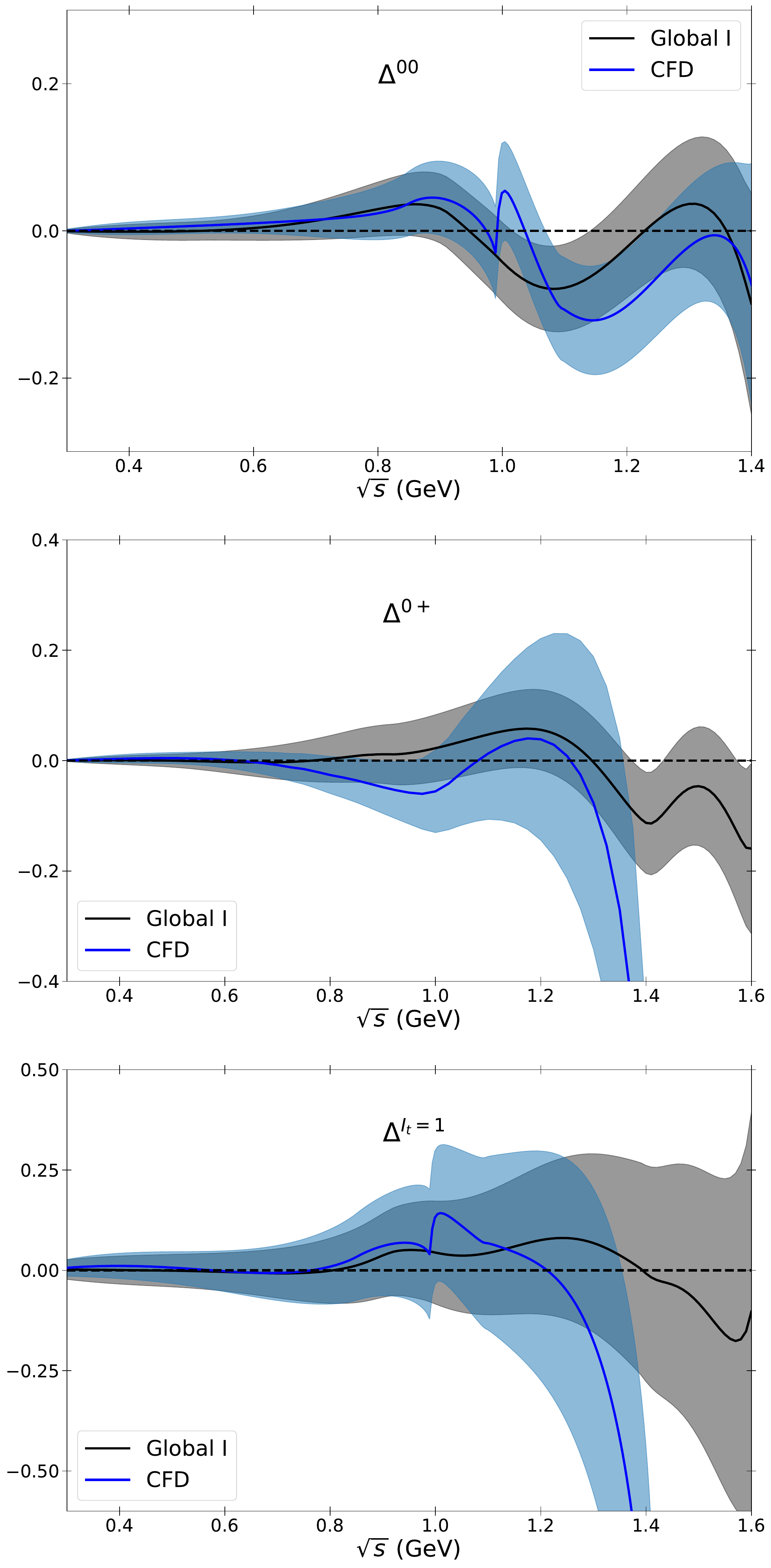}
\vspace*{-0.3cm}
\caption{Comparison between the FDR error bands when Global Fit I or
the CFD in~\cite{GarciaMartin:2011cn} are used as input. We show the differences $\Delta^i$ for $i=00,0+, I_t=1$.
Note that, the new error bands are smaller between 0.9 and 1.4 GeV, particularly 
for the $\Delta^{0+}$, dominated by the new P-wave analysis.
Also, the instability of the $0+$ and $I_t=1$ FDRs at 1.4 GeV, (the FDRs last point in~\cite{GarciaMartin:2011cn}), which was an artifact of the poor matching, has been removed. The fulfillment of $\Delta^{00}$ between 1 and 1.2 GeV has visibly improved too. Note that our $0+$ and $I_t=1$ FDRs now extend up to 1.6 GeV.}\label{fig:FDRerrors} 
\end{figure}

It is important to emphasize that the error bands in the 0.9 to 1.4 GeV region of the FDRs
for the Global Fits obtained in this work are smaller than
those of previous works~\cite{GarciaMartin:2011cn,Pelaez:2019eqa}.
This is clearly seen in Fig.~\ref{fig:FDRerrors}, where
we plot the $\Delta^i(s), i=00,0+, I_t=1$ differences defined in Eqs.~\eqref{eq:FDRsym} and~\eqref{eq:FDRanti} both for our Global Fit I and the CFD parametrizations in~\cite{GarciaMartin:2011cn}. 
The new bands have become visibly smaller, mostly due
to our much-improved P wave above the $\pi \omega$ threshold, and to a lesser extent, to the  slight improvement in the D0 wave. 
The largest uncertainty reduction occurs in $\Delta^{0+}$,
which is dominated by the P wave, followed by the $\Delta^{I_t=1}$, where the P wave also has a large contribution. The $\Delta^{00}$
uncertainty band, where the P wave does not contribute, is only slightly smaller than in previous works.
For this reason, the FDRs in the 0.9 to 1.4 GeV region, now extended to 1.6 GeV for two of them, have become even more stringent constraints than in previous works, and their compliance is a more notable feature.

The average fulfillment is remarkably good when looking at
the two large energy intervals below and above 1 GeV.
Nevertheless, in Fig.~\ref{fig:FDRerrors} we can see that there is a smaller interval, roughly 1.02 to 1.14 GeV, where 
$\Delta^{00}$ lies slightly beyond one deviation from zero,
although never beyond 1.4 deviations. In that interval $\bar d^2_{00}=1.53$, so it is not too worrisome. 
Actually, this has been an improvement, because as seen in Fig.~\ref{fig:FDRerrors} and already remarked in~\cite{Pelaez:2022qby} (see Fig.~1 there),
$\Delta^{00}$ for the CFD does not vanish within uncertainties in an even larger region between 1.07 and 1.24 GeV, where the discrepancy reached a maximum of 1.8 deviations at one point.
The same happens for an even smaller discrepancy in an even smaller interval near 1.4 GeV 
and in the last point for $\Delta^{0+}$ at 1.6 GeV.
In any case, Figs.~\ref{fig:FDRs} and~\ref{fig:FDRerrors}
illustrate that the Global Fit I, despite its smaller uncertainties, satisfies the three FDRs very well and in a much more uniform way than the old CFD.

\subsubsection{The three constrained Global Fits.}

So far, we have illustrated the FDR fulfillment with the constrained Global Fit I. Thus, in Table~\ref{tab:FDRs} we now collect the values of $\bar d_i^2$ in the two regions, above and below 1 GeV, not only for Global Fit I but also for Global Fits II and III. They also satisfy the FDRs well $\bar d_i^2<1$ in the two regions, although with somewhat larger $\bar d_i^2$.

As with Global Fit I, Global Fits II and III also have smaller regions where the $\Delta^{00}$ discrepancy is slightly beyond one deviation away from 0. Namely, Global Fit II
has $\bar d^2_{00}=1.4$ in the 1.10 to 1.18 GeV interval, and Global Fit III has $\bar d^2_{00}=2$ between 1.09 and 1.19 GeV. 
Moreover, Global Fit II between 0.7 and 1 GeV --- a 300 MeV region--- has $\bar d^2_{00}=1.23$.
In this sense, although Global Fits II and III also 
improve their CFD counterparts in~\cite{GarciaMartin:2011cn}, their fulfillment of the $F^{00}$ FDR is slightly worse than for Global Fit I.

\begin{figure}[h]
\centering
\includegraphics[width=0.46\textwidth, height=0.82\textwidth]{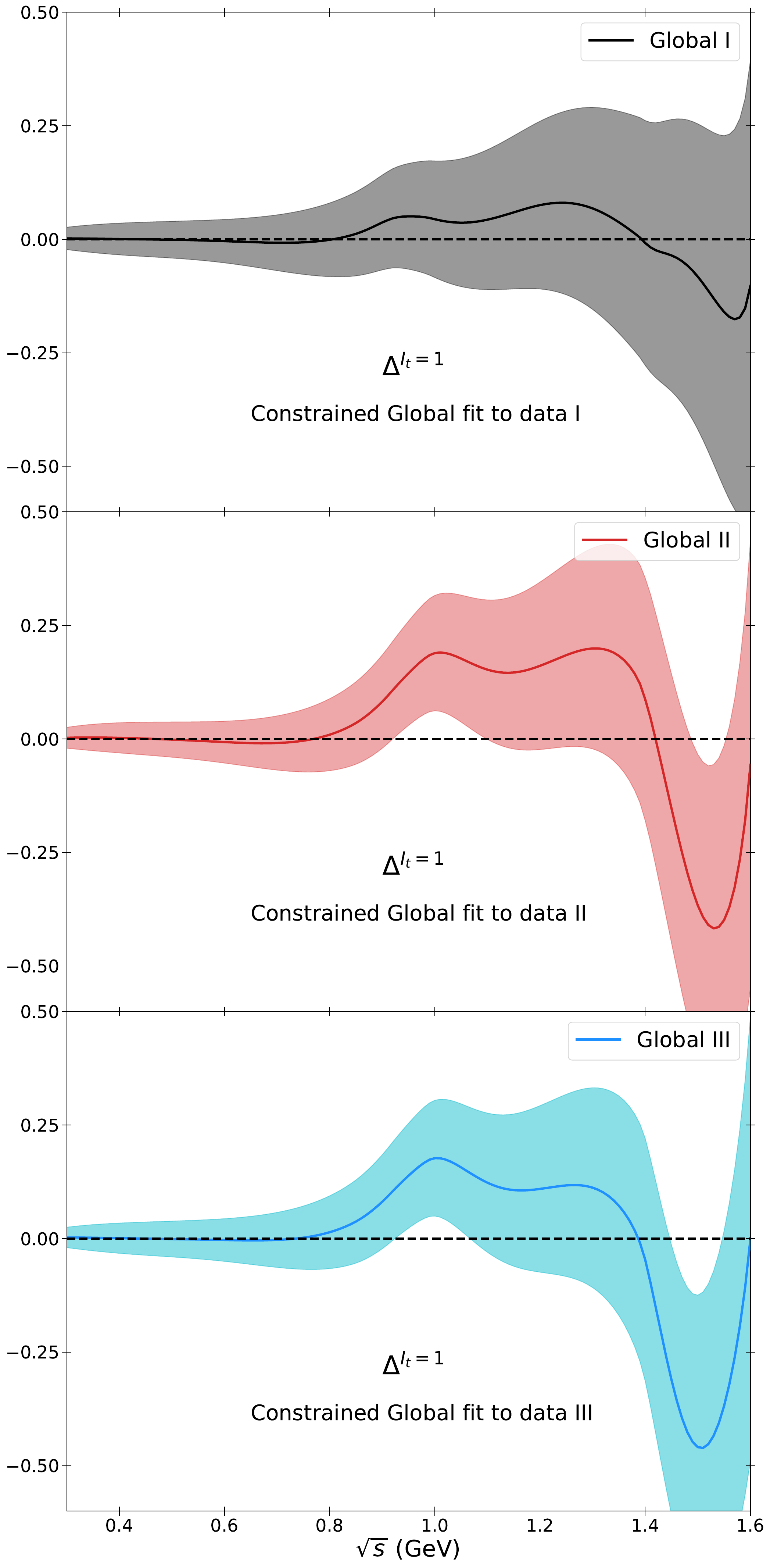}
\caption{Detail of the fulfillment of the $F^{I_t=1}$ FDR for the three Global Fits. We plot the difference between dispersive and direct calculations $\Delta^{I_t=1}$ given in Eq.~\eqref{eq:FDRanti}, and its uncertainty band. 
Global Fits II and III are slightly disfavored relative to Global Fit I because they satisfy this relation slightly worse and less uniformly. In particular,
their $\Delta^{I_t=1}$ is more than one deviation away from zero in two regions 
wider than 100 MeV around 1 GeV and 1.5 GeV, which does not happen for Global Fit I (see also Table~\ref{tab:FDRs}).}\label{fig:Olsson}
\end{figure}

The situation regarding the $F^{I_t=1}$ FDR is more telling, since Global Fits II, and III satisfy the $I_t=1$
FDR in a clearly less uniform way than Global Fit I.
This is shown in Fig.~\ref{fig:Olsson},
where we plot the $\Delta^{I_t=1}$ differences and their uncertainties for the three Global Fits.  Although the average $\bar d_{I_t=1}^2<1$, both below and above 1 GeV, only the Global Fit I has $\bar d_{I_t=1}^2<1$ everywhere. In contrast, for Global Fits II and III, $\Delta^{I_t=1}$ lies beyond one deviation from zero in two regions wider than 100 MeV, around 1 GeV and 1.5 GeV.
As it can be seen in Table~\ref{tab:FDRs}, for these two constrained Global Fits, 
the $F^{I_t=1}$ FDR yields $\bar d^2> 1.5$ between $0.93$ and $1.06 \gev$. In addition, Global Fit III yields $\bar d^2\simeq 1.5$ from
$1.46$ to $1.56\gev$. The $0.93-1.06 \gev$ region 
is of particular interest because it is where most waves become inelastic.
Actually, these small discrepancies in Global Fits II and III can be amended if we set the opening of their D0-wave inelasticity below the $K\bar K$ threshold, as done for Global Fit I. 
However, as explained in previous sections, this is not how the data for Solutions II and III were obtained. 
Hence, in this respect,  Global Fits II and III are, once again, slightly disfavored compared to Global Fit I, and in particular, the opening of the D0-wave inelasticity below 1 GeV is strongly favored.



\begin{table}
\renewcommand{\arraystretch}{1.3}
\centering 
\begin{tabular}{c|cccc} 
\multicolumn{5}{c}{$\bar d^2_i$ FDRs}\\
\toprule
$\bar d^2_i$ & $\sqrt{s}$ & \hspace{0.2mm} Global I & \hspace{0.2mm} Global II  & \hspace{0.2mm} Global III \\
\toprule
\multirow{2}{*}{$\pi^0 \pi^0$} & $<1\,$GeV &0.23 &0.69 &0.26  \\
&  $\left[1,1.4\right]\,$GeV& 0.63 & 0.58 &0.68\\
\colrule
\multirow{2}{*}{$\pi^+ \pi^0$} & $<1\,$GeV & 0.02&0.06 & 0.03 \\
& $\left[1,1.6\right]\,$GeV& 0.54 &0.88 &0.81\\
\colrule
\multirow{4}{*}{$I_{t}=1$} & $<1\,$GeV& 0.03&0.24 &0.21  \\
& $\left[1,1.6\right]\,$GeV & 0.09& 0.83&0.67\\
& $\left[0.93,1.06\right]\,$GeV& 0.13 & 1.81 &1.56\\
& $\left[1.46,1.56\right]\,$GeV&0.09& 1.04&1.47\\
\botrule
\end{tabular} 
\caption{Average discrepancies $\bar d^2_i$ after constraining Global Fits I, II, and III with the three FDR $i=00,0+, I_t=1$. For all the FDRs we separate 
two large regions above and below 1 GeV, where their average 
fulfillment is good, i.e. $\bar d^2_i<1$. Still, Global Fits II and III always perform somewhat worse than Global Fit I. In particular, in two smaller segments, but still about 100 MeV wide, Global Fits II and III do not satisfy so well the $F^{I_t=1}$ FDR, i.e. $\bar d^2_i>1$.
}
\label{tab:FDRs} 
\end{table}

\subsubsection{Improved matching with the high-energy regime}

\begin{figure}
\centering
\includegraphics[width=0.48\textwidth]{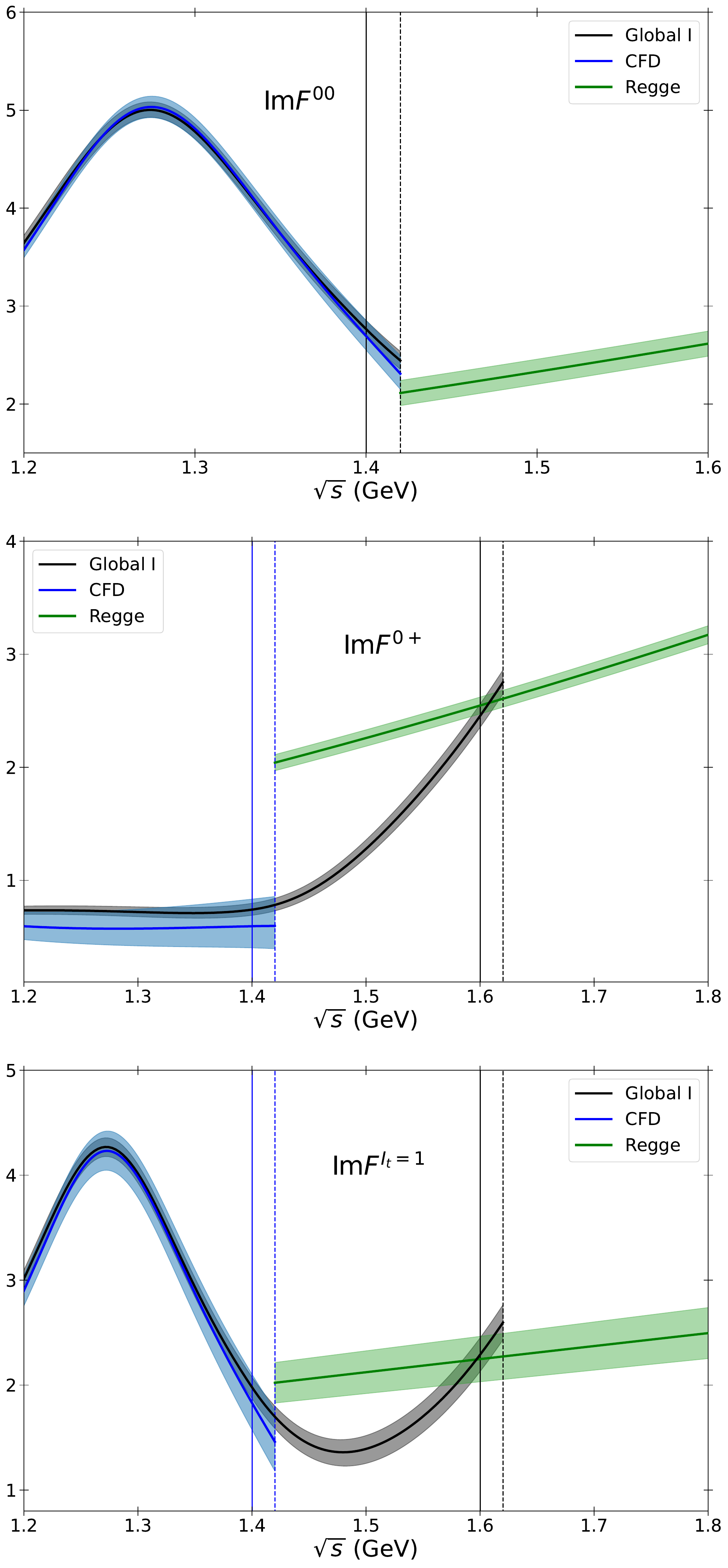}
\vspace*{-.5cm}
\caption{ $\im F^{00}, \im F^{0+}$ and $\im F^{I_t=1}$ in the 1.2 to 1.8 GeV region. 
Since Ref.~\cite{GarciaMartin:2011cn} was focused on energies below 1.1 GeV, 
a crude matching between the CFD and the Regge description at 1.42 GeV (blue-dotted vertical line)
was considered good enough. For a precise description up to higher energies, we now use as input the Global Fits up to 1.62 GeV (black-dotted vertical line) for $F^{0+}$ and $F^{I_t=1}$, and Regge above. This matching point displacement clearly improves the matching with the Regge parametrization. 
Note that, for safety, we only consider as constraints the FDRs up to 20 MeV below the matching point (continuous versus dotted vertical lines). We use Global Fit I for illustration, but the situation is similar for Global Fits II and III.
}\label{fig:Regge} 
\end{figure}

Following~\cite{Pelaez:2004vs,Kaminski:2006qe,Kaminski:2006yv,Yndurain:2007qm,GarciaMartin:2011cn}, the input for the FDRs at high energies is a relatively simple parametrization of data on $NN$, $\pi N$ and $\pi\pi$ total cross sections (see~\cite{Pelaez:2003ky} for the formalism, and~\cite{GarciaMartin:2011cn} for the updated treatment). It is important to remark that Regge theory is just a semi-local approach expected to provide an average description, particularly once resonances start to appear. 
Therefore, at any given energy, it is not expected to match exactly the amplitude reconstructed from partial waves.
Since the interest of \cite{Pelaez:2004vs,Kaminski:2006qe,Kaminski:2006yv,GarciaMartin:2011cn,Yndurain:2007qm,GarciaMartin:2011jx} was mostly
in the low-energy region or in resonances up to 1 GeV (external $s$ variable in Eqs.~\eqref{eq:FDRsym} and~\eqref{eq:FDRanti}), the Regge regime was set to start at 1.42 GeV for all FDRs (internal $s'$ variable in Eqs.~\eqref{eq:FDRsym} and~\eqref{eq:FDRanti}).
As seen in Fig.~\ref{fig:Regge}, no particular attention was paid to provide a smooth matching between the partial-wave reconstruction of $\Im F^i$ below 1.42 GeV and the Regge regime above.
Actually, the matching in the $F^{0+}$ case was pretty bad (green versus blue at 1.42 GeV). Such discontinuities in the integrand could give rise to nearby artifacts in the FDR output, as seen in Fig.~\ref{fig:FDRerrors} for the CFD. Of course,  with the focus in the region below 1.1 GeV, this was not a big concern for~\cite{GarciaMartin:2011cn}.
However, in this work, we aim to obtain a precise description up to energies of $1.4\,$GeV and above.

For this reason, we have decided to move higher the matching point with the Regge regime; from 1.42 GeV to the round energy value close to the point where the $F^i$ amplitudes, reconstructed as a sum of partial waves, match the Regge description within uncertainties. 
These plots justify why the partial-wave input is matched with the Regge description at 1.42, 1.62, and 1.62 GeV for the $F^{00}, F^{0+}$ and $F^{I_t=1}$ FDRs, respectively.
Still, as also done~\cite{GarciaMartin:2011cn}, for safety we only use the output of FDRs up to an energy about 20 MeV below the matching point.  Namely, the FDRs are applied as constraints only up to 1.4, 1.6, and 1.6 GeV, respectively.

Nevertheless, we have found that the final Global Fit does not depend much on the precise matching point. This justifies why we have chosen the matching energy with the Regge representation to be at the 
  same round numbers, 1.42 or 1.62 GeV, for all Global Fits. We explored the possibility of fine-tuning these matching points for each Global Fit, but there is no real gain, just a proliferation of fine-tuned parameters.

In summary, by choosing a better matching point between the amplitude reconstructed from partial waves and the Regge description, we hope to have soothed any concern about the appearance of unphysical artifacts.
Moreover, it has allowed us to impose two FDRs up to 200 MeV above their previous region of application.

\subsection{Roy-like dispersion relations}

\begin{figure*}
\centering
\includegraphics[width=0.47\textwidth]{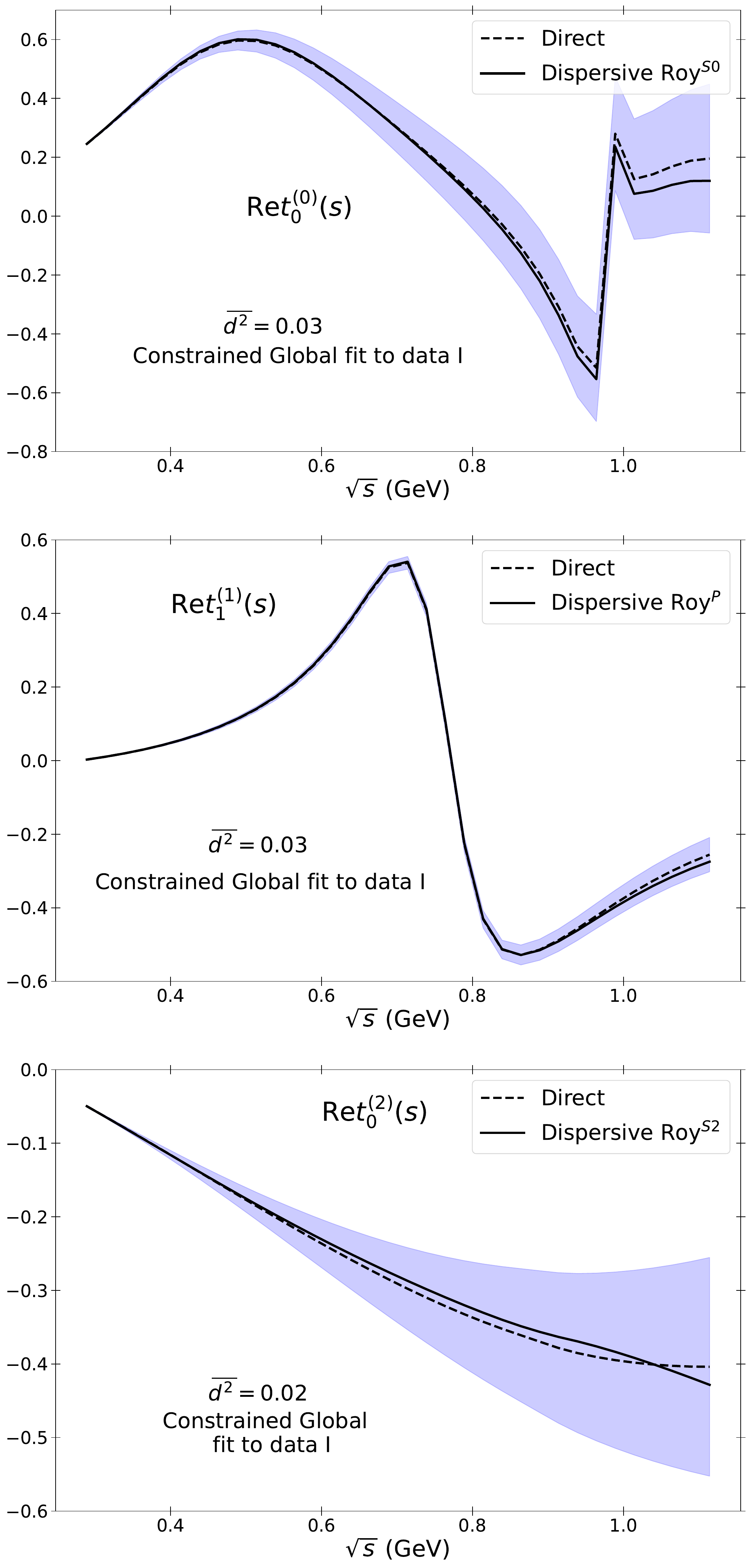}
\includegraphics[width=0.47\textwidth]{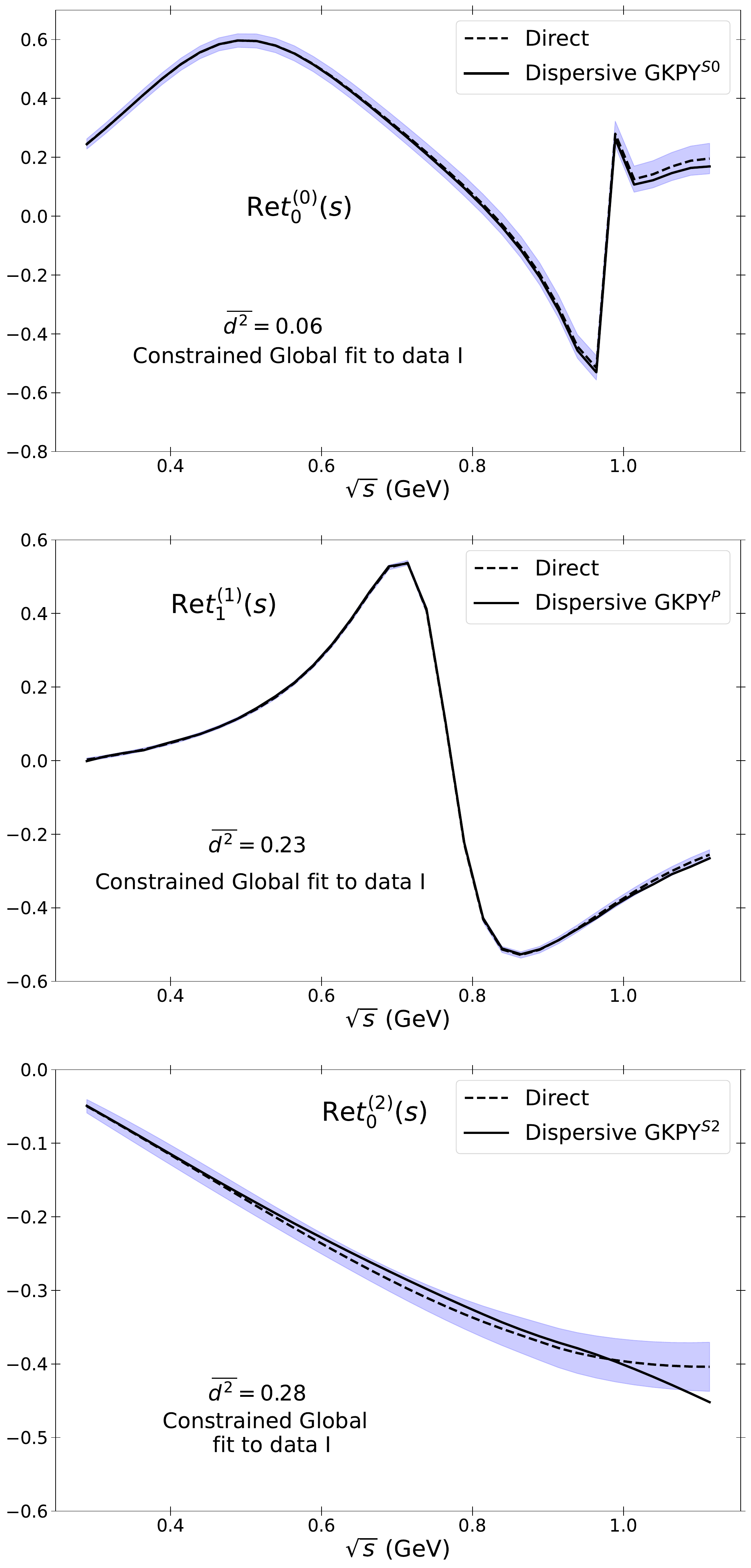}
\caption{Fulfillment of Roy and GKPY equations
(left and right columns, respectively) by the constrained Global Fit I. Note that the error band of each Roy equation is smaller than that of the corresponding GKPY relation at low energies but larger at high energies.
The bands cover the uncertainty in the difference between the direct calculation from the Global Fit and 
the calculation with the dispersive integral. We attach it to the direct calculation for illustration.}\label{fig:Roy-like}
\end{figure*}

Roy equations for $\pi\pi$ scattering~\cite{Roy:1971tc}
(see~\cite{Ananthanarayan:2000ht,Colangelo:2001df,DescotesGenon:2001tn,Kaminski:2002pe,GarciaMartin:2011cn,Pelaez:2015qba,Pelaez:2021dak} for recent analyses and reviews)
are partial-wave dispersion relations where crossing symmetry has been used to 
rewrite the left-cut (crossed-channel) contributions in terms of s-channel partial waves. 
Thus, they form a coupled system for the infinite tower of partial waves. At sufficiently low energies, one can concentrate on those with the lowest angular momentum, typically, the S0, S2, and P waves (and occasionally, D and F waves, see~\cite{Kaminski:2011vj}). In such a case, higher partial waves and high energy information are provided as input and gathered in the so-called driving terms.
In their most common form, which we use here, their derivation implies the integration in $t$ of fixed-$t$ dispersion relations to obtain partial-wave dispersion relations. This means their applicability range is limited to an energy of around $1.14$ GeV.
The original Roy equations were derived with two subtractions, but there is also a once-subtracted version called GKPY equations~\cite{GarciaMartin:2011cn}. We refer to both of them as ``Roy-like" equations. As we will see below, this different number of subtractions implies that, when calculated from the same data without further theoretical input, Roy equations are more precise than GKPY equations at low energies and vice versa at high energies.
This is because  Roy uncertainties grow rapidly with energy, whereas
GKPY uncertainties do not grow so fast. In practice, Roy uncertainties are smaller than those of GKPY below the resonance region, i.e., $s^{1/2}\lesssim 0.5\gev$, and larger above. Hence, Roy equations are particularly powerful in constraining threshold parameters~\cite{Ananthanarayan:2000ht,Colangelo:2001df} or the low-energy 
constants of ChPT. Actually, when supplemented with ChPT constraints, Roy equations can be solved for the S and P waves and provide predictions below some matching point, or even for the lightest resonances~\cite{Ananthanarayan:2000ht,Colangelo:2001df,DescotesGenon:2001tn,Caprini:2005zr}.
However, here we simply adhere to a data-driven formalism and use the dispersion relations as constraints on the fits.
Nevertheless, we refer to ~\cite{Ananthanarayan:2000ht,Colangelo:2001df,DescotesGenon:2001tn,Kaminski:2002pe,GarciaMartin:2011cn,Pelaez:2015qba,Pelaez:2021dak} for explicit expressions and further details.

In the previous section, we imposed Forward Dispersion Relations to obtain constrained Global Fits from the threshold to 1.4 GeV.
Recall that below 0.9 GeV we are mimicking the CFD parametrizations of~\cite{GarciaMartin:2011cn}. They were obtained as data fits constrained to satisfy the three FDRs, as well as Roy and GKPY equations for the S0, P, and S2 waves. The only significant improvements in the present work happen above 0.9 GeV, and, for several waves, 
these changes are small until 1.4 GeV. Consequently, Roy and GKPY equations are already very well satisfied within uncertainties by just imposing FDRs. The only exception is the S2 GKPY equation above 1 GeV, whose output comes about 1.5 deviations away for the Global Fits constrained only with FDRs.
This mismatch requires a small fix in our Global Fits.

\begin{table}[h]
\renewcommand{\arraystretch}{1.3}
\centering 
\begin{tabular}{cccc} 
\multicolumn{4}{c}{$\bar d^2_i$ Roy-like Equations}\\
\toprule
$\bar d^2_i$ & \hspace{0.2mm} Global I & \hspace{0.2mm} Global II  & \hspace{0.2mm} Global III \\
\toprule
 Roy S0& 0.03& 0.02&0.03\\
 GKPY S0& 0.06& 0.01&0.02\\
\colrule
Roy P& 0.03 & 0.05&0.06\\
 GKPY P& 0.23&0.25 &0.27\\
 \colrule
 Roy S2& 0.02& 0.02&0.03\\
 GKPY S2& 0.28& 0.14&0.26\\
\botrule
\end{tabular} 
\caption{Average quadratic distances of Roy and GKPY equations for the 
three Global Fits. These ``Roy-like" dispersion relations, reaching only up to 1.1 GeV, are remarkably well described within uncertainties for the three fits.}\label{tab:Roy-like}
\end{table}

To that end, we have also imposed Roy and GKPY equations
together with FDRs, starting from the Global Fits obtained after minimizing the FDRs first. The changes are very small but amend the small deviation in the S2 GKPY equation.
As a matter of fact, all the plots for the constrained Global Fits that we have provided before are fully constrained with three FDRs, three Roy, and three GKPY equations. In Fig.~\ref{fig:Roy-like} we 
illustrate the nice fulfillment of Roy and GKPY equations by the Global Fit I. In those plots, we can see that, as discussed above, the Roy equation uncertainties are smaller than those of GKPY at low energies but larger in the resonance region. We can also see that both Roy and GKPY equations are well satisfied within errors.
The only point slightly beyond 1 deviation is the highest one at 1.114 GeV, for the GKPY S2 wave.

The situation is very similar for Global Fits II and III. Actually, we provide in Table~\ref{tab:Roy-like} the values of the average quadratic distances (defined as in Eq.~\eqref{eq:distances})
between output and input of Roy and GKPY equations.
The three Global Fits satisfy
remarkably well all Roy-like dispersion relations
in all regions of interest.

\section{Summary and Discussion}
\label{sec:discussion}

In this work, we have provided Global Fits, which are relatively simple sets of dispersively constrained $\pi\pi\rightarrow\pi\pi$ fits to data on scattering partial waves. Specifically, we describe the P, S2, D0, D2, F, G0, and G2 waves. The S0 wave parametrization is not revisited here because it was already studied in great detail in~\cite{Pelaez:2019eqa} together with the P wave that we have improved here above $\sim0.9\,$GeV. We only discuss in an appendix the small changes in the S0 parameters induced by the larger changes in the other waves.

By ``Global" we mean that the parametrizations are continuous in the whole energy region where they are defined, and also have a continuous derivative, except as required by thresholds.
Moreover, they
extend from the $\pi\pi$ threshold up to 1.8 GeV or somewhat beyond, depending on where the data on each partial wave cease to exist.
They are dispersively constrained because, when fitting the data, we impose as penalty functions the averaged quadratic distances between the direct and dispersive calculation of nine dispersion relations. These are: three Forward Dispersion Relations (FDRs) for the full amplitude as well as three Roy equations and three GKPY equations (one each for the S0, P, and S2 partial waves). Roy and GKPY equations are used up to 1.1 GeV, which is their applicability limit.
 The $\pi^0\pi^0$ FDR, dominated by the S0 wave, is applied up to 1.4 GeV as was done in~\cite{Pelaez:2019eqa}. In addition, we have extended up to 1.6 GeV the two FDRs that involve the P wave.

To this end, we have revisited the P-wave parametrization, slightly updating the input in the elastic region with the recent pion form factor data analysis in~\cite{Colangelo:2018mtw}, and 
improving the description above 0.9 GeV. The latter has been achieved by allowing its inelasticity to open at the $\pi \omega$ threshold and using better parametrizations that do not make the uncertainty grow artificially with the energy.
This wave has suffered the largest update, but the parametrizations of other waves have also been modified to become inelastic somewhat below the $K\bar K$ threshold and to have more uniform, smaller, and more realistic uncertainties. Moreover, we have also 
provided new parametrizations for the F and G0 waves, to reproduce their resonant behavior, as well as the G2 wave, since they are needed to test FDRs up to higher energies than before. For this same reason, we have extended the description of all the waves beyond 1.4 GeV, often with more flexible parametrizations.
As a result, we have improved the accuracy of our partial waves and reduced the uncertainty band in the dispersion relations output. Consequently, a better matching with the Regge description than in previous works is required. This is achieved by increasing the matching point, which also allows us to extend the FDR applicability region.
Simultaneously, the stability 
of the fits is improved and the uncertainties near the matching point become smaller.
All in all, the new constrained Global Fits satisfy very well the FDRs even though their uncertainty bands are considerably smaller, particularly for the two relations involving the P wave.

In addition, let us note that we have studied separately three different sets of data, called Solutions I, II, and III here, which give rise to three Global Fits. Their S2, D2, and G waves as well as all other waves up to energies of around $0.9\gev$ are almost indistinguishable, which means that the three of them satisfy Roy and GKPY equations very well.
Above that energy, the situation changes from wave to wave.
For the P wave, the three fits are qualitatively similar but incompatible up to 1.4 GeV, where they start to differ widely. For the F wave, they are qualitatively similar but incompatible within their uncertainties in the whole energy region. The three fits for the D0 wave are indistinguishable for the phase shift up to 1.3 GeV but the elasticity is very different from one another from 0.9 GeV. 

The original data in Solutions II and III had very tiny or lacked uncertainties in some waves.
Even  assuming errors similar to Solution I, 
the average FDR fulfillment is somewhat worse and much less homogeneous for Global Fits II and III. In particular, these two fits have two regions $\sim$100 MeV wide, where it has not been possible to make them satisfy the $I_t=1$ FDR within uncertainties. The first of those regions, sitting around 1 GeV, favors 
the onset of the D0-wave inelasticity at $\sim$0.9 GeV instead of the original onset at the
$K\bar K$ threshold. In addition, the S0-wave elasticity of Solutions II and III  prefers the ``non-dip" scenario above 1 GeV, known to be disfavored by GKPY equations, although we have made them follow the ``dip" scenario. Moreover, when FDRs are imposed, Global Fits II and III deviate somewhat further 
from their original data set than Global Fit I. 
None of these single caveats is enough, by itself, to discard either Global Fit II or III. That is why we have provided them together with Global Fit I. Nevertheless, Global Fit I, free of these caveats, seems to be slightly favored against the other two, and we have used it to illustrate our constrained fitting procedure.

Finally, it is important to emphasize that all the nominal resonance parameters in our parametrizations, ( like $\hat m_\rho$, $m_{f_2}$, $m_{\rho_3}$,  $m_{f_4}$, $\Gamma_{\rho_3}$, $\Gamma_{f_4}$, etc.), are merely parameters of phenomenological parametrizations intended for use on the real axis.
Resonances are rigorously characterized by the position and residue of their associated $T$-matrix pole in the contiguous Riemann sheet.
The naive extrapolation of our phenomenological formulas into the complex $s$-plane to determine resonance pole parameters is highly model-dependent.
This model dependence can be eliminated---or at least significantly reduced---by employing proper analytic continuation techniques, such as partial-wave dispersion relations or other analytic continuation methods applied to the dispersive results on the real axis. 
Such analyses, however, lie beyond the scope of this manuscript and may be pursued in future work along the lines of Refs.~\cite{GarciaMartin:2011jx,Pelaez:2022qby}.

In summary, in this work we provide precise constrained Global Fits that describe the available $\pi\pi\rightarrow\pi\pi$ scattering data up to 1.8 GeV or more, with realistic uncertainty bands. 
In addition, they simultaneously 
satisfy six partial-wave dispersion relations up to 1.1 GeV, the forward dispersion relation for the $\pi^0\pi^0$ 
amplitude up to 1.4 GeV and two other independent forward dispersion relations up to 1.6 GeV.
This is done with relatively simple but well-behaved functions, which we hope make these partial-wave sets
a useful tool to study $\pi\pi$ scattering by itself, but also implement easily and reliably $\pi\pi$ interactions in other hadronic processes.

\begin{acknowledgments} 
We thank, G. Colangelo, C. Hanhart,  M. Hoferichter, B. Kubis, and P. Stoffer for discussions and comments, particularly pointing out possible improvements upon previous works by JRP, JRE, and collaborators. We are particularly indebted to G. Colangelo, M. Hoferichter, and P. Stoffer for providing their results for the pion vector form factor. JRP thanks the kind hospitality of the Helmholtz-Institut f\"ur Strahlen-und Kernphysik (Theorie), Universit\"at Bonn, Germany, where he spent three months, while working on this manuscript, as an invited visiting researcher under the MCIN ``Salvador de Madariaga" Program Grant PRX22/00129.
This work is part of the Grant PID2022-136510NB-C31 funded by MCIN/AEI/ 10.13039/501100011033, it has also received funding from the European Union’s Horizon 2020 research and innovation program under grant agreement No.824093. P. R. is supported
by the MIU (Ministerio de Universidades, Spain) fellowship FPU21/03878, and J.R.E. by the Ram\'on y Cajal program (RYC2019-027605-I) of the Spanish MICIU/AEI. 
\end{acknowledgments}

\appendix

\section{Threshold parameters}\label{app:Threshold}

Our aim in this work has been to obtain simple and global parametrizations
of partial-wave data covering the energy range from threshold to at least 1.8 GeV,
simultaneously consistent with the dispersive constraints.
We have made them continuous and with a continuous derivative (except as required by thresholds).

Previous dispersive analyses were more focused on the low-energy region, particularly below 1.1 GeV, and, up to 0.9 GeV, we have basically mimicked them. Of course, to cover a much wider energy region with just one parametrization, and being just a mimic of the CFD at low energies,
we have sacrificed the precision attained by the CFD. For the $\pi\pi$ threshold parameters, the old CFD is still more accurate, and even more 
when combined with sum rules, as done in~\cite{GarciaMartin:2011cn}.

Nevertheless, we have checked that our threshold parameters are consistent with the values in~\cite{GarciaMartin:2011cn}.
Thus, we have gathered in
Table~\ref{tab:ThresholdParameters} 
the values of threshold parameters for the S, P, D, F, and G waves
that result from our constrained Global Fits. 
We also provide the values used as input, obtained from the CFD parametrization in~\cite{GarciaMartin:2011cn}, except for those of the G waves, which are taken from~\cite{Kaminski:2006qe}.
In addition, we are listing the best values in~\cite{GarciaMartin:2011cn}. These are the most reliable
because they are not only obtained from the CFD but from the use of 
sum rules, and therefore have a much better precision and are parametrization independent.

\begin{widetext}
\begin{center}
 \vspace*{-.5cm}
 \begin{table}[h]
 \renewcommand{\arraystretch}{1.3}
  \centering
  \begin{tabular}{lccccc}
  \multicolumn{6}{c}{Threshold parameters}\\
\toprule
&Global I & Global II &  Global III & Input values \cite{GarciaMartin:2011cn} & Best values \cite{GarciaMartin:2011cn}\\
\toprule
$a^{(0)}_0(\times m_{\pi})$  & $0.225 \pm 0.016$& $0.226 \pm 0.010$&$0.230 \pm 0.011$&$0.221 \pm 0.009$& $0.220 \pm 0.008$\\
$b^{(0)}_0(\times m_{\pi}^3)$ & $0.276 \pm 0.011$& $0.271 \pm 0.008$& $0.271 \pm 0.008$&$0.278 \pm 0.007$& $0.278 \pm 0.005$\\
$a^{(2)}_0(\times m_{\pi})$ & $-0.044 \pm 0.008$& $-0.041 \pm 0.008$&$-0.041 \pm 0.007$&$-0.043 \pm 0.008$&$-0.042 \pm 0.004$ \\
$b^{(2)}_0(\times m_{\pi}^3)$ & $-0.080 \pm 0.009$ &$-0.081 \pm 0.010$ &$-0.079 \pm 0.009$ &$-0.080 \pm 0.009$ &$-0.082 \pm 0.004$\\
$a^{(1)}_1(\times10^3 m_{\pi}^3)$ & $38.7 \pm 1.2$ &$37.6 \pm 1.1$ & $37.8 \pm 1.2$&$38.5 \pm 1.2$ &$38.1 \pm 0.9$\\
$b^{(1)}_1(\times10^3 m_{\pi}^5)$ & $4.9 \pm 0.6$& $4.7 \pm 0.7$& $4.7 \pm 0.7$&$5.07 \pm 0.26$ &$5.37 \pm 0.14$\\
$a^{(0)}_2(\times10^4 m_{\pi}^5)$ & $19.1 \pm 0.5$ &$18.6 \pm 0.4$ &$18.5 \pm 0.4$ &$18.8 \pm 0.4$ &$17.8 \pm 0.3$\\
$b^{(0)}_2(\times10^4 m_{\pi}^7)$ & $-4.4 \pm 0.3$ & $-4.1 \pm 0.3$&$-4.1 \pm 0.3$ &$-4.2 \pm 0.3$ &$-3.5 \pm 0.2$\\
$a^{(2)}_2(\times10^4 m_{\pi}^5)$ & $3.3 \pm 1.3$ &$3.3\pm 1.3$ &$3.1 \pm 1.1$ &$2.8 \pm 1.0$ &$1.85 \pm 0.18$\\
$b^{(2)}_2(\times10^4 m_{\pi}^7)$ & $-3.9 \pm 1.7$&$-3.7 \pm 1.5$ & $-3.5 \pm 1.3$&$-2.8 \pm 0.8$ &$-3.3 \pm 0.1$\\
$a^{(1)}_3(\times10^5 m_{\pi}^7)$ & $5.5 \pm 1.5$ & $5.8 \pm 1.6$&$4.7 \pm 1.1$ &$5.1 \pm 1.3$ & $5.65 \pm 0.21$\\
$b^{(1)}_3(\times10^5m_{\pi}^9)$ & $-4.7 \pm 3.0$ &$-3.8 \pm 2.3$ & $-2.6 \pm 1.5$&$-4.6 \pm 2.5$ & $-4.06 \pm 0.27$\\
$a^{(0)}_4(\times10^6 m_{\pi}^9)$ & $8 \pm 16$ &$8 \pm 16$ &$8 \pm 16$ &$8 \pm 2$ &$8.0 \pm 0.4$ \cite{Kaminski:2006qe}\\
$a^{(2)}_4(\times10^6 m_{\pi}^9)\,\,$ & $6.5 \pm 1.7$& $6.4 \pm 1.5$&$6.1 \pm 1.6$ &$4.5 \pm 1.0 $&$4.5 \pm 0.2$ \cite{Kaminski:2006qe}\\
\botrule
  \end{tabular}
  \caption{Threshold parameters in $m_\pi$ units. Despite the aim of the Global Fits is not precision at low energies, it can be checked that their threshold values are compatible with the input from the CFD and even with the best values in~\cite{GarciaMartin:2011cn,Kaminski:2006qe}. 
  Recall we have used the CFD values of~\cite{GarciaMartin:2011cn} as input for our Global Fits. They do not exist for the G waves, for which we have taken the values from sum rules obtained in \cite{Kaminski:2006qe}
  but with uncertainties enlarged by a factor of 5 following the pattern of the F wave.
}
  \label{tab:ThresholdParameters}
    \vspace*{-.7cm}
\end{table}   
\end{center}
\end{widetext}

Thus, by looking at Table~\ref{tab:ThresholdParameters}, we confirm that the values that result from our Global Fits are all very compatible with their input values. Of course, being a mimic of the CFD, they are not competitive with the original input, and even less competitive 
compared to the best values in~\cite{GarciaMartin:2011cn}.
Still,  the table shows perfect consistency of the Global Fits with the best low-energy information.

\section{The S0-wave parametrization}
\label{app:S0wave}

Throughout this paper, we have kept the very same global parametrization of the S0 wave provided in~\cite{Pelaez:2019eqa}. The reasons are that it was already global, there are no novelties in the data for this wave (contrary to the P wave), the existing parametrizations were sufficiently flexible, and the $K\bar K$ channel dominates the opening of the inelasticity. However, since we have changed the other waves, when imposing the FDRs we have allowed small variations in the S0-wave parameters within uncertainties. As a result, these parameters change slightly, particularly, those affecting the S0 wave above 1.4 GeV. We have gathered the new parameters in Tables~\ref{tab:S0_I},~\ref{tab:S0_II}~and~\ref{tab:S0_III}.
Of course, we use them with the very same parametrization provided in~\cite{Pelaez:2019eqa} (check the erratum too).

\begin{table}[ht]
\renewcommand{\arraystretch}{1.3}
\centering 
\begin{tabular}{c c | c c | c c} 
\hline\hline  
\multicolumn{2}{c}{$t^0_{0,\text{conf}}$} & \multicolumn{2}{c}{$t^0_{f_0}$} & \multicolumn{2}{c}{$\sqrt{s}>1.4 \gev$} \\
\toprule
\rule[-0.05cm]{0cm}{.35cm}$B_0$ & 11.4$\pm$0.3 & $K_0$ &   5.04$\pm$0.28 & $d_0$ &   $-$11.9$\pm$3.7\\
\rule[-0.05cm]{0cm}{.35cm}$B_1$ & $-$0.6$\pm$1.1 & $K_1$ &  $-$4.36$\pm$0.16 &  $d_1$ & $\equiv0$  \\ 
\rule[-0.05cm]{0cm}{.35cm}$B_2$ & 18.6$\pm$2.7  & $K_2$ &     $-$0.05$\pm$0.16 & $d_2$ & $\equiv0$\\ 
\rule[-0.05cm]{0cm}{.35cm}$B_3$ & $-$6.7$\pm$3.1 & $K_3$ &   $-$0.28$\pm$0.06 &  $\epsilon_2$ &   13.7$\pm$4.0 \\ 
\rule[-0.05cm]{0cm}{.35cm}$B_4$ & $-$20.2$\pm$3.7 & &  & $\epsilon_3$ & $\equiv0$\\ 
\rule[-0.05cm]{0cm}{.35cm}$B_5$ &  5.6$\pm$4.8 & $\re \sqrt{s_p}$  & $\equiv \hspace{2.5mm} 0.996$ \gev& $\epsilon_4$ & $\equiv0$\\ 
\rule[-0.05cm]{0cm}{.35cm}$z_0$ &  0.137$\pm$0.028 \gev& $\im \sqrt{s_p}$  & $\equiv -0.025$ \gev& &\\ 
\botrule
\end{tabular} 
\caption{S0-wave parameters of the constrained Global Fit~I, to be used with the parametrization provided in~\cite{Pelaez:2019eqa}.}
\label{tab:S0_I} 
\end{table}

\begin{figure}[H]
\centering
\includegraphics[width=0.48\textwidth]{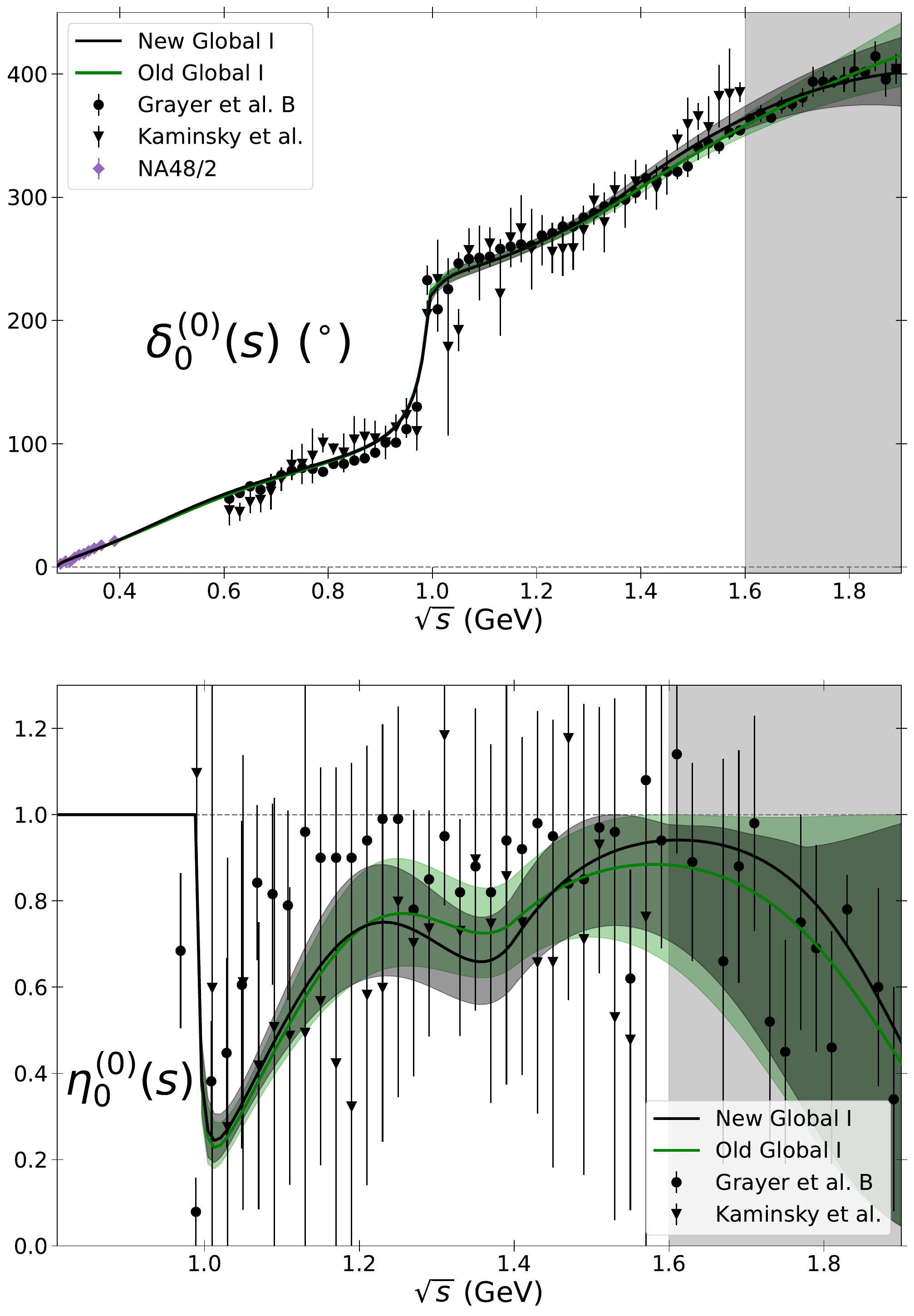}
 \vspace*{-.7cm}
\caption{ S0-wave phase shift (top) and elasticity (bottom). We show the result of our dispersively constrained Global Fit I versus the ``Old Global I" parametrization from~\cite{Pelaez:2019eqa}. Above 1.6 GeV (shaded region) there are no dispersive constraints. The data comes from the Solution B of Grayer et al.~\cite{Grayer:1974cr}, Kaminski et al.~\cite{Kaminski:1996da}, and NA48/2~\cite{Batley:2010zza}.}\label{fig:S0}
\end{figure}

In Fig.~\ref{fig:S0}, we show our slightly updated Global Fit I S0 wave compared with the Global I in~\cite{Pelaez:2019eqa}. As expected, the changes are very small and affect mostly the region above 1.4 GeV.
A similar situation occurs with Global Fits II and III and their counterparts in~\cite{Pelaez:2019eqa}.

\begin{figure}[H]
\centering
\includegraphics[width=0.48\textwidth]{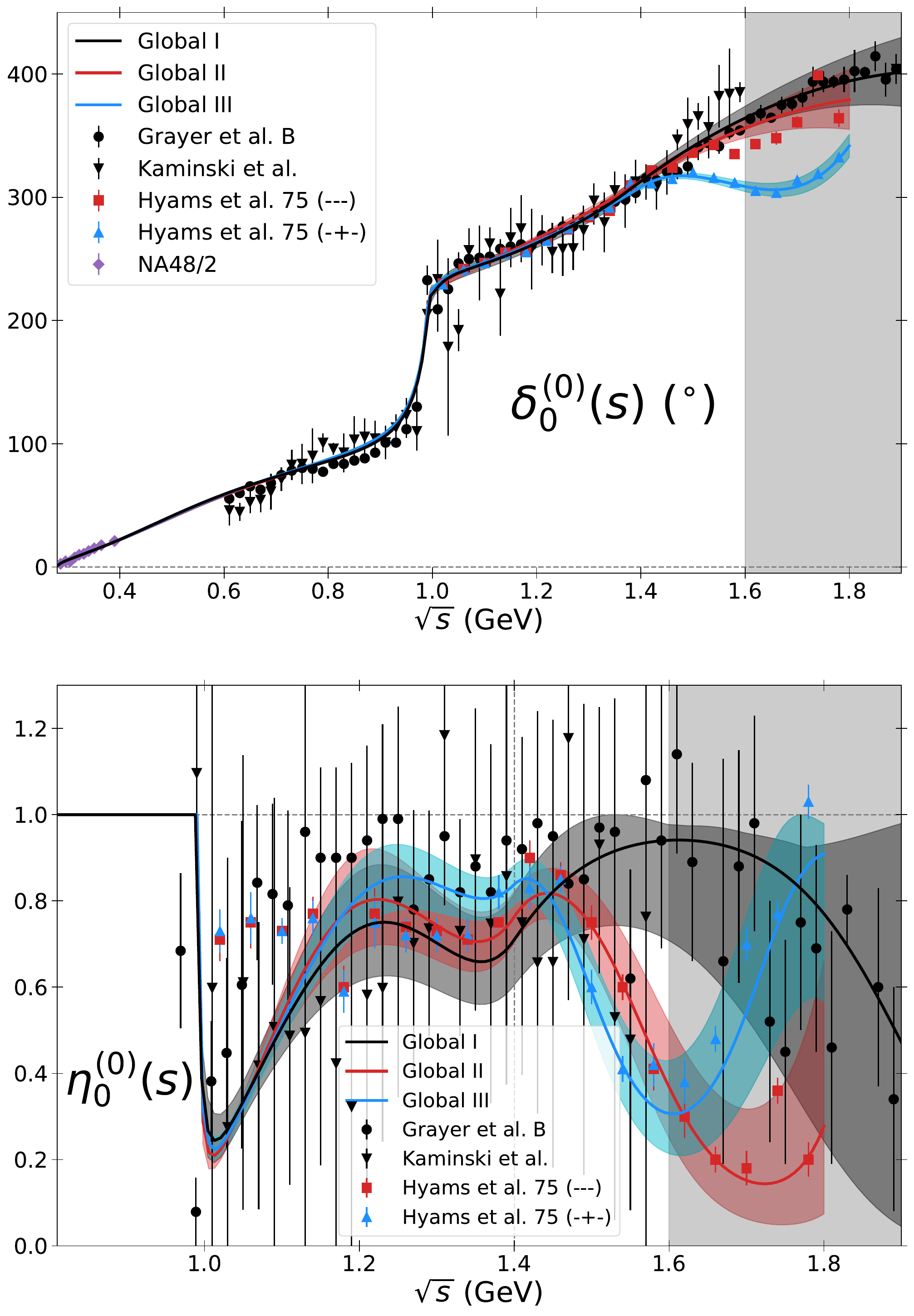}
\caption{ We compare the three S0-wave Global Fits.
They are almost identical to the results obtained in \cite{Pelaez:2019eqa} with only small differences mostly above 1.4 GeV.
Their phase shifts are almost indistinguishable up to 1.4 GeV, where the Global Fit III deviates strongly from the other two.
Their elasticity is compatible up to roughly 1.4 GeV, but then, Global Fits II and III deviate strongly from Global Fit I, becoming much more inelastic. Above 1.6 GeV (shaded region) there are no dispersive constraints. Data for Global Fit I come from Grayer et al.~\cite{Grayer:1974cr} and Kaminski et al.~\cite{Kaminski:1996da}, whereas for 
Global Fits II and III come from Hyams et al. \cite{Hyams:1975mc}.}\label{fig:S0_sols}
\end{figure}

Finally, we show the three Global Fits in Fig.~\ref{fig:S0_sols}. As usual, they are fairly compatible up to 1.4 GeV but differ widely above.
For us here, it is interesting to note that between 1 and 1.10 GeV, the data Solutions II and III of Hyams et al. 75~\cite{Hyams:1975mc} prefer an elasticity $\eta\simeq 0.75$. This value corresponds to the ``non-dip" solution, which is strongly disfavored by Roy and GKPY analyses~\cite{GarciaMartin:2011cn,Moussallam:2011zg}.
Thus, in this work, as in~\cite{Pelaez:2019eqa}, we are imposing in the Global Fits the ``dip-solution" from the CFD in~\cite{GarciaMartin:2011cn}. As a result, the Global Fits II and III have such a dip, although their data do not call for it, while still describing the rest of data from~\cite{Hyams:1975mc} above 1.15 GeV.

\begin{table}[ht]
\renewcommand{\arraystretch}{1.3}
\centering 
\begin{tabular}{c c | c c | c c} 
\hline\hline
\multicolumn{2}{c}{$t^0_{0,\text{conf}}$} & \multicolumn{2}{c}{$t^0_{f_0}$} & \multicolumn{2}{c}{$\sqrt{s}>1.4 \gev$} \\
\hline\hline  
\rule[-0.05cm]{0cm}{.35cm}$B_0$ & 12.1$\pm$0.3 & $K_0$ &   5.02$\pm$0.08 & $d_0$ &   $-$11.1$\pm$6.2\\
\rule[-0.05cm]{0cm}{.35cm}$B_1$ & $-$1.4$\pm$0.8 & $K_1$ &  $-$4.71$\pm$0.08 &  $d_1$ & $\equiv0$  \\ 
\rule[-0.05cm]{0cm}{.35cm}$B_2$ & 14.8$\pm$1.5  & $K_2$ &     0.01$\pm$0.18 & $d_2$ & $\equiv0$\\ 
\rule[-0.05cm]{0cm}{.35cm}$B_3$ & $-$4.9$\pm$1.5 & $K_3$ &   $-$0.36$\pm$0.04 &  $\epsilon_2$ &   81.7$\pm$2.4 \\ 
\rule[-0.05cm]{0cm}{.35cm}$B_4$ & $-$18.9$\pm$1.3 & &  & $\epsilon_3$ & $-$183.8$\pm$8.5\\ 
\rule[-0.05cm]{0cm}{.35cm}$B_5$ &  0.5$\pm$4.5 & $\re \sqrt{s_p}$  &  $\equiv \hspace{2.5mm} 0.996$ \gev &$\epsilon_4$ & $-$51$\pm$25\\ 
\rule[-0.05cm]{0cm}{.35cm}$z_0$ &  0.137$\pm$0.028 \gev& $\im \sqrt{s_p}$  & $\equiv -0.025$ \gev& &\\ 
\hline\hline
\end{tabular} 
\caption{S0-wave parameters of the constrained Global Fit II, to be used with the parametrization provided in~\cite{Pelaez:2019eqa}.}
\label{tab:S0_II} 
\end{table}

\begin{table}[ht]
\renewcommand{\arraystretch}{1.3}
\centering 
\begin{tabular}{c c | c c | c c} 
\hline\hline
\multicolumn{2}{c}{$t^0_{0,\text{conf}}$} & \multicolumn{2}{c}{$t^0_{f_0}$} & \multicolumn{2}{c}{$\sqrt{s}>1.4 \gev$} \\
\hline\hline  
\rule[-0.05cm]{0cm}{.35cm}$B_0$ & 11.9$\pm$0.3 & $K_0$ &   5.28$\pm$0.08 & $d_0$ &   73.3$\pm$1.5\\
\rule[-0.05cm]{0cm}{.35cm}$B_1$ & $-$1.0$\pm$0.9 & $K_1$ &  $-$4.64$\pm$0.04 &  $d_1$ & 27.4$\pm$0.4  \\ 
\rule[-0.05cm]{0cm}{.35cm}$B_2$ & 16.5$\pm$1.7  & $K_2$ &     0.18$\pm$0.07 & $d_2$ & $-$0.27$\pm$0.20\\ 
\rule[-0.05cm]{0cm}{.35cm}$B_3$ & $-$5.3$\pm$1.6 & $K_3$ &   $-$0.37$\pm$0.04 &  $\epsilon_2$ &   171.7$\pm$2.0 \\ 
\rule[-0.05cm]{0cm}{.35cm}$B_4$ & $-$22.7$\pm$1.2 & &  & $\epsilon_3$ & $-$1041$\pm$8\\ 
\rule[-0.05cm]{0cm}{.35cm}$B_5$ &  5.6$\pm$2.8 & $\re \sqrt{s_p}$  &  $\equiv \hspace{2.5mm} 0.996$ \gev& $\epsilon_4$ & 1678$\pm$31\\ 
\rule[-0.05cm]{0cm}{.35cm}$z_0$ &  0.137$\pm$0.028 \gev& $\im \sqrt{s_p}$  &$\equiv -0.025$ \gev& &\\ 
\hline\hline
\end{tabular} 
\caption{S0-wave parameters of the constrained Global Fit III, to be used with the parametrization provided in~\cite{Pelaez:2019eqa}.}
\label{tab:S0_III} 
\end{table}

\vfill\null 

\bibliographystyle{apsrev4-2}
\bibliography{largebiblio.bib}

\end{document}